\documentclass[12pt]{article}

\usepackage[compact]{titlesec}
\usepackage{times}
\newcommand\spacingset[1]{\renewcommand{\baselinestretch}%
{#1}\small\normalsize}

\usepackage[utf8]{inputenc}
\usepackage{natbib}
\usepackage[subrefformat=parens,labelformat=parens]{subfig}
\usepackage{amsmath, comment}
\usepackage{amsfonts}
\usepackage{amsthm}
\usepackage{xargs, xstring}
\usepackage{xcolor}
\usepackage{setspace}
\usepackage{bm}
\usepackage{graphicx, float}
\usepackage[inline]{enumitem}
\usepackage[toc,page,header]{appendix}
\usepackage{minitoc}

\usepackage{longtable}
\usepackage{array}
\newcolumntype{L}[1]{>{\raggedright\let\newline\\\arraybackslash\hspace{0pt}}m{#1}}
\newcolumntype{C}[1]{>{\centering\let\newline\\\arraybackslash\hspace{0pt}}m{#1}}
\newcolumntype{R}[1]{>{\raggedleft\let\newline\\\arraybackslash\hspace{0pt}}m{#1}}

\usepackage{titlesec}
\titleformat*{\section}{\large\bfseries}
\titleformat*{\subsection}{\normalsize\bfseries}
\titleformat{\subsubsection}[runin]
  {\normalfont\normalsize\bfseries}{\thesubsubsection}{1em}{}
\titleformat*{\paragraph}{\normalsize\bfseries}
\titleformat*{\subparagraph}{\normalsize\bfseries}

\makeatletter
\newcommand*\bigcdot{\mathpalette\bigcdot@{.5}}
\newcommand*\bigcdot@[2]{\mathbin{\vcenter{\hbox{\scalebox{#2}{$\m@th#1\bullet$}}}}}
\makeatother

\title{\sc \large {\Large Causal Inference with Spatio-temporal Data:} \\[10pt]
Estimating the Effects of Airstrikes on Insurgent Violence in Iraq\thanks{
This material is based upon work partially supported by the National Science Foundation under Grant No. 2124124, 2124463, and 2124323.
Lyall gratefully acknowledges financial support from the Air Force Office of Scientific Research (Grant $\#$FA9550-14-1-0072).  The findings and conclusions reached here do not reflect the official views or policy of the United States Government or Air Force.  In addition, Imai thanks the Sloan Foundation (\# 2020–13946) for financial support.  The authors would also like to thank Soubhik Barari, Iavor Bojinov, Naoki Egami, Connor Jerzak, Sayar Karmakar, and Neil Shephard for their constructive comments.}}

\author{
Georgia Papadogeorgou\thanks{Assistant Professor, Department of Statistics, University of Florida, Gainesville FL 32611. Email: \href{mailto:gpapadogeorgou@ufl.edu}{gpapadogeorgou@ufl.edu}, URL: \url{https://gpapadogeorgou.netlify.com}} \and Kosuke Imai\thanks{Professor, Department of Government and Department of Statistics, Harvard University. 1737 Cambridge Street, Institute for Quantitative Social Science, Cambridge MA, 02138. Email: \href{mailto:imai@Harvard.Edu}{imai@Harvard.Edu}, URL: \url{https://imai.fas.harvard.edu}} \and Jason Lyall\thanks{James Wright Chair in Transnational Studies and Associate Professor, Department of Government, Dartmouth College, Hanover, NH 03755. Email: \href{mailto:jason.lyall@dartmouth.edu}{jason.lyall@dartmouth.edu}, URL: \url{www.jasonlyall.com}} \and Fan Li\thanks{Professor, Department of Statistical Science, Duke University, Durham, NC 27708. Email: \href{mailto:fl35@duke.edu}{fl35@duke.edu}, URL: \url{http://www2.stat.duke.edu/\~fl35}}}

\date{\today}

\newcommand{\independent}{\perp\!\!\!\perp}
\newcommand\numberthis{\addtocounter{equation}{1}\tag{\theequation}}

\newcommand{\onecov}{X}
\newcommand{\covs}{{\bm \onecov}}
\newcommand{\Var}{\text{Var}}
\newcommand{\bW}{\bm{W}}
\newcommand{\bw}{\bm{w}}
\newcommand{\asymvar}{v}
\newcommand{\tor}{^{(r)}}

\newcommand{\alltimes}{\mathcal{T}}
\newcommand{\alltrt}{\mathcal{W}}
\newcommand{\allout}{\mathcal{Y}}
\newcommand{\allcovs}{\mathcal{\covs}}

\newcommand{\lag}{M}
\newcommandx{\bound}[1][1=Y]{\delta_{#1}}
\newcommandx{\complement}{^{\mathsf{c}}}
\newcommandx{\boundary}[1][1=B]{\partial #1}

\newcommandx{\anyhist}[2][1=t,2=Y]{\overline{#2}_{#1}}
\newcommandx{\whist}[1][1=t]{\anyhist[#1][\bw]}
\newcommandx{\Whist}[1][1=t]{\anyhist[#1][\bW]}
\newcommandx{\history}[1][1=t-1]{\overline{H}_{#1}}

\newcommandx{\sparseset}[2][1=w,2=t]{S_{#1_{#2}{\IfEqCase{#1}{ {Y}{(\whist[#2])}}}}}

\newcommandx{\trtintensity}{h}
\newcommandx{\intervset}{L}
\newcommandx{\notintervset}{{\intervset\complement}}

\newcommandx{\intervdist}[3][1=F,2={},3={}]{ \IfEqCase{#1}{
{F}{F_{\trtintensity_{#2}}^{\includeM[#3]}}
{T}{F_{\mathbf{\trtintensity}{#2}}} }}

\newcommandx{\interv}[3][1=F,2={},3={}]{\intervdist[#1][#2][#3]}

\newcommandx{\intervdistf}[3][1={},2=T,3=t]{f_{\trtintensity_{#1}} \IfEqCase{#2}{{T}{(W_{#3})}} }

\newcommandx{\numpoints}[2][1=t,2=over]{
    \IfEqCase{#2}{{over}{N} {hat}{\widehat{N}}}_{#1}}
\newcommandx{\effpoints}[2][1=t,2=F]{\IfEqCase{#2}{{F}{\tau} {T}{\widehat{\tau}}}_{#1}}
\newcommandx{\ytopower}[1][1=t]{\widehat{Y}_{#1}}
\newcommandx{\includeM}[1][1=1]{\IfEqCase{#1}{{1}{} {#1}{{#1}}}}

\newcommandx{\avgout}[5][1=F, 2={}, 3={},4=t,5=B]{\numpoints[#5 #4][over](\interv[#1][#2][#3])}
\newcommandx{\tempavgout}[4][1=F, 2={}, 3=\lag,4=B]{\numpoints[#4][over](\interv[#1][#2][#3])}

\newcommandx{\effect}[5][1=F,2=F,3=',4='',5=\lag]{\effpoints[B t][#1] (\interv[#2][#3][#5], \interv[#2][#4][#5])}
\newcommandx{\tempeffect}[6][1=F,2=F,3=',4='',5=\lag,6=B]{ \effpoints[#6][#1] (\interv[#2][#3][#5], \interv[#2][#4][#5])}

\newcommandx{\estimatort}[4][1=F,2={},3={},4=F]{ \ytopower[t] (\interv[#1][#2][#3]  \IfEqCase{#4}{{T}{;\omega}})}
\newcommandx{\estimatorNt}[4][1=F,2={},3={},4=B]{\numpoints[#4 t][hat](\interv[#1][#2][#3])}
\newcommandx{\estimator}[4][1=F,2={},3={},4=F]{\ytopower[] (\interv[#1][#2][#3] \IfEqCase{#4}{{T}{;\omega}})}
\newcommandx{\estimatorN}[4][1=F,2={},3=\lag,4=B]{\numpoints[#4][hat] ( \interv[#1][#2][#3])}

\newcommandx{\propscore}[3][1=t,2={},3=w]{e_{#1}({#3}_{#2})}
\newcommandx{\SApropscore}[3][1=t,2={},3=w]{e_{#1}^*({#3}_{#2})}
\newcommandx{\pspar}[1][1=T]{\IfEqCase{#1}{{T}{\bm \gamma} {F}{\gamma}}}
\newcommandx{\parpropscore}[4][1=t,2={},3=w,4=F]{e_{#1}({#3}_{#2}; \IfEqCase{#4}{{T}{\pspar_0} {F}{\pspar}})}
\newcommandx{\scorefun}[5][1=F,2=F,3={},4={},5=t]{\bm \psi_{#4} \big(
    \IfEqCase{#1}{{F}{w_{#5}} {T}{W_{#5}}},
    \IfEqCase{#2}{{F}{\anyhist[#5-1][h]} {T}{\history[#5-1]}};
    \pspar_{#3} \big)}

\newcommandx{\allpars}[1][1=T]{\IfEqCase{#1}{{T}{\bm \theta} {F}{\theta}}}
\newcommandx{\estimeq}[2][1=F,2={}]{s_{#2}(\history[t-1], W_t, Y_t \ ; \ \IfEqCase{#1}{{T}{\allpars_0} {F}{\allpars}})}
\newcommandx{\estimeqsmall}[2][1=F,2={}]{s_{#2}(\anyhist[t-1][h], w_t, y_t \ ; \ \IfEqCase{#1}{{T}{\allpars_0} {F}{\allpars}})}

\usepackage{mathtools}
\DeclarePairedDelimiter{\ceil}{\lceil}{\rceil}

\newcommandx{\quant}[2][1=1,2=t]{A_{#1 #2}}
\newcommandx{\quantt}[1][1=1]{B_{#1}}
\newcommandx{\quanttt}[2][1=1,2=1]{C_{#1 t}^{#2}}
\newcommandx{\error}[1][1=t]{err_{#1}}
\newcommandx{\filtration}[1][1=t-1]{\mathcal{F}_{#1}}
\newcommandx{\spoint}[1][1=in]{s_{Y_t}^{#1}}

\newcommandx{\D}[1][1={}]{\mathcal{D}_{#1}}

\newtheorem{theorem}{Theorem}
\newtheorem{corollary}{Corollary}
\newtheorem{prop}{Proposition}
\newtheorem{lemma}{Lemma}

\newtheorem{assumption}{Assumption}
\newtheorem{remark}{Remark}

\newtheorem*{claim*}{Claim}

\usepackage[colorlinks=true,allcolors=blue]{hyperref}
\usepackage[capitalize,noabbrev]{cleveref}
\crefformat{equation}{#2\textup{(#1)}#3}
\crefformat{section}{\S#2#1#3}
\crefformat{subsection}{\S#2#1#3}
\crefformat{subsubsection}{\S#2#1#3}
\crefformat{assumption}{Assumption #1}
\crefformat{prop}{Proposition #1}
\crefformat{claim}{Claim #1}

\begin{document}
\spacingset{1}

\maketitle
\thispagestyle{empty}
\setcounter{page}{0}

\begin{abstract}
Many causal processes have spatial and temporal dimensions. Yet the classic causal inference framework is not directly applicable when the treatment and outcome variables are generated by spatio-temporal point processes. We extend the potential outcomes framework to these settings by formulating the treatment point process as a stochastic intervention. Our causal estimands include the expected number of outcome events in a specified area under a particular stochastic treatment assignment strategy. Our methodology allows for arbitrary patterns of spatial spillover and temporal carryover effects.  Using martingale theory, we show that the proposed estimator is consistent and asymptotically normal as the number of time periods increases.  We propose a sensitivity analysis for the possible existence of unmeasured confounders, and extend it to the H\'ajek estimator.  Simulation studies are conducted to examine the estimators' finite sample performance.  Finally, we illustrate the proposed methods by estimating the effects of American airstrikes on insurgent violence in Iraq from February 2007 to July 2008.  Our analysis suggests that increasing the average number of daily airstrikes for up to one month may result in more insurgent attacks. We also find some evidence that airstrikes can displace attacks from Baghdad to new locations up to 400 kilometers away.


\end{abstract}

\newpage
\spacingset{1.83}
\section{Introduction}

Many causal processes involve both spatial and temporal dimensions. Examples include the environmental impact of newly constructed factories, the economic and social effects of refugee flows, and the various consequences of disease outbreaks. These applications also illustrate key methodological challenges.  First, when the treatment and outcome variables are generated by spatio-temporal processes, there exists an infinite number of possible treatment and event locations at each point in time.  In addition, spatial spillover and temporal carryover effects are likely to be complex and may not be well understood.

Unfortunately, the classical causal inference framework that dates back to \citet{neym:23} and \citet{fish:35} is not directly applicable to such settings.  Indeed, standard causal inference approaches assume that the number of units that can receive the treatment is finite \citep[e.g.,][]{rubi:74a,robi:97}.  Although a small number of studies develop a continuous time causal inference framework, they do not incorporate a spatial dimension \citep[e.g.,][]{gill:robi:01,zhan:joff:small:11}.  In addition, causal inference methods have been used for analyzing functional magnetic resonance imaging (fMRI) data, which have both spatial and temporal dimensions. For example, \cite{luo2012inference} apply randomization-based inference, while  \citet{sobel2014causal} employ structural modelling. We instead focus on data generated by different underlying processes, leading to new estimands and estimation strategies.

Specifically, we consider settings in which the treatment and outcome events are assumed to be generated by spatio-temporal point processes (Section~\ref{sec:estimands}).  The proposed method is based on a single time series of spatial patterns of treatment and outcome variables, and builds upon three strands of the causal inference literature: interference, stochastic interventions, and time series.

First, we address the possibility that treatments might affect outcomes at a future time period and at different locations in arbitrary ways.  Although some researchers have considered unstructured interference, they assume non-spatial and cross-sectional settings \citep[see e.g.,][and references therein]{bass:airo:18,savj:aron:hudg:19}.  In addition, \citet{aronow2019design} study spatial randomized experiments in a cross-sectional setting, and under the assumption that the number of potential intervention locations is finite and their spatial coordinates are known and fixed. By contrast, our proposed spatio-temporal causal inference framework allows for {\it temporally and spatially unstructured interference} over an infinite number of locations.

Second, instead of separately estimating the causal effects of treatment received at each location, we consider the impacts of different {\it stochastic treatment assignment strategies}, defined formally as the intervention distributions over treatment point patterns. Stochastic interventions have been used to estimate effects of realistic treatment assignment strategies \citep{Diaz2012population, young2014identification, papadogeorgou2019causal} and to address challenging causal inference problems including violation of the positivity assumption \citep{kennedy2019nonparametric}, interference \citep{Hudgens2008,imai:jian:mala:21}, mediation analysis \citep{lok:16, diaz2019causal}, and multiple treatments \citep{imai:jian:19}.  We show that this approach is also useful for causal inference with spatio-temporal treatments and outcomes.

Finally, our methodology allows for arbitrary patterns of spatial and temporal interference. As such, our estimation method does not require the separation of units into minimally interacting sets \citep[e.g.,][]{tchetgen2017auto}. Nor does it rely on an outcome modelling approach that entails specifying a functional form of spillover effects based on, for example, geographic distance. Instead, we view our data as a single time series of maps, which record the locations of treatment and outcome realizations as well as the geographic coordinates of other relevant events. Our estimation builds on the time-series causal inference approach pioneered by \citet{boji:shep:19}.

We propose a spatially-smoothed inverse probability weighting estimator that is consistent and asymptotically normal under a set of reasonable assumptions, regardless of whether the propensity scores are known, or estimated from a correctly specified model (Section~\ref{sec:estimation}). To do so, we establish a new central limit theorem for martingales that can be widely used for causal inference in observational, time series settings.  We also show that the proposed estimator based on the estimated propensity score has a lower asymptotic variance than when the true propensity score is known. This generalizes the existing theoretical result under the independently and identically distributed setting \citep{Hirano2003efficient} to the spatially and temporally dependent setting.  Finally, to assess the potential impact of unobserved confounding, we develop a sensitivity analysis method by generalizing the sensitivity analysis of \citet{rosenbaum2002observational} to our spatio-temporal context and to the H\'ajek estimator with standardized weights (Section~\ref{sec:SA}). We conduct simulation studies to assess the finite sample performance of the proposed estimators (Section~\ref{sec:simulations}).

Our motivating illustration is the evaluation of the effects of American airstrikes on insurgent violence in Iraq from February 2007 to July 2008 (Section~\ref{sec:iraq}). We consider all airstrikes during each day anywhere in Iraq as a {\it treatment pattern}. Instead of focusing on the causal effects of each airstrike, we estimate the effects of different {\it airstrike strategies}, defined formally as the distributions of airstrikes throughout Iraq (Section~\ref{sec:application}). The proposed methodology enables us to capture spatio-temporal variations in treatment effects, shedding new light on how airstrikes affect the location, distribution, and intensity of insurgent violence.

Specifically, under a set of assumptions, our analysis suggests that a higher number of airstrikes, without modifying their spatial distribution, may increase the number of insurgent attacks, especially near Baghdad, Mosul, and the roads between them.  We also find that changing the focal point of airstrikes to Baghdad without modifying the overall frequency can shift insurgent attacks from Baghdad to Mosul and its environs. Under our assumptions, these findings suggest that airstrikes can increase insurgent attacks {\it and} disperse them over considerable distances.  Furthermore, our analysis shows that increasing the number of airstrikes may initially reduce attacks but ultimately increase them over the long run.  Our sensitivity analysis indicates, however, that these findings are somewhat sensitive to the potential existence of unmeasured confounders.  Thus, further analyses are necessary in order for us to reach more definitive conclusions about the impacts of airstrikes.

The proposed methodology has a wide range of applications beyond the specific example analyzed in this paper. For example, the causal effects of pandemics and crime on a host of economic and social outcomes could be evaluated using our methodology. With the advent of massive and granular data sets, we expect the need to conduct causal analysis of spatio-temporal data will only continue to grow.

\section{Motivating Application: Airstrikes and Insurgent Activities in Iraq}
\label{sec:iraq}

Airstrikes have emerged as a principal tool for fighting against insurgent and terrorist organizations in civil wars around the globe. In the past decade alone, the United States has conducted sustained air campaigns in at least six different countries, including Afghanistan, Iraq, and Syria. Although it has been shown that civilians have all-too-often borne the brunt of these airstrikes \citep{Lyall:19b}, we have few rigorous studies that evaluate the impact of airstrikes on subsequent insurgent violence. Even these studies have largely reached opposite conclusions, with some claiming that airpower reduces insurgent attacks while others arguing they spark escalatory spirals of increased violence \citep[e.g.,][]{Lyall:19a,Mir:19,Dell:18,Kocher:11}.  

Moreover, all existing studies have two interrelated methodological shortcomings: they carve continuous geographic space into discrete, often arbitrary, units, and they make simplifying assumptions about patterns of spatial and temporal interference. \citet{mir_moore_19}, for example, argue that drone strikes in Pakistan have reduced terrorist violence. But they use a coarse estimation strategy that bins average effects of drone strikes into broad half-year increments over entire districts that cannot capture local spatial and temporal dynamics. Similarly, \citet{rigterink_21} draws on 443 drone strikes to estimate airstrike effects on 13 terrorist groups in Pakistan, concluding that they have mixed effects. Yet her group-month estimation strategy cannot detect spillover effects nor accurately capture the timing of insurgent responses. In short, we need a flexible methodological approach that avoids the pitfalls of binning treatment and outcome measures into too-aggregate, possibly misleading, temporal and spatial units. 

We enter this debate by examining the American air campaign in Iraq. We use declassified US Air Force data on airstrikes and shows of force (simulated airstrikes where no weapons are released) for the February 2007 to July 2008 period. The period in question coincides with the ``surge'' of American forces and airpower designed to destroy multiple Sunni and Shia insurgent organizations in a bid to turn the war's tide.

Aircraft were assigned to bomb targets via two channels. First, airstrikes were authorized in response to American forces coming under insurgent attack. These close air support missions represented the vast majority of airstrikes in 2007--08. Second, a small percentage (about 5\%) of airstrikes were pre-planned against high-value targets, typically insurgent commanders, whose presence had been detected from intercepted communications or human intelligence.  In each case, airstrikes were driven by insurgent attacks that were either ongoing or had occurred in the recent past in a given location. As a result, the models used later in this paper adjust for prior patterns of insurgent violence in a given location for several short-term windows.

We also account for prior air operations, including shows of force, by American and allied aircraft. Insurgent violence in Iraq is also driven by settlement patterns and transportation networks.  Our models therefore include population size and location of Iraqi villages and cities as well as proximity to road networks, where the majority of insurgent attacks were conducted against American convoys. Finally, prior reconstruction spending might also drive the location of airstrikes. Aid is often provided in tandem with airstrikes to drive out insurgents, while these same insurgents often attack aid locations to derail American hearts-and-minds strategies. Taken together, these four factors---recent insurgent attacks, the presence of American forces, settlement patterns, and prior aid spending---drove decisions about the location and severity of airstrikes.  We emphasize that we may not observe all factors used for decisions on airstrikes.  We will address this limitation by developing and applying a sensitivity analysis.

\begin{figure}[!t]
\centering \vspace{-.5in}
\subfloat[Airstrikes over time]{
\includegraphics[width=0.5\textwidth]{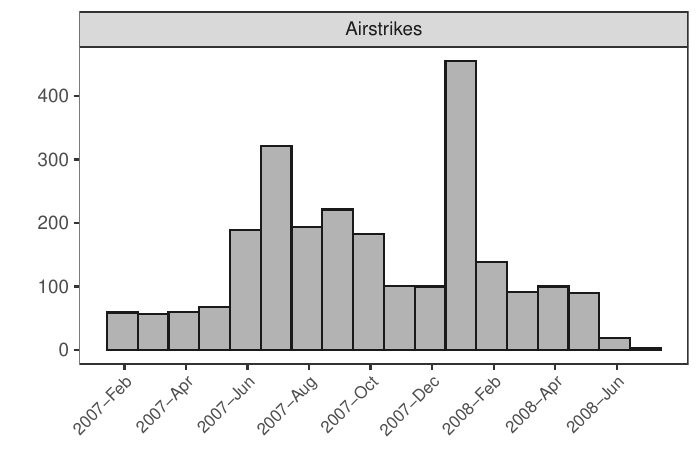}\label{fig:air_time}}
\hspace{0.25in}
\subfloat[Airstrikes over space]{
\includegraphics[width=0.35\textwidth]{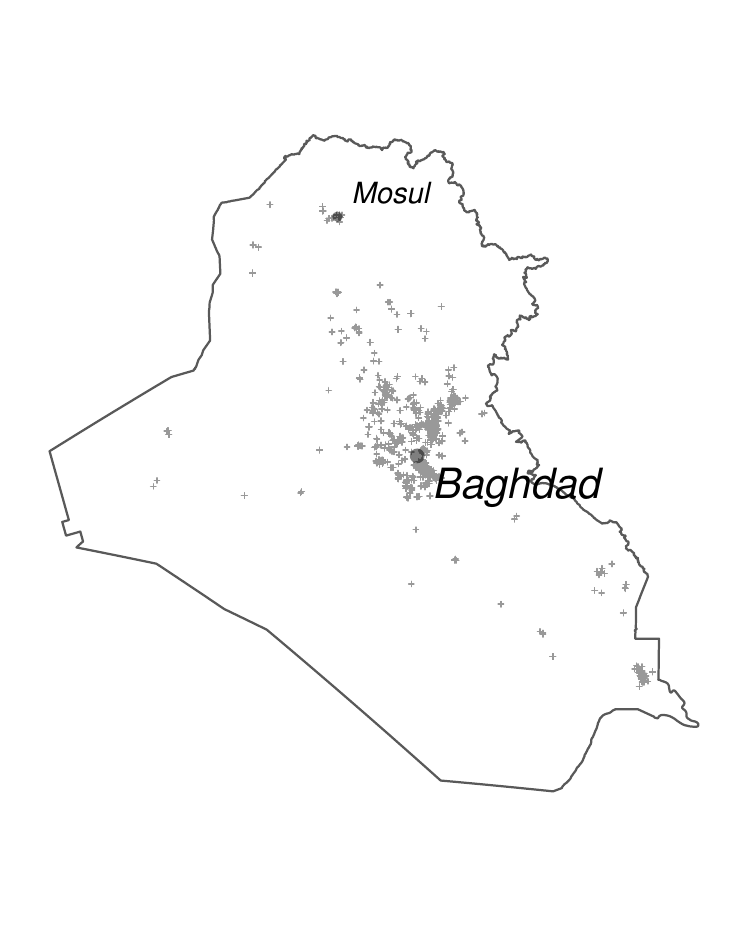}\label{fig:air_space}
}\\
\subfloat[Insurgent violence over time]{
\includegraphics[width=0.5\textwidth]{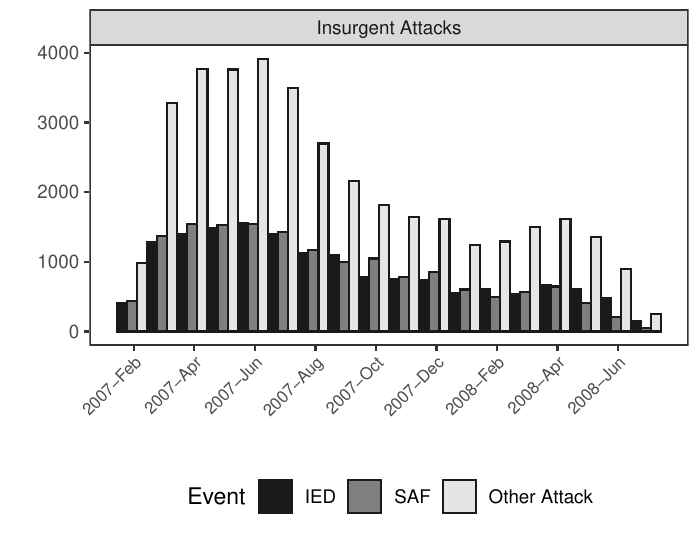}\label{fig:ins_time}}
\hspace{0.25in}
\subfloat[Insurgent violence over space]{
\includegraphics[width=0.35\textwidth]{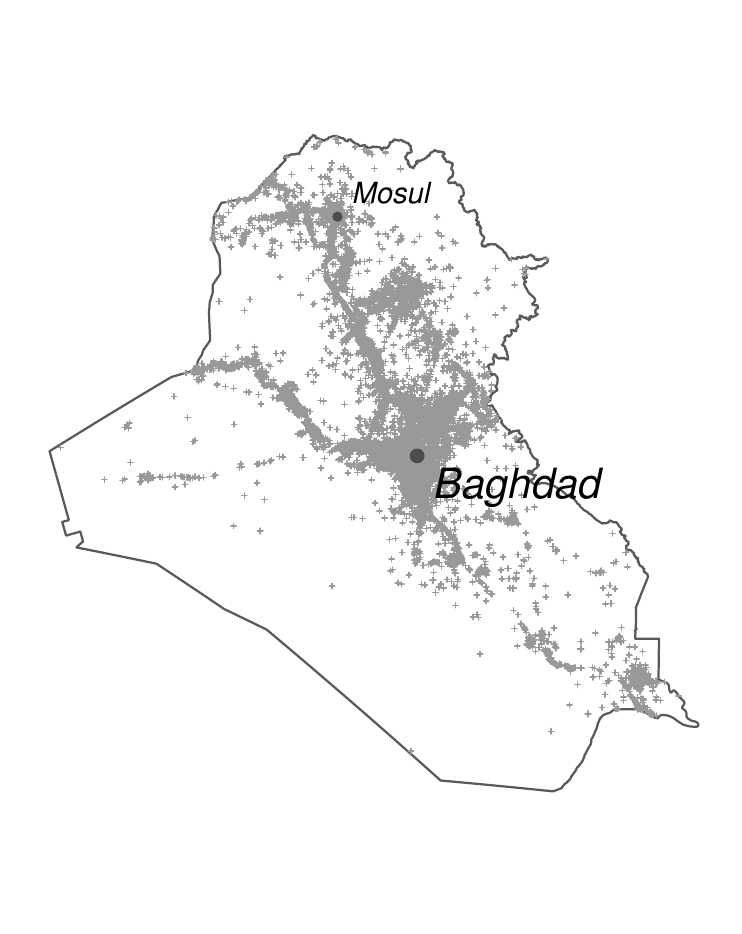}
\label{fig:ins_space}}
\caption{Distribution of the treatment and outcome point processes over time and space. Plots~(b)~and~(d) show the locations of airstrikes and insurgent attacks, respectively, during the time period February 23, 2007 to July 05, 2008.  Insurgent attacks are sorted into one of three categories: Improvised Explosive Devices (IEDs), Small Arms Fire (SAF), and other attacks.}
\label{fig:observed_intensities}
\end{figure}

Figure~\ref{fig:observed_intensities} summarizes the spatial and temporal distributions of airstrikes (treatment variable) and insurgent violence (outcome variable). Figure~\ref{fig:air_time} presents the temporal distribution of airstrikes recorded by the US Air Force each month. There were a total of 2,246 airstrikes during this period. Figure~\ref{fig:air_space} plots the spatial density of these airstrikes across Iraq, with spatial clustering observed around Baghdad and the neighboring ``Sunni Triangle,'' a hotspot of insurgency. Figure~\ref{fig:ins_time} plots the monthly distribution of insurgent attacks by type: Improvised Explosive Devices (IEDs), small arms fire (SAF), and other attacks. A total of 68,573 insurgent attacks were recorded by the US Army's CIDNE database during this time period.  Finally, Figure~\ref{fig:ins_space} plots the locations of insurgent attacks across Iraq. Baghdad, the Sunni Triangle, and the highway leading north to Mosul are all starkly illustrated.

\section{Causal Inference Framework for Spatio-temporal Data}
\label{sec:estimands}

In this section, we propose a causal inference framework for spatio-temporal point processes.  We describe the setup, and define causal estimands based on stochastic interventions.

\subsection{The Setup}

We represent the locations of airstrikes for each time period (e.g., day) as a spatial point pattern measured at time $t \in \alltimes = \{1, 2, \dots, T\}$ where $T$ is the total number of the discrete time periods.  Let $W_t(s)$ denote the binary treatment variable at location $s$ for time period $t$, indicating whether or not the location receives the treatment during the time period.  We use $W_t$ as a shorthand for $W_t(\Omega)$, which evaluates the binary treatment variable $W_t(s)$ for each element $s$ in a set $\Omega$. The set $\Omega$ is {\it not} assumed to be a finite grid, but it is allowed to include an infinite number of locations that may receive the treatment. In addition, $\alltrt$ represents the set of all possible point patterns at each time period where, for simplicity, we assume that this set does not vary across time periods, i.e., $W_t \in \alltrt$ for each $t$.  The set of {\it treatment-active locations}, i.e., the locations that receive the treatment, at time $t$ is denoted by $\sparseset[W] = \{s \in \Omega : W_t(s) = 1\}$.  We assume that the number of treatment-active locations is finite for each time period, i.e., $|\sparseset[W]| < \infty$ for any $t$.  In our study, the treatment-active locations correspond to the set of coordinates of airstrikes. Finally, $\Whist = (W_1, W_2, \dots, W_t)$ denotes the collection of treatments over the time periods $1, 2, \dots, t$.

We use $w_t$ to represent a realization of $W_t$ and $\whist = (w_1, w_2, \dots, w_t)$ to denote the history of treatment point pattern realizations from time 1 through time $t$.  Let $Y_t(\whist)$ represent the potential outcome at time $t \in \alltimes$  for any given treatment sequence $\whist \in \alltrt^t = \alltrt \times \cdots \times \alltrt$, depending on {\it all} previous treatments.  Similar to the treatment, $Y_t(\whist)$ represents a point pattern with locations $\sparseset[Y]$, which are referred to as the \textit{outcome-active locations}. In our study, $\sparseset[Y]$ represents the locations of insurgent attacks if the patterns of airstrikes had been $\whist[t]$.  Let $\anyhist[T][\allout] = \{ Y_{t}(\whist[t]): \whist[t] \in \alltrt^{t}, t \in \alltimes \}$ denote the collection of potential outcomes for all time periods and for all treatment sequences.

Among all of these potential outcomes for time $t$, we only observe the one corresponding to the observed treatment sequence, denoted by $Y_t=Y_t(\Whist)$. We use $\anyhist[t][\bm Y] = \{Y_{1}, Y_2, \ldots, Y_{t}\}$ to represent the collection of observed outcomes up to and including time period $t$.
In addition, let $\covs_t$ be the set of possibly time-varying confounders that are realized prior to $W_t$ but after $W_{t-1}$.  No assumption is necessary about the temporal ordering of any variables in $\covs_t$ and $Y_{t-1}$.
Let $\anyhist[T][\allcovs] = \{\covs_t(\whist[t-1]): \whist[t-1] \in \alltrt^{t-1}, t \in \alltimes \}$ be the set of potential values of $\covs$ under any possible treatment history and for all time periods.
We also assume that the observed covariates correspond to the covariates under the observed treatment path, $\covs_t = \covs_t(\Whist[t-1])$, and use $\anyhist[t][\covs] = (\covs_1, \covs_2,\ldots,\covs_t)$ to denote the collection of observed covariates over the time periods $1, 2, \ldots, t$.  Finally, we use $\history[t] = \{\anyhist[t][\bW], \anyhist[t][\bm Y], \anyhist[t+1][\covs] \}$ to denote all observed history preceding the treatment at time $t + 1$.

Since our statistical inference is based on a single time series, we consider all potential outcomes and potential values of the time-varying confounders as {\it fixed}, pre-treatment quantities. Then, the randomness we quantify is with respect to the assignment of treatment $W_t$ given the complete history including all counterfactual values $\history[t-1]^*$ where $\history[t]^* = \{\Whist[t], \anyhist[T][\allout], \anyhist[T][\allcovs] \}$ and $\history[t] \subset \history[t]^*$.

\subsection{Causal Estimands under Stochastic Interventions}

A notion central to our proposed causal inference framework is \emph{stochastic intervention}. Instead of setting a treatment variable to a fixed value, a stochastic intervention specifies the probability distribution that generates the treatment under a potentially counterfactual scenario.  Although our framework accommodates a large class of intervention distributions, for concreteness, we consider intervention distributions based on Poisson point processes, which are fully characterized by an intensity function $\trtintensity: \Omega \rightarrow [0, \infty)$.  For example, a homogeneous Poisson point process with $\trtintensity(s) = \trtintensity$ for all $s \in \Omega$, implies that the number of treatment-active locations follows a $\text{Poisson}(\trtintensity |\Omega|)$ distribution, with locations distributed independently and uniformly over $\Omega$. In general, the specification of stochastic intervention should be motivated by policy or scientific objectives. Such examples in the context of our study are given in Section~\ref{subsec:study_interventions}.

Our causal estimands are the expected number of (potential) outcome-active locations under a specific stochastic intervention of interest, and the comparison of such quantities under different intervention distributions.  We begin by defining the causal estimands for a stochastic intervention taking place over a single time period. Let $\intervdist[F]$ denote the distribution of a spatial point process with intensity $\trtintensity$. Also, let $N_B(\cdot)$ denote a counting measure on a region $B \subset \Omega$. Then, we can define the expected number of outcome-active locations for a region $B$ at time $t$ as
\begin{equation}
\small
\begin{aligned}
\avgout[F][][1] & \ = \ \int_{\alltrt} N_B\left( Y_t \left(\Whist[t-1], w_t \right)\right) \ \mathrm{d} \intervdist(w_t)
 \ = \ \int_{\alltrt} \left| S_{Y_t \left(\Whist[t-1], w_t\right)} \cap B \right| \ \mathrm{d} \intervdist(w_t).
\end{aligned}
\label{eq:avgout_k1}
\end{equation}
In our application, this quantity represents the expected number of insurgent attacks within a region of Iraq $B$ if the airstrikes at time $t$ were to follow the point process specified by $\intervdist$, given the observed history of airstrikes up to time $t - 1$. The region $B$ does not need to be defined as a connected subset of $\Omega$, and it can be the union of potentially non-bordering sets (for example, the suburbs of two cities).

\begin{figure}[!b]
\centering
\includegraphics[width=\textwidth]{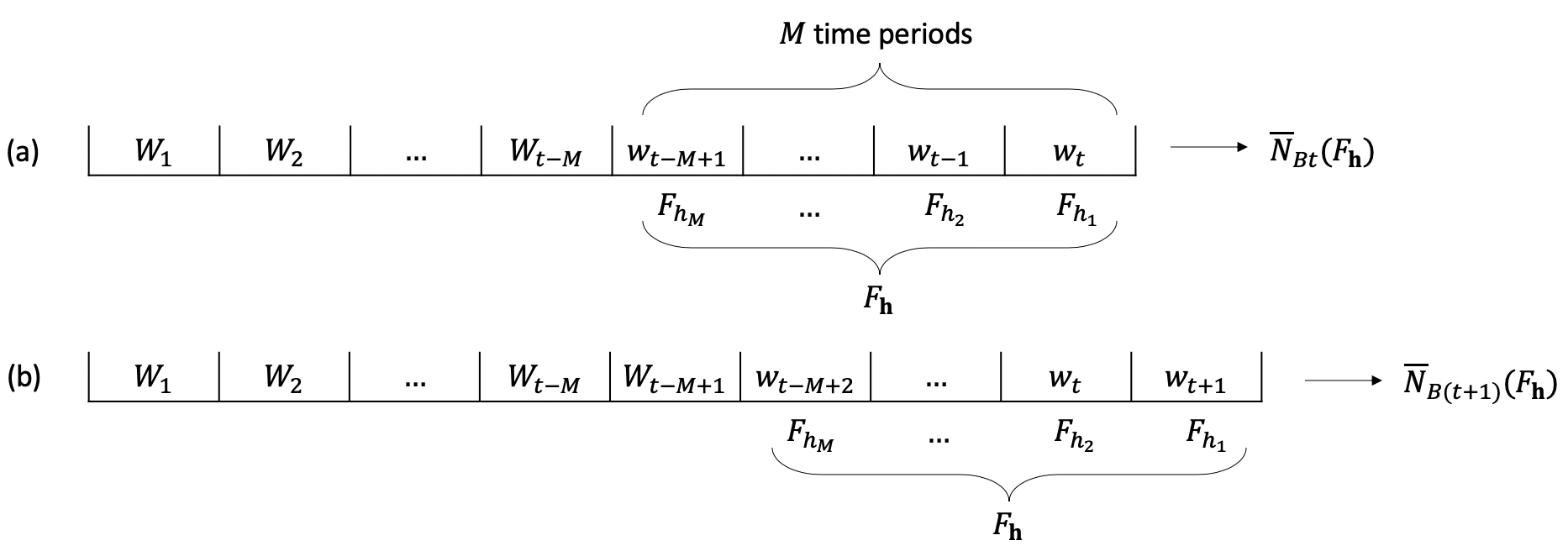}
\caption{Graphical Illustration of Stochastic Intervention over Multiple Time Periods for Time Period $t$ and $t + 1$. Under intervention $\interv[T]$, treatments during time periods $t - \lag + 1, \dots, t-1, t$ are assigned according to distributions $\intervdist[F][\lag], \dots, \intervdist[F][2], \intervdist[F][1]$.}
\label{fig:trt_intervention}
\end{figure}

We can extend the above estimand to an intervention taking place over $\lag$ consecutive time periods. Consider an intervention, denoted by $\intervdist[T] = \intervdist[F][1] \times \dots \times \intervdist[F][\lag]$, under which the treatment at time $t$ is assigned according to $\intervdist[F][1]$, at time $t - 1$ according to $\intervdist[F][2]$, continuing until time period $t - \lag + 1$ for which treatment is assigned according to $\intervdist[F][\lag]$. A treatment path based on this intervention is displayed in \cref{fig:trt_intervention}(a). Then, we define a general estimand as
\begin{equation}
\begin{aligned}
    \avgout[T]
    & \ = \ \int_{\alltrt^\lag} N_B\left( Y_t \left(\Whist[t-\lag], w_{t-\lag+1}, \dots, w_t \right)\right) \ \mathrm{d} \intervdist[F][1](w_t) \cdots \mathrm{d} \intervdist[F][\lag](w_{t-\lag+1}) \\
    & \ = \ \int_{\alltrt^\lag} \left| S_{Y_t\left(\Whist[t-\lag], w_{t-\lag+1}, \dots, w_t \right)} \cap B \right| \ \mathrm{d} \intervdist[F][1](w_t) \cdots \mathrm{d} \intervdist[F][\lag](w_{t-\lag+1}).
\end{aligned}
\label{eq:avgout_k}
\end{equation}
This quantity represents the expected number of outcome events within region $B$ and at time $t$ if the treatment point pattern during the previous $\lag$ time periods was to follow the stochastic intervention with distribution $\intervdist[T]$. Treatments during the initial $t-\lag$ time periods were the same as observed. A special case of $\interv[T]$ assumes that treatments during the $\lag$ time periods are independent and identically distributed draws from the same distribution $\intervdist$, which we denote by $\intervdist[T] = \intervdist[F][][\lag]$.

Given the above setup, we define the average treatment effect of stochastic intervention $\intervdist[T][']$ versus $\intervdist[T]['']$ for a region $B$ at time $t$ as
\begin{equation}
\effect[F][T] \ = \ \avgout[T][''] - \avgout[T]['],
\label{eq:effect_k}
\end{equation}
where ${\mathbf{\trtintensity}}' = (\trtintensity_1', \trtintensity_2', \dots, \trtintensity_\lag')$ represents a collection of treatment intensities over $\lag$ consecutive time periods (similarly for $\mathbf \trtintensity''$).

We further consider the average, over time periods $t=\lag, \lag +1, \ldots, T$, of the expected potential outcome for region $B$ at each time period if treatments during the $\lag$ proceeding time periods arose from $\interv[T]$. This quantity is defined as
\begin{equation}
    \tempavgout[T] \ = \ \frac1{T - \lag + 1} \sum_{t = \lag}^T \avgout[T].
    \label{eq:tempavgout_k}
\end{equation}
\cref{fig:trt_intervention} shows two of the terms averaged in Equation~\cref{eq:tempavgout_k}, i.e., $\avgout[T]$ and $\avgout[T][][][(t+1)]$. For $\avgout[T]$, treatments up to $t-\lag$ are set to their observed values, and treatments at time periods $t-\lag +1, \dots, t$ are drawn from $\interv[T]$. The same definition applies to $\avgout[T][][][(t +1)]$, but intervention time periods are shifted by 1: treatments up to $t - \lag + 1$ are set to their observed values, while treatments during time periods $t - \lag + 2, \dots, t + 1$ are drawn from $\interv[T]$. In Equation~\cref{eq:tempavgout_k}, the summation starts at $t = \lag$ since the quantity $\avgout[T]$ assumes that there exist $\lag$ prior time periods during which treatments are intervened on. We suppress the dependence of $\tempavgout[T]$ on $T$ for notational simplicity.

Similarly, based on $\tempavgout[T]$, we define the causal effect of intervention $\interv[T][']$ versus $\interv[T]['']$ as
\begin{equation}
\begin{aligned}
    \tempeffect[F][T][']['']
    & \ = \ \tempavgout[T][''] - \tempavgout[T]['] \ = \ \frac1{T - \lag + 1} \sum_{t = \lag}^T \effect[F][T].
    \label{eq:temp_effect_k}
\end{aligned}
\end{equation}
This estimand represents the average, over time periods $t=\lag, \lag + 1, \ldots, T$, of the expected change in the number of points at each time period when the observed treatment path $\Whist[T]$ was followed until $t - \lag$ with subsequent treatments $W_{t - \lag +1}, \ldots, W_t$ arising according to $\interv[T][']$ versus $\interv[T]['']$.

The effect size of a point pattern treatment would depend on $\lag$, and a greater value of $\lag$ allows one to study slow-responding outcome processes. Moreover, specifying $\interv[T][']$ and $\interv[T]['']$ such that they are identical except for the assignment at $\lag$ time periods prior, $\trtintensity_\lag', \trtintensity_\lag''$, yields the lagged effect of a treatment change, which resembles the lagged effects defined by \cite{boji:shep:19} for binary treatments and non-stochastic interventions.

The above estimands are defined while conditioning on the treatments of all previous time periods. This is important because we do not want to restrict the range of temporal carryover effects.  Although the proposed estimand is generally data-dependent, the quantity becomes fixed under some settings. For example, if the potential outcomes at time $t$ are restricted to depend at most on the latest $L$ treatment point patterns, then the estimands for stochastic interventions that take place over $\lag \geq L$ time periods will no longer depend on the observed treatment path.

\section{Estimation and Inference}
\label{sec:estimation}

In this section, we introduce a set of causal assumptions and the proposed estimator that combines inverse probability of treatment weighting with kernel smoothing. We then derive its asymptotic properties. All proofs are given in \cref{app_sec:proofs}.

\subsection{The Assumptions}

Similar to the standard causal inference settings, variants of the unconfoundedness and overlap assumptions based on stochastic interventions are required for the proposed methodology. For simplicity, we focus on stochastic interventions with identical and independent distribution over $\lag$ periods, $\interv[T] = \interv[F][][\lag]$, and intensity $\trtintensity$. Our theoretical results, however, extend straightforwardly to stochastic interventions with non-i.i.d. treatment patterns.

\begin{assumption}[Unconfoundedness]\spacingset{1.25}
The treatment assignment at time $t$ does not depend on any, past or future, potential outcomes and potential confounders conditional on the observed history of treatments, confounders and outcomes up to time $t-1$:
\[
f(W_t \mid \Whist[t-1], \anyhist[T][\allout], \anyhist[T][\allcovs]) \ = \ f(W_t \mid \history).
\]
\label{ass:unmeasured_conf}
\end{assumption}
\vspace{-30pt}
\noindent \cref{ass:unmeasured_conf} resembles the sequential ignorability assumption in the standard longitudinal settings \citep{robins1999association, robins2000marginal}, but it is more restrictive. The assumption requires that the treatment assignment does not depend on both past and future potential values of the time-varying confounders as well as those of the outcome variable, conditional on their past observed values. In contrast, the standard sequential ignorability assumption only involves future potential outcomes.

Unfortunately, sequential ignorability would not suffice in the current setting. The reason is that we utilize data from a {\it single} unit measured repeatedly over many time periods to draw causal conclusions.  This contrasts with the typical longitudinal settings where data are available on a large number of independent units over a short time period. Our assumption is similar to the non-anticipating treatment assumption of \cite{boji:shep:19} for binary non-stochastic treatments, while explicitly showing the dependence on the time-varying confounders. By requiring the treatment to be conditionally independent of the time-varying confounders, we assume that all ``back-door paths'' from treatment to either the outcome or the time-varying confounders are blocked \citep{pearl2000causality}.

Next, we consider the overlap assumption, also known as positivity, in the current setting.  We define the probability \textit{density} of treatment realization $w$ at time $t$ given the history, $\propscore = f(W_t = w \mid \history)$, as the propensity score at time period $t$. Also, let $\intervdistf[][F]$ denote the probability density function of the stochastic intervention $\interv[F]$.  The assumption requires the ratio of propensity score over the density for the stochastic intervention, rather than the propensity score itself, is bounded away from zero.  
\begin{assumption}[Bounded relative overlap] \spacingset{1.25}
There exists a constant $\bound[W] > 0$ such that $\propscore > \bound[W] \cdot \intervdistf[][F](w)$  for all $w \in \alltrt$.
\label{ass:positivity}
\end{assumption}
\noindent Assumption~\ref{ass:positivity} ensures that all the treatment patterns which are possible under the stochastic intervention of interest can also be observed. This assumption enforces that the support of the intervention distribution has to be included in the support of the propensity score, and does not allow for interventions that assign positive mass to fixed treatments $w$.

\subsection{The Propensity Score for Point Process Treatments}
\label{sec:balance}

The propensity score plays an important role in our estimation.  Here, we show that the propensity score for point process treatments has two properties analogous to those of the standard propensity score \citep{rosenbaum1983central}.  That is, the propensity score is a balancing score, and under \cref{ass:unmeasured_conf} the treatment assignment is unconfounded conditional on the propensity score.
\begin{prop} \spacingset{1.25}
The propensity score $\propscore$ is a balancing score. That is, 
$f(W_t = w \mid \propscore, \history) \ = \ f(W_t = w \mid \propscore)$ holds for all $t$.
\label{theorem:balancing_score}
\end{prop}
\noindent In practice, \cref{theorem:balancing_score} allows us to empirically assess the propensity score model specification by checking the predictive power of covariates in $\history$ for the treatment $W_t$ conditional on the propensity score. For example, if a covariate significantly improves prediction in a point process model of $W_t$ after adjusting for the estimated propensity score, then the covariate is not balanced and the propensity score model is likely to be misspecified. 

\begin{prop} \label{theorem:ps_unconfoundedness}\spacingset{1.25}
Under \cref{ass:unmeasured_conf}, the treatment assignment at time $t$ is unconfounded given the propensity score at time $t$, that is, given
\( \displaystyle
f(W_t \mid \Whist[t-1], \anyhist[T][\allout], \anyhist[T][\allcovs]) \ = \ f(W_t \mid \history),
\)
we have
\begin{equation*}
f(W_t \mid \Whist[t-1], \anyhist[T][\allout], \anyhist[T][\allcovs]) \ = \ f(W_t \mid \propscore[t][t][W]).
\end{equation*}
\end{prop}
\noindent \cref{theorem:ps_unconfoundedness} shows that the potentially high-dimensional sets, $\history^*$ and $\history$, can be reduced to the one dimensional propensity score $\propscore$ as a conditioning set sufficient for estimating the causal effects of $W_t$. 

\subsection{The Estimators}

To estimate the causal estimands defined in Section~\ref{sec:estimands}, we propose propensity-score-based estimators that combine the inverse probability of treatment weighting (IPW) with the kernel smoothing of spatial point patterns.  The estimation proceeds in two steps. First, at each time period $t$, the surface of outcome-active locations is spatially smoothed according to a chosen kernel. Then, this surface is weighted by the relative density of the observed treatment pattern under the stochastic intervention of interest and under the actual data generating process.

An alternative approach would be the direct modelling of the outcome.  For example, one would model the outcome point process as a function of the past history following the $g$-computation in the standard longitudinal settings \citep{robins1986new}.  However, such an approach would require an accurate specification of spatial spillover and temporal carryover effects.  This is a difficult task in many applications.  Instead, we focus on modelling the treatment assignment mechanism.

Formally, consider a univariate kernel $K: [0,\infty) \rightarrow [0, \infty)$ satisfying $\int K(u) \mathrm{d}u = 1$. Let $K_b$ denote the scaled kernel defined as $K_b(u) = b^{-1} K(u / b)$ with bandwidth parameter $b$. 
We define $\estimatort[F][][\lag]: \Omega \rightarrow \mathbb{R}^{+}$ as
\begin{equation}
\begin{aligned}
\estimatort[F][][\lag][T] \ = \
\prod_{j = t - \lag + 1}^t
\frac{\intervdistf[][T][j]}{\propscore[j][j][W]}
\Bigg[ \sum_{s \in \sparseset[{}Y][t]} K_b(\|\omega - s\|) \Bigg],
\end{aligned}
\label{eq:estimatort}
\end{equation}
where $\|\cdot\|$ denotes the Euclidean norm. The summation represents the spatially-smoothed version of the outcome point pattern at time period $t$.  The product of ratios represents a weight similar to those in the marginal structural models \citep{robins2000marginal}, but in accordance with the stochastic intervention $\interv[F][][\lag]$: each of the $\lag$ terms represents the likelihood ratio of treatment $W_j$ in the counterfactual world of the intervention $\interv$ versus the actual world with the observed data at a specific time period.

Assuming that the kernel $K$ is continuous, the estimator given in Equation~\cref{eq:estimatort} defines a continuous surface over $\Omega$. The continuity of $\estimator[F][][\lag]$ allows us to use it as an intensity function when estimating causal quantities. This leads to the following estimator for the expected number of outcome-active locations in any region $B$ at time $t$, defined in Equation~\cref{eq:avgout_k},
\begin{equation}
\estimatorNt[F][][\lag] \ = \ \int_B \estimatort[F][][\lag][T] \ \mathrm{d} \omega.
\label{eq:estimatorNt}
\end{equation}
We can now construct the following estimator for the temporally-expected average potential outcome defined in Equation~\cref{eq:tempavgout_k},
\begin{equation}
\estimatorN[F][][\lag] \ = \
\frac1{T-\lag+1} \sum_{t = \lag}^T \estimatorNt[F][][\lag].
\label{eq:estimatorN}
\end{equation}
We estimate the causal contrast between two interventions $\interv[F][1][\lag]$ and $\interv[F][2][\lag]$ defined in Equation~\cref{eq:temp_effect_k} as,
\begin{equation}
\tempeffect[T][F][1][2][\lag] \ = \ \estimatorN[F][2][\lag] - \estimatorN[F][1][\lag].
\label{eq:estimator_effect}
\end{equation}

An alternative estimator of $\avgout[F][][\lag]$ could be obtained by replacing the kernel-smoothed version of the outcome in Equation~\cref{eq:estimatort} with the number of observed outcome active locations in $B$ at time $t$. Even though this estimator has the same asymptotic properties discussed below, the kernel-smoothing of the outcome ensures that, for a specific intervention $\interv[F][][\lag]$, once the surface in Equation~\cref{eq:estimatort} is calculated, it can then be used to estimate the temporally-expected effects defined in Section~\ref{sec:estimands} for any $B \subset \Omega$. In addition, it allows for the visualization of the outcome surface under an intervention, making it easier to identify the areas of increased or decreased activity as illustrated in Section~\ref{sec:application}.

In the next section we establish the asymptotic properties of the proposed IPW estimators. In our simulations (Section~\ref{sec:simulations}) and empirical study in (Section~\ref{sec:application}), we also use the H\'ajek estimator, which standardizes the IPW weights and replaces Equation~\cref{eq:estimatorN} with
\begin{equation}
\estimatorN[F][][\lag] \ = \
\sum_{t = \lag}^T \estimatorNt[F][][\lag] \ \Big/ \ 
\sum_{t = \lag}^T \left\{ \prod_{j = t - \lag + 1}^t
\frac{\intervdistf[][T][j]}{\propscore[j][j][W]} \right\}.
\label{eq:hajek_estimator}
\end{equation}
We find that this H\'ajek estimators outperform the corresponding IPW estimators in finite samples, mirroring the existing results under other settings \citep[e.g.,][]{liu2016inverse, cole2021comparing}.

\subsection{The Asymptotic Properties of the Proposed IPW Estimators}

Below, we establish the asymptotic properties of the proposed IPW estimators. Our results differ from the existing asymptotic normality results in the causal inference literature in several ways. First, our inference is based on a single time series of point patterns that are both spatially and temporally dependent. Second, we employ a kernel-smoothed version of the outcome. Third, using martingale theory, we derive a new central limit theorem in time-dependent, observational settings. We now present the main theoretical results.  All proofs are given in \cref{app_sec:proofs}.
\begin{theorem}[Asymptotic Normality] \spacingset{1.25}
Suppose that Assumptions~\ref{ass:unmeasured_conf}~and~\ref{ass:positivity} as well as the regularity conditions (Assumption~\ref{ass:regularity_conditions}) hold.  Then, if the bandwidth $b_T \rightarrow 0$ and as $T \rightarrow \infty$, we have that
$$\sqrt{T}(\estimatorN - \tempavgout) \overset{d}{\rightarrow} \mathcal{N}(0, \asymvar),$$
where $\asymvar$ represents the probability limit of $(T-M+1)^{-1}\sum_{t=M}^T \asymvar_t$ as $T \to \infty$ with $$\asymvar_t \ = \ \Var\left[ \prod_{j = t - \lag + 1}^t \frac{\intervdistf[][T][j]}{\propscore[j][j][W]} N_B(Y_t) \mid \history[t-\lag]^* \right] \quad \textrm{for } t \geq \lag.$$
\label{theorem:normality}
\end{theorem}
\noindent
The key idea of our proof is to separate the estimation error arising due to the treatment assignment $W_t$ given the complete history $\history[t-1]^*$, from the error due to spatial smoothing. Using martingale theory, we show that the former is $\sqrt{T}$-asymptotically normal, where the temporal dependence is controlled based on Assumption~\ref{ass:unmeasured_conf}. The latter is shown to converge to zero at a rate faster than $1/\sqrt{T}$.

According to Theorem~\ref{theorem:normality}, the knowledge of $\asymvar$ would enable asymptotic inference about the temporally-expected potential outcome. The variance $\asymvar$ is the converging point of $(T-M+1)^{-1}\sum_{t=M}^T \asymvar_t$ where $\asymvar_t$ represents a time period-specific variance. Unfortunately, since we only observe one treatment path for each time period $t$, we cannot directly estimate the time-specific variances, $\asymvar_t$, and thus $\asymvar$, without additional assumptions.

We circumvent this problem by using an upper bound of $v$, a quantity which we can consistently estimate. Specifically, let $\asymvar_t^* = \text{E}\left\{ [\estimatorNt[F][][\lag]]^2 \mid \history[t-\lag]^* \right\}$. For $\asymvar^*$ such that $(T-\lag+1)^{-1}\sum_{t = \lag}^T \asymvar_t^* \overset{p}{\rightarrow}  \asymvar^*$, we have $\asymvar \leq \asymvar^*$.  Then, an $\alpha$-level confidence interval for $\tempavgout$ based on the asymptotic variance bound $\asymvar^*/T$ will achieve the nominal asymptotic coverage. Although $\asymvar^*$ cannot be directly calculated either, there exists a consistent estimator of its upper bound, as stated in the following lemma:
\begin{lemma}[Consistent Estimation of Variance Upper bound] \spacingset{1.25} \label{lemma:consistent_variance}
Suppose that Assumptions~\ref{ass:unmeasured_conf}~and~\ref{ass:positivity} and the regularity conditions (Assumption~\ref{ass:regularity_conditions}) hold. Then, as $b_T \rightarrow 0$ and $T \rightarrow \infty$, we have
$$\frac{1}{T-\lag+1} \sum_{t = \lag}^T \left[ \estimatorNt[F][][\lag]^2 - \asymvar_t^* \right] \overset{p}{\rightarrow} 0.$$
\end{lemma}
\noindent
In \cref{subsec:proofs_truePS} we extend the above results to the estimator $\tempeffect[T][F][1][2]$.

So far, all of the theoretical results presented above have been established with the true propensity score $\propscore$. However, in practice, the propensity score is unknown and must be estimated. The next theorem shows that, when the propensity score is estimated under the correct model specification, the proposed estimator maintains its consistency and asymptotic normality. To prove this result, we extended classic M-estimation theory to multivariate martingale difference series, established a new central limit theorem for time series data, and derived the properties of the propensity score models under the spatio-temporal settings. To our knowledge, these results are new even though related results exist under the continuous time setting \citep{kuchler1999note, crimaldi2005convergence}.  We believe that our results may be useful when studying the asymptotic properties of causal estimators in other dependent, observational settings (see \cref{app_subsec:asym_estps} for more details).

\begin{theorem}[Asymptotic Normality Using the Estimated Propensity Score] \spacingset{1.25}
Suppose that Assumptions~\ref{ass:unmeasured_conf} and \ref{ass:positivity} as well as the regularity conditions (Assumptions \ref{ass:regularity_conditions}, \ref{ass:propensity_score}, \ref{app_ass:estimator_ps}) hold.  If the bandwidth $b_T \rightarrow 0$, then as $T \rightarrow \infty$, we have
$$\sqrt{T}(\estimatorN - \tempavgout) \overset{d}{\rightarrow} \mathcal{N}(0, \asymvar^e).$$
\label{theorem:normality_estps}
\end{theorem}

Next, we show that using the estimated propensity scores from a correctly specified model yields more efficient estimates than using the true propensity scores. This generalizes the well-known analogous result proved for the independent and identically distributed setting \citep[e.g.,][]{Hirano2003efficient} to the spatially and temporally dependent case \citep[see][for a similar result in a different dependent setting]{zeng2021propensity}. Thus, even with the estimated propensity score, we can make asymptotically conservative inference based on the variance upper bound derived above.

\begin{theorem}[Asymptotic Efficiency under the Estimated Propensity Score] \spacingset{1.25}
The estimator in Equation~\cref{eq:estimatorN} based on the estimated propensity score from a correctly specified parametric model has asymptotic variance that is no larger than the asymptotic variance of the same estimator using the known propensity score.  That is, for $\asymvar$ in \cref{theorem:normality} and $\asymvar^e$ in \cref{theorem:normality_estps}, we have $\asymvar^e \leq \asymvar$.
\label{theorem:smaller_asymvar}
\end{theorem}

The asymptotic results presented here require the area of interest, $\Omega$, to be fixed while the number of time periods $T$ increases. We note that point pattern treatments and outcomes might also arise in situations where the number of time periods $T$ is fixed, but the area under study $\Omega$ grows to include more regions.
In \cref{app_subsec:sprawl-asymptotics}, we provide an alternative causal inference framework for point pattern treatments under this new design by extending our causal estimands, estimation and asymptotic results to the spatio-temporal setting with an area consisting of a growing number of independent regions.


\section{Sensitivity Analysis}
\label{sec:SA}

The validity of our estimators critically relies upon the assumption of no unmeasured confounding (\cref{ass:unmeasured_conf}).  We develop a sensitivity analysis to address the potential violation of this key identification assumption.  Specifically, we extend the sensitivity analysis pioneered by \citet{rosenbaum2002observational} to the spatio-temporal context and to the H\'ajek estimator with standardized weights, which we consider in our simulation and empirical studies.

Suppose there exists an unmeasured, potentially time-varying confounder $U_t$.  We assume that the unconfoundedness assumption holds only after conditioning on the realized history of this unobserved confounder as well as $\history$, i.e.,
\begin{align*}
f(W_t \mid \Whist[t-1], \anyhist[T][\allout], \anyhist[T][\allcovs], \anyhist[T][\mathcal U]) \ = \ f(W_t \mid \history, \anyhist[t][U]), 
\end{align*}
where $\anyhist[T][\mathcal U]$ represents the collection of all potential values of $U_t$ across all time points $t=1,2,\ldots,T$ whereas $\anyhist[t][U]$ represents the history of realized but unmeasured confounder $U$ up to time $t$.  Note that $U_t$ can be correlated with the observed confounders.

The existence of an unmeasured confounder invalidates the inference based on the propensity score with observed covariates alone $\propscore$ because the true propensity score, denoted by $\SApropscore = f_{W_t}(w \mid \history[t-1], \anyhist[t][U])$, conditions on the history of the unmeasured confounder $\anyhist[t][U]$. To develop a sensitivity analysis, we assume the ratio of estimated versus true propensity scores for the realized treatment $W_t$ is bounded by a value $\Gamma (\geq 1)$,
\begin{equation*}
\frac{1}{\Gamma} \ \leq \ \rho_t  = \frac{\propscore[t][][W_t]}{\SApropscore[t][][W_t]} \ \leq \ \Gamma. \label{eq:Gamma}
\end{equation*}
A larger value of $\Gamma$ allows a greater degree of violation of the unconfoundedness assumption. 

In our application, we use the H\'ajek-version of the proposed estimator, which we find to be more stable than the IPW estimator (see Section~\ref{sec:simulations}).  Thus, to develop a sensitivity analysis, we derive an algorithm for bounding the H\'ajek estimator for stochastic interventions for each fixed value of $\Gamma$ (see \cref{app_sec:SA} for the sensitivity analysis of the IPW estimator). 
Specifically, for all values of $\bm \rho = (\rho_1, \rho_2, \dots, \rho_T) \in [\Gamma^{-1}, \Gamma]^T$, we wish to bound the following two quantities:
\begin{align*}
\estimatorN[F][][][\bm \rho] &= \frac {\sum_{t = 1}^T \rho_t \ w_t(\interv) \ \widetilde N_B(Y_t) }{ \sum_{t = 1}^T \rho_t \ w_t(\interv) }, \quad \text{and} \\
\tempeffect[T][F][1][2][1][\bm \rho] &=\frac {\sum_{t = 1}^T \rho_t \ w_t(\interv[F][2]) \ \widetilde N_B(Y_t) }{ \sum_{t = 1}^T \rho_t \ w_t(\interv[F][2]) } -
\frac {\sum_{t = 1}^T \rho_t \ w_t(\interv[F][1]) \ \widetilde N_B(Y_t) }{ \sum_{t = 1}^T \rho_t \ w_t(\interv[F][1]) }
\end{align*}
where
\begin{equation*}
w_t(\interv) = \frac{\intervdistf[][T]}{\propscore[t][t][W]} \quad \text{and} \quad
\widetilde N_B(Y_t) =
\int_B \sum_{s \in \sparseset[{}Y][t]} K_b(\|\omega - s\|) \mathrm{d} \omega.
\end{equation*}
Below we show how to formulate the bounding problem for $\estimatorN[F][][][\bm \rho]$ as a linear program, and how to use the bounds for $\estimatorN[F][][][\bm \rho]$ to also bound the effect estimator
$\tempeffect[T][F][1][2][1][\bm \rho]$. 
\begin{theorem}[Bounding the Causal Quantities] \spacingset{1.25} $\ $
\begin{enumerate}[leftmargin=*]
\item
The problem of maximizing $\estimatorN[F][][][\bm \rho]$ over $\bm \rho \in [\Gamma^{-1}, \Gamma]^T$ is equivalent to the following linear program,
\begin{align*}
\textrm{maximize}_{\bm\rho^\ast} & \ \sum_{t=1}^T \rho_t^* \ w_t(\interv) \ \widetilde N_B(Y_t) \\
 & \textrm{subject to} \quad
\frac{\kappa}{\Gamma} \ \le \ \rho_t^* \ \le \ \Gamma \kappa,\ \sum \rho_t^* \ w_t(\interv) = 1,\ \text{and} \ \kappa \geq 0.
\end{align*}
where $\bm \rho=\bm \rho^* / \kappa$.

\item Suppose that $\bm \rho^{\max}_j$ and $\bm \rho^{\min}_j$ represent the values of $\bm \rho$ that maximize and minimize $\estimatorN[F][j][][\bm \rho]$, respectively, for $j = 1, 2$. Then, the bounds for the causal effect are obtained as,
\begin{equation}
\estimatorN[F][2][][{\bm \rho^{min}_2}]  -
\estimatorN[F][1][][{\bm \rho^{max}_1}] \ \le \ \tempeffect[T][F][1][2][1][\bm \rho] \ \le \
\estimatorN[F][2][][{\bm \rho^{max}_2}]  -
\estimatorN[F][1][][{\bm \rho^{min}_1}].
\label{eq:bound_effect}
\end{equation}
\end{enumerate}

\label{theorem:SA_bound_both}
\end{theorem}
\noindent The proof for bounding $\estimatorN[F][][][\bm \rho]$ is based on the Charnes-Cooper transformation of linear fractionals \citep{charnes1962programming}, and the proof for bounding $\tempeffect[T][F][1][2][1][\bm \rho]$ is given in \cref{app_sec:SA}.
For bounding $\estimatorN[F][][][\bm \rho]$, this proposition allows us to use a standard linear algorithm to obtain the optimal solution for $(\bm \rho^*, \kappa)$ and transform it back to the optimal solution $\bm \rho$.
Then, we can use these bounds to also acquire bounds on the effect estimator.
Since all bounds are wider for a greater value of $\Gamma$, the estimated effects are robust to propensity score misspecification up to the smallest value of $\Gamma$ for which the interval of bounds in Equation~\cref{eq:bound_effect} includes 0. Due to the standardization of weights in the H\'ajek estimator, the bound in Equation~\cref{eq:bound_effect} is conservative, in the sense that, if the causal estimate is shown to be robust up to some value $\Gamma$, then it is robust up to an even greater degree of propensity score model misspecification $\Gamma^* \geq \Gamma$. Similar bounds can be derived for the stochastic interventions that take place over multiple time periods (see \cref{app_sec:SA} for details). 

The propensity score modelling in our spatio-temporal setting is much more complex with an infinite number of potential treatment locations than in the conventional cross-sectional setting.  As a result, the modelling uncertainty for the propensity score is much greater. This makes it difficult to compare the scale of $\Gamma$ between the spatio-temporal and conventional cross-section settings.  In particular, the value of $\Gamma$ is expected to be much closer to the null value of one in the spatio-temporal context.  

\section{Simulation Studies}
\label{sec:simulations}

We conduct simulation studies to empirically investigate several key theoretical properties of the proposed methodology: 
\begin{enumerate*}[label=(\alph*)]
\item the performance of our estimator under different stochastic interventions and as the number of time periods increases,
\item the accuracy of the asymptotic approximation,
\item the difference between the theoretical variance bound and the actual variance,
\item the performance of the inferential approach based on the estimated asymptotic variance bound,
\item the relative efficiency of the estimator when using the true and estimated propensity scores, and
\item the balancing properties of the estimated propensity score.
\end{enumerate*}
We use the \texttt{spatstat} R package \citep{Baddeley2015spatial} to generate point patterns from Poisson processes and fit Poisson process models to the simulated data.

\subsection{The Study Design}

To construct a realistic simulation design, we base our data generating process on the observed data from our application. We consider a time series of point patterns of length $T \in \{200, 400, 500\}$. For each time series length $T$, 200 data sets are generated. The scenario with $T = 500$ closely resembles our observed data, which have $T = 469$.

\subsubsection*{Time-varying and time-invariant confounders.} 
\begin{figure}[!t]
\centering
\subfloat[Time-invariant confounder $\onecov^1(\omega)$]{\includegraphics[width = 0.3\textwidth,trim=60 70 10 60, clip]{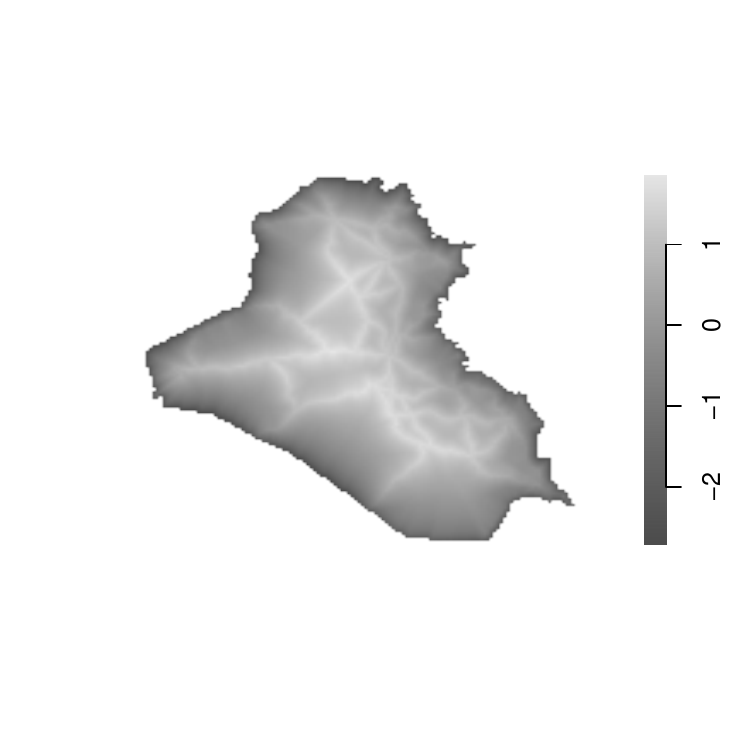} \label{fig:sims_iraq_roads}}
\hspace{.2in}\subfloat[Realization of time-varying confounder $\onecov_t^3(\omega)$]{ \includegraphics[width = 0.3\textwidth,trim=60 70 10 60, clip]{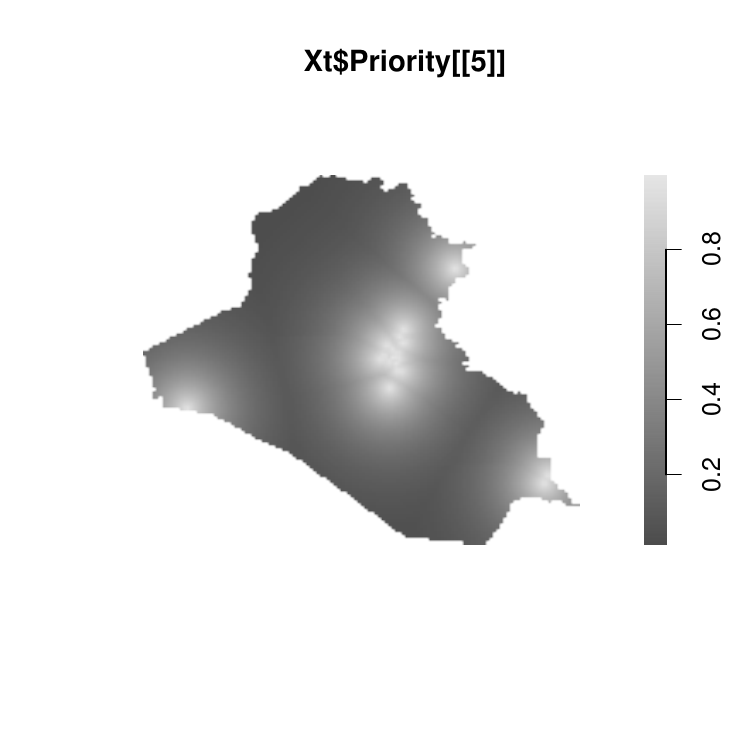} \label{fig:sims_iraq_Xt_one}}
\hspace{.2in}\subfloat[Distribution of treatment point patterns]{ \includegraphics[width = 0.3\textwidth,trim=60 70 10 60, clip]{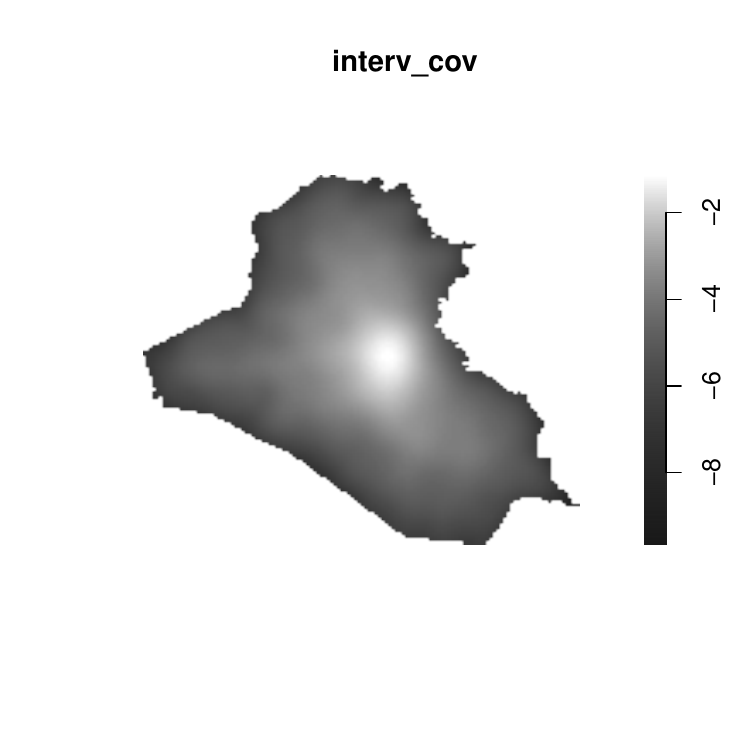} \label{fig:sims_iraq_interv_cov}}
\caption{Simulated Data.
Panel (a) shows one of the two time-invariant confounders representing the exponential decay of distance to the road network.
Panel (b) shows one realization for one of the time-varying confounders. After generating points from a non-homogeneous Poisson process, which depends on the observed airstrike density, we define the time-varying confounder as the exponential decay of distance to these points.
Panel (c) shows the estimated log-density of treatment patterns, which is used as the density $\phi$ in the definition of stochastic interventions.
}
\end{figure}

Our simulation study includes two time-invariant and two time-varying confounders. We base the first time-invariant confounder on the distance from Iraq's road network and its borders, by defining its value at location $\omega \in \Omega$ as
$\onecov^1(\omega) =  \exp\{- 3 D_1(\omega)\} + \log(D_2(\omega))$,  where $D_1(\omega)$is the distance from $\omega$ to the closest road, and $D_2(\omega)$ is the distance to the country's border. This covariate is shown in \cref{fig:sims_iraq_roads}. The second covariate is defined similarly as $\onecov^2(\omega) = \exp\{- D_3(\omega)\}$ where $D_3(\omega)$ is the distance from $\omega$ to Baghdad. 

We generate the time-varying confounders, $\onecov_t^3(\omega)$ and $\onecov_t^4(\omega)$, using the kernel-smoothed density of the observed airstrike and attack patterns. Specifically, we pool all airstrike locations across time and estimate the density of airstrike patterns, denoted by $\widehat f(\omega)$ at location $\omega$ (shown in the right plot of \cref{fig:air_intens_pre_blackout}).  Based on this density, we draw a point pattern from a non-homogeneous Poisson point process with intensity function
$\lambda^{X^3}(\omega) = \exp \big\{ \rho_0 + \rho_1 \widehat f(\omega) \big\}$, for $\rho_0 \approx - 2.7$ and $\rho_1 = 8$, and define $\onecov^3(\omega)$ as $\exp\{- D_4(\omega)\}$, where $D_4(\omega)$ is the distance from location $\omega$ to the closest point. We generate $\onecov^4(\omega)$ similarly based on the estimated density for insurgent attacks, and for corresponding values $\rho_0 \approx -3.2$ and $\rho_1 = 7$. \cref{fig:sims_iraq_Xt_one} shows one realization of $\onecov_t^3(\omega)$.

\subsubsection*{Spatio-temporal point processes for treatment and outcome variables.}

For each time period $t \in \alltimes$, we generate $W_t$ from a non-homogeneous Poisson process that depends on all confounders $\covs_t(\omega) = (\onecov^1(\omega), \onecov^2(\omega), \onecov_t^3(\omega), \onecov_t^4(\omega))^\top$, as well as the previous treatment and outcome realizations, $W_{t - 1}$ and $Y_{t - 1}$.  The intensity of this process is given by
\begin{equation}
\lambda_t^W (\omega) \ = \ \exp\big\{ \alpha_0 + \bm{\alpha}_{\bm{X}}^\top \covs_t(\omega) + \alpha_W W_{t-1}^*(\omega) + \alpha_Y Y_{t - 1}^*(\omega) \big\},\label{eq:sims_iraq_treatment_intensity}
\end{equation}
where $W_{t - 1}^*(\omega)=\exp \{-2 D_W(\omega)\}$ and $Y_{t - 1}^*(\omega)=\exp \{-2 D_Y(\omega)\}$ with $D_W(\omega)$ and $D_Y(\omega)$ being the minimum distance from $\omega$ to the points in $\sparseset[W][t-1]$ and $\sparseset[Y{}][t-1]$, respectively.

Similarly, we generate $Y_t$  from a non-homogeneous Poisson process with intensity
\begin{equation}
\lambda_t^Y(\omega) \ = \ \exp\left\{ \gamma_0 + \bm{\gamma}_{\bm{X}}^\top \covs_t(\omega) +\gamma_2 \onecov_{t-1}^2(\omega) + \gamma_W W_{(t-3):t}^*(\omega) + \gamma_Y Y_{t - 1}^*(\omega) \right\},\label{eq:outcome_intensity}
\end{equation}
where $W_{(t-3):t}^*(\omega)=\exp \{-2 D_W^\ast(\omega)\}$ with $D_W^\ast(\omega)$ being the distance from $\omega$ to the closest points in $\bigcup_{j = t-3}^t \sparseset[W][j]$. This specification imposes a lag-three  dependence of the outcome on the lagged treatment process. The model leads to an average of 5.5 treatment-active locations and 31 outcome-active locations within each time period, resembling the frequency of events in our observed data. The spatial distribution of generated treatment point patterns also resembles the observed one (compare \cref{fig:sims_iraq_interv_cov} to the right plot of \cref{fig:air_intens_pre_blackout}). The simulated and observed outcome point patterns also have similar distributions.  

\subsubsection*{Stochastic interventions.}

We consider stochastic interventions of the form $\interv[F][][\lag]$ for a  non-homogeneous Poisson process with intensity $h$, which is defined as $h(\omega) = c \phi(\omega)$ for $c$ ranging from $3$ to 8, and surface $\phi$ set to the density shown (in logarithm) in \cref{fig:sims_iraq_interv_cov}. This definition of stochastic intervention based on the treatment density aligns with the specification in our study in Section~\ref{sec:application}.  We consider varying the intervention duration by setting $M \in  \{1, 3, 7, 30\}$.  We also examine lagged interventions over three time periods, i.e., $\interv[T]=\intervdist[F][3] \times \intervdist[F][2] \times \intervdist[F][1]$. The intervention for the first time period $\intervdist[F][3]$ is a Poisson process with intensity $h_3(\omega) = c \phi(\omega)$ for $c$ ranging from 3 to 7, whereas $\intervdist[F][2] = \intervdist[F][1]$ is a non-homogeneous Poisson process with intensity $5\phi(\omega)$.  For each stochastic intervention, we consider three regions of interest, $B$, of different sizes, representing the whole country, the Baghdad administrative unit, and a small area in northern Iraq which includes the town of Mosul.

\subsubsection*{Approximating the true estimands.}

Equation~\cref{eq:outcome_intensity} shows that the potential outcomes depend on the realized treatments during the last four time periods as well as the realized outcomes from the previous time period. This implies that the estimands for all interventions, even for $M > 4$, depend on the observed treatment and outcome paths and are therefore not constant across simulated data sets.  Therefore, we approximate the true values of the estimands in each data set in the following manner. For each time period $t$, and each $r = 1, 2, \ldots, R$ repetition, we generate realizations $w_{t - \lag + 1}\tor, \dots, w_{t - 1}\tor, w_t\tor$ from the intervention distribution $\intervdist[T]$. Based on the treatment path $(\Whist[t - \lag], w_{t - \lag + 1}\tor, \dots, w_t\tor)$, we generate outcomes $y_{t-\lag + 1}\tor, \dots, y_t\tor$ using Equation~\cref{eq:outcome_intensity}. This yields $\sparseset[y\tor][t]$, which contains the outcome-active locations based on one realization from the stochastic intervention. Repeating this process $R$ times and calculating the average number of points that lie within $B$ provides a Monte Carlo approximation of $\avgout[T]$, and further averaging these over time gives an approximation of $\tempavgout[T]$.

\subsubsection*{Estimation.}

We estimate the expected number of points $\tempavgout[T]$ and the effect of a change in the intervention on this quantity $\tempeffect[F][T]$ using the following estimators:
\begin{enumerate*}[label=(\alph*)]
\item the proposed estimators defined in Equations~\cref{eq:estimatorN}~and~\cref{eq:estimator_effect} with the true propensity scores;
\item the same proposed estimators with the estimated propensity scores based on the correctly-specified model;
\item the above two estimators with the H\'ajek-type standardization in \cref{eq:hajek_estimator}; and
\item the unadjusted estimator based on the propensity score model using a homogeneous Poisson process with no predictor.
\end{enumerate*}

All estimators utilize the smoothed outcome point pattern. Spatial smoothing is performed using Gaussian kernels with standard deviation equal to $10 T^{-2/3} \delta$, which is decreasing in $T$, and for $\delta$ scaling the bandwidth according to the size of the geometry under study. We choose this bandwidth such that for $T = 500$ (the longest time series in our simulation scenario) the bandwidth is approximately 0.5, slightly smaller than the size of the smallest region of interest $B$ (square with edge equal to 0.75). We discuss the choice of the bandwidth in Section~\ref{subsec:application_bandwidth}.

\subsubsection*{Theoretical variance and its upper bound.}
\cref{theorem:normality} provides the expression for the asymptotic variance of the proposed IPW estimator. We compute Monte Carlo approximations to this variance and its upper bound. Specifically, for each time period $t$ and each replication $r$, the computation proceeds as follows:
\begin{enumerate*}[label=\arabic*)]
\item we generate treatment and outcome paths $w_{t - \lag + 1}\tor, y_{t - \lag + 1}\tor, \dots, w_t\tor, y_t\tor$ using the distributions specified in Equations~\cref{eq:sims_iraq_treatment_intensity}~and~\cref{eq:outcome_intensity},
\item using the data $(w_{t - \lag + 1}\tor, \dots, w_t\tor)$ and the outcome $y_t\tor$, we compute the estimator according to Equations~\cref{eq:estimatort}~and~\cref{eq:estimatorNt}, and finally
\item we calculate the variance and the second moment of these estimates over $R$ replications, which can be used to compute the asymptotic variance and variance bound of interest.
\end{enumerate*}
Their averages over time give the desired Monte Carlo approximations.  We use a similar procedure to approximate the theoretical variance and variance bound of $\tempeffect[T][T]$.

\subsubsection*{Estimating the variance bound and the resulting inference.}

We use \cref{lemma:consistent_variance} to estimate the variance bound. This estimated variance bound is then used to compute the confidence intervals and conduct a statistical test of whether the causal effect is zero. Inference based on the H\'ajek estimator is discussed in \cref{app_sec:hajek}.

\subsubsection*{Balancing property of the propensity score.}

Using the correctly specified model, we estimate the propensity score at each time period $t$. The inverse of the estimated propensity score is then used as the weight in the weighted Poisson process model for $W_t$ with the intensity specified in Equation~\cref{eq:sims_iraq_treatment_intensity}. We compare the statistical significance of the predictors between the weighted and unweighted model fits. Large $p$-values under the weighted model would suggest that the propensity score adequately balances the confounding variables.

\subsubsection*{Relative efficiency of estimators based on the true and estimated propensity score.}

According to \cref{theorem:smaller_asymvar}, the asymptotic variance of the estimator based on the true propensity score is at least as large as that of the estimator based on the estimated propensity score. We investigate the relative magnitude of the Monte Carlo approximations of the corresponding two variances.

\subsection{Simulation Results}

\begin{figure}[p]
\centering
\includegraphics[width=0.8\textwidth,trim=0 40 0 0, clip]{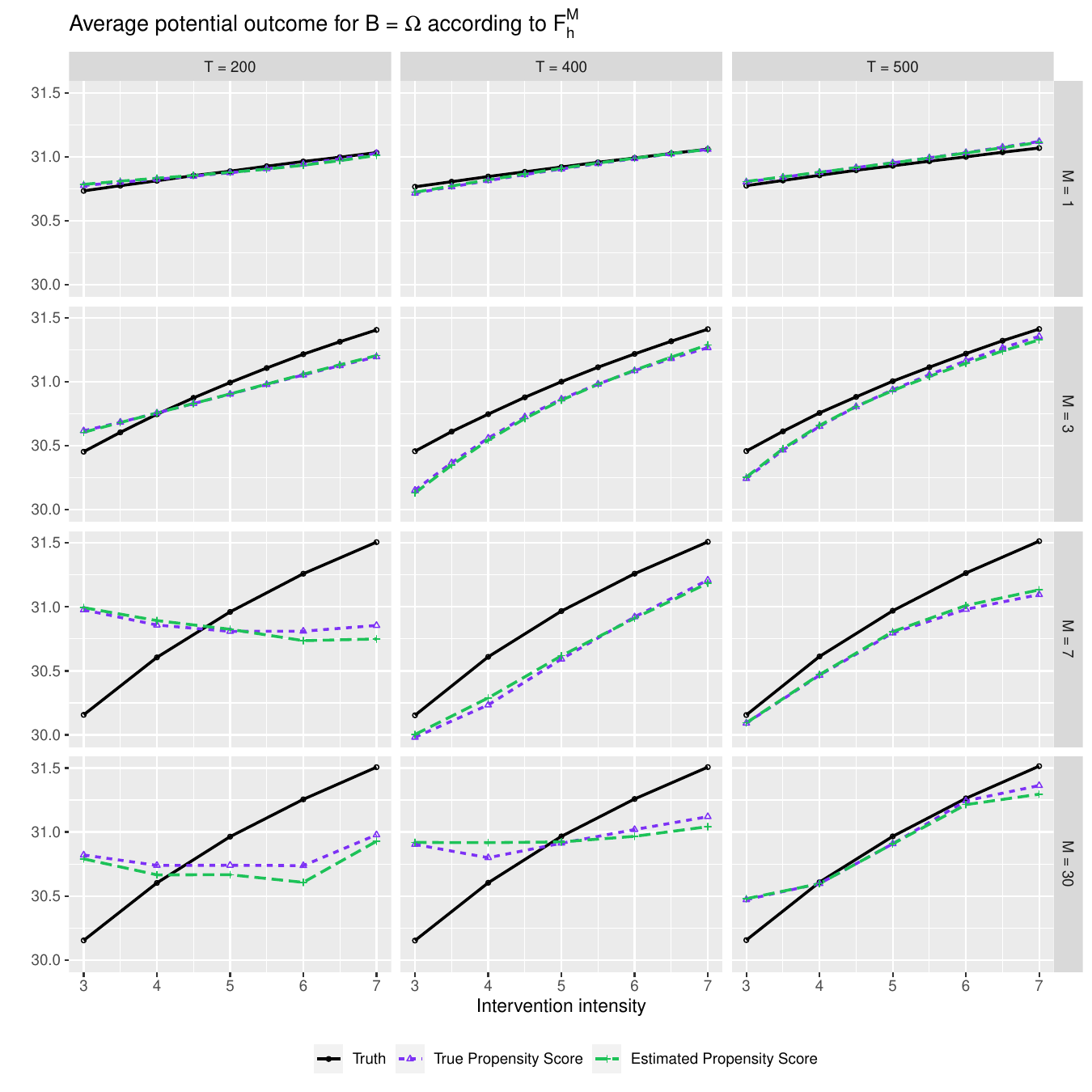}\\
\includegraphics[width=0.8\textwidth]{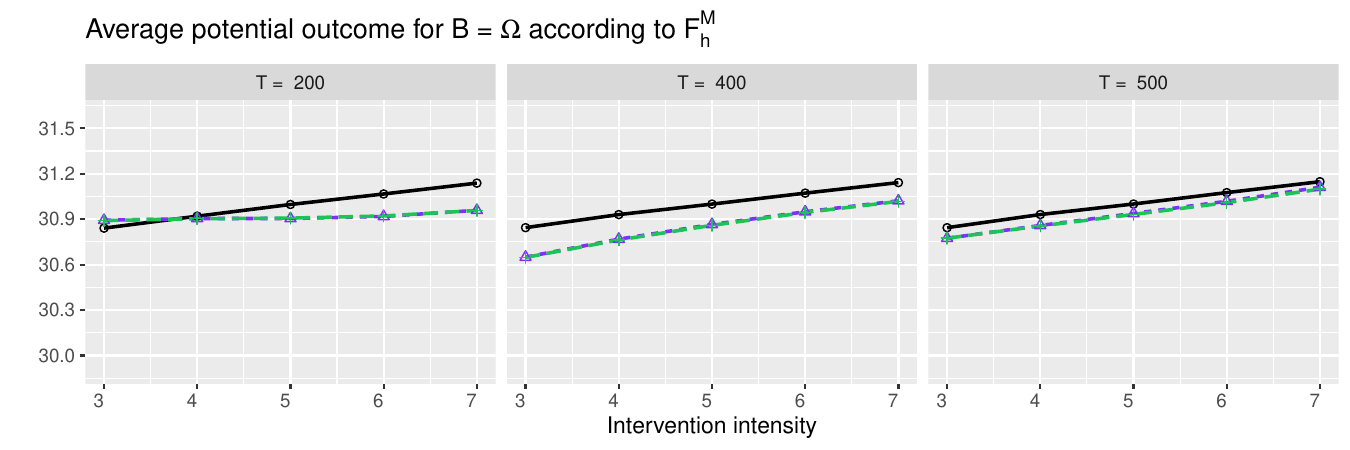}\\[-10pt]
\includegraphics[width=\textwidth,trim=0 0 0 610, clip]{5res3_y_int1256_B1.pdf}
\vspace{-10pt}
\caption{Simulation Results for the Average Potential Outcomes. In the top four rows, we present the true and estimated average potential outcomes in $B = \Omega$ under interventions $F_h^\lag$ with the varying intensity (horizontal axis) and $\lag \in \{1, 3, 7, 30\}$ (rows), respectively. In the bottom row, we consider the average potential outcome for the lagged intervention over three time periods $\interv[T]$, with the varying intensity of $\interv[F][3]$ shown on the horizontal axis. The black lines with solid circles represent the truths, the H\'ajek estimator based on the true propensity score is shown in purple, and the H\'ajek estimator based on the estimated propensity score is in green.}
\label{fig:5res3_estimates}
\end{figure}

\cref{fig:5res3_estimates} presents the results for all the stochastic interventions that were considered. The top four rows show how the (true and estimated) average potential outcomes in the whole region ($B = \Omega$) change as the intensity varies under interventions $\interv[F][][\lag]$ for $\lag \in \{1, 3, 7, 30\}$, respectively.  The last row shows how the true and estimated average potential outcomes in the same region change under the three time period {\it lagged} interventions when the intensity at three time periods ago ranges from 3 to 7. For both simulation scenarios, we vary the length of the time series from 200 (left column) to 500 (right column).

The unadjusted estimator returned values that are too far from the truth and are not shown here. We find that the accuracy of the proposed estimator improves as the number of time periods increases.  Notice that the convergence is slower for larger values of $\lag$. This is expected because the uncertainty of the treatment assignment is greater for a stochastic intervention with a longer time period.  We find that the H\'ajek estimator performs well across most simulation scenarios even when $T$ is relatively small and $M$ is large. The IPW estimator (investigated more thoroughly in \cref{app_sec:add_square_sims}) tends to suffer from extreme weights because the weights are multiplied over the intervention time periods as shown in Equation~\cref{eq:estimatort}.  These results indicate a deteriorating performance of the IPW estimator as the value of $M$ increases, whereas the standardization of weights used in the H\'ajek estimator appears to partially alleviate this issue. Results were comparable for the two other sets $B$.

\begin{figure}[!t]
\centering
\includegraphics[width = \textwidth, trim=0 55 0 0, clip]{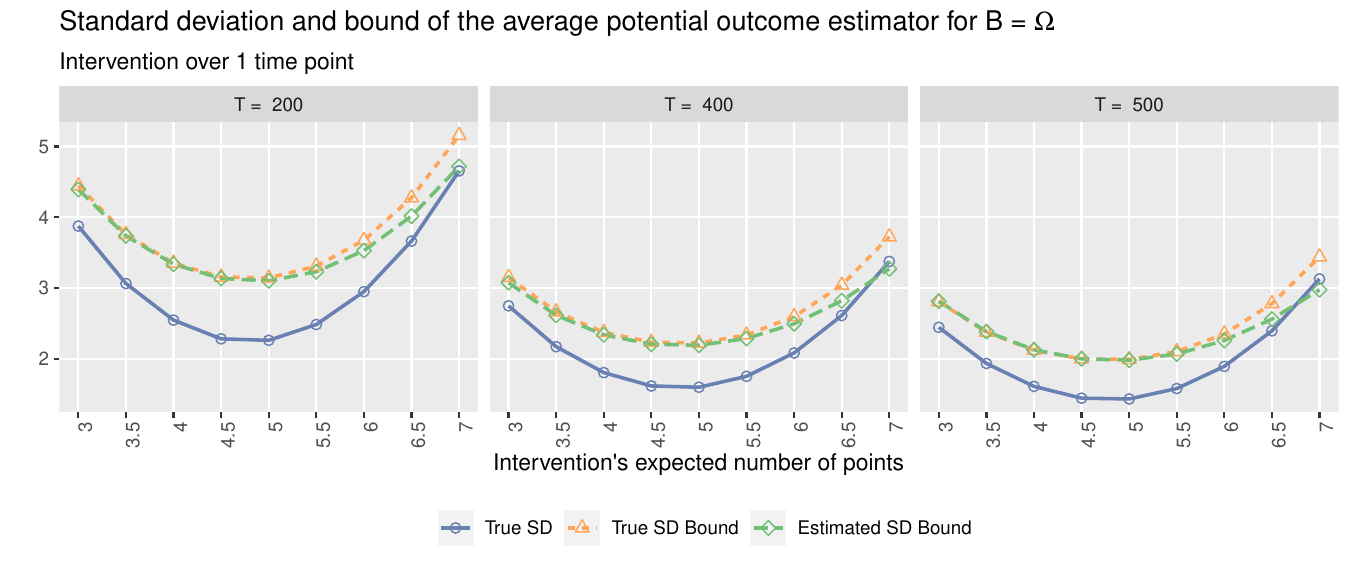} \\[10pt]
\includegraphics[width = \textwidth, trim=0 0 0 25, clip]{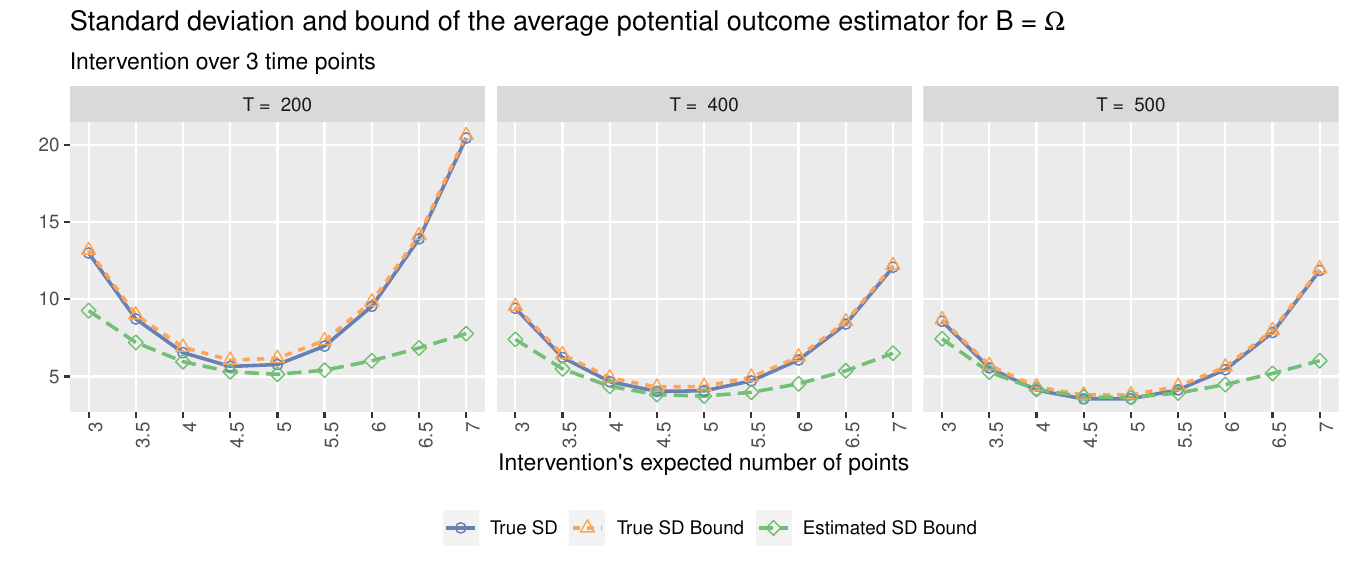}
\caption{Comparison of the Asymptotic Standard Deviation with the True and Estimated Asymptotic Standard Deviation Bound for the Average Potential Outcome Estimator. The comparison is based on the varying number of expected points (horizontal axis) under the stochastic intervention $\interv[F][][\lag]$ taking place over $\lag = 1$ (top row) and $\lag = 3$ (bottom row) time periods. The columns correspond to a simulation setting with a different time series length.}
\label{fig:sims_iraq_var_y_int1_B1}
\end{figure}

Next, we compare the true theoretical variance, $\asymvar / T$, with the variance bound $\asymvar^* / T$ and its consistent estimator (see \cref{lemma:consistent_variance}).  We assess the conservativeness of the theoretical variance bound by focusing on the proposed estimators with the true propensity score. \cref{fig:sims_iraq_var_y_int1_B1} shows the results of an intervention $\interv[F][][M]$ for $\lag \in \{1, 3\}$, and for region $B = \Omega$. The results for the other regions are similar and hence omitted.

First, we focus on the theoretical variance and variance bound (blue line with open circles, and orange dotted lines with open triangles, respectively). As expected, the true variance decreases as the total number of time periods increases, and the theoretical variance bound is at least as large as the variance. In the setting with $\lag = 3$, the theoretical variance follows the variance closely, evident by the fact that the two lines are essentially indistinguishable. We have found this to be the case in all scenarios with higher uncertainty, indicating that the theoretical variance bound is not overly conservative.  Indeed, the variance bound is visibly larger than the true variance only in the low-variance scenarios of interventions over a single time period, as shown in the top row of \cref{fig:sims_iraq_var_y_int1_B1} (and in \cref{app_subsec:add_sims_iraq_variance}).

Second, we compare the theoretical variance bound with the estimated variance bound (green dashed lines with open rhombuses). As the length of time series increases, the estimated variance bound more closely approximates its theoretical value (consistent with \cref{lemma:consistent_variance}). Furthermore, the estimated variance bound is close to its theoretical value under low uncertainty scenarios and when the intervention intensity more closely resembles that of the actual data generating process.  However, we find that the estimated variance bound underestimates the true variance bound in high uncertainty scenarios, and convergence to its true value is slower for larger values of $\lag$ (see \cref{app_subsec:add_sims_iraq_variance}).

\begin{table}[!b]
\centering
\caption{Variance Ratio of the Proposed Estimator based on the True Propensity Score over the Proposed Estimator based on the Estimated Propensity Score.  The results are based on Monte Carlo approximation with $T = 500$.  The estimated propensity score is obtained from the correctly specified model.  If the ratio is greater than 1, the estimated propensity score yields more efficient estimator than the true propensity score. We consider interventions that are constant over all intervention time periods, $\interv[F][][\lag]$ for $\lag \in \{1, 3, 7, 30\}$ (top four rows), and the lagged intervention over three time periods $\interv[T]=\intervdist[F][3] \times \intervdist[F][2] \times \intervdist[F][1]$ (bottom row). }  \label{tab:rel_var}
\begin{tabular}{lrrrrr}
  \hline
 & \multicolumn{5}{c}{Expected number of treatment active} \\
 & \multicolumn{5}{c}{locations under the intervention} \\ & $c = 3$ & $c = 4$ & $c = 5$ & $c = 6$ & $c = 7$ \\
  \hline
$\lag = 1\phantom{0}$ & 1.24 & 1.38 & 1.41 & 1.32 & 1.08 \\
$\lag = 3\phantom{0}$ & 1.09 & 1.18 & 1.24 & 1.14 & 1.06 \\
$\lag = 7\phantom{0}$ & 1.07 & 0.85 & 1.08 & 0.61 & 0.54 \\
$\lag = 30 $ & 0.60 & 0.75 & 0.87 & 0.58 & 0.75 \\  \hline
Lagged \hspace{20pt} & 1.08 & 1.21 & 1.24 & 1.20 & 1.11 \\
   \hline
\end{tabular}
\end{table}

We also compare the variance of the estimator based on the true propensity score with that of the estimator based on the estimated propensity score. \cref{tab:rel_var} shows the ratio of the Monte Carlo variances which, according to \cref{theorem:smaller_asymvar}, should be larger than 1, asymptotically. Consistent with the above simulation results, we find that the ratio is above 1 for interventions over one and three time periods. In addition, the ratio is largest in the low uncertainty scenarios where either the number of intervention periods, $\lag$, is small, or the expected number of points is near the observed value under the intervention ($c \approx 5$).  In contrast, in the high uncertainty situations with longer intervention periods, e.g. $\lag \in \{7, 30\}$, the ratio remains below 1, implying that the asymptotic approximation may not be sufficiently accurate for the sample sizes considered.

In Appendices~\ref{app_subsec:add_sims_iraq_coverage}~and~\ref{app_subsec:add_sims_iraq_uncertainty}, we also investigate the performance of the inferential procedure based on the true variance, true variance bound, and estimated variance bound, for both the IPW and H\'ajek estimators. The confidence interval for the IPW estimator tends to yield coverage close to its nominal level only for the interventions over a small number of time periods.   In contrast, the confidence interval for the H\'ajek estimator has good coverage probability even for the interventions over a larger number of time periods. Partly based on these findings, we use the H\'ajek estimator and its associated confidence interval in our empirical application (see Section~\ref{sec:application}).

Finally, we evaluate the performance of the propensity score as a balancing score (\cref{theorem:balancing_score}). In \cref{app_subsec:add_sims_iraq_balance}, we show that the p-values of the previous outcome-active locations variable ($Y_{t - 1}^*$ in Equation~\cref{eq:sims_iraq_treatment_intensity}) are substantially greater in the weighted propensity score model than in the unweighted model, where the weights are equal to the inverse of the estimated propensity score. These findings are consistent with the balancing property of the propensity score.

In \cref{app_sec:add_square_sims} we present an alternative simulation study, though all qualitative conclusions remain unchanged.

\section{Empirical Analyses}
\label{sec:application}

In this section, we present our empirical analyses of the datasets introduced in Section~\ref{sec:iraq}.  We first describe the airstrike strategies of interest and then discuss the causal effect estimates obtained under those strategies.

\subsection{Airstrike Strategies and Causal Effects of Interest}
\label{subsec:study_interventions}

\begin{figure}[!t]
\centering
\includegraphics[width = 0.7\textwidth]{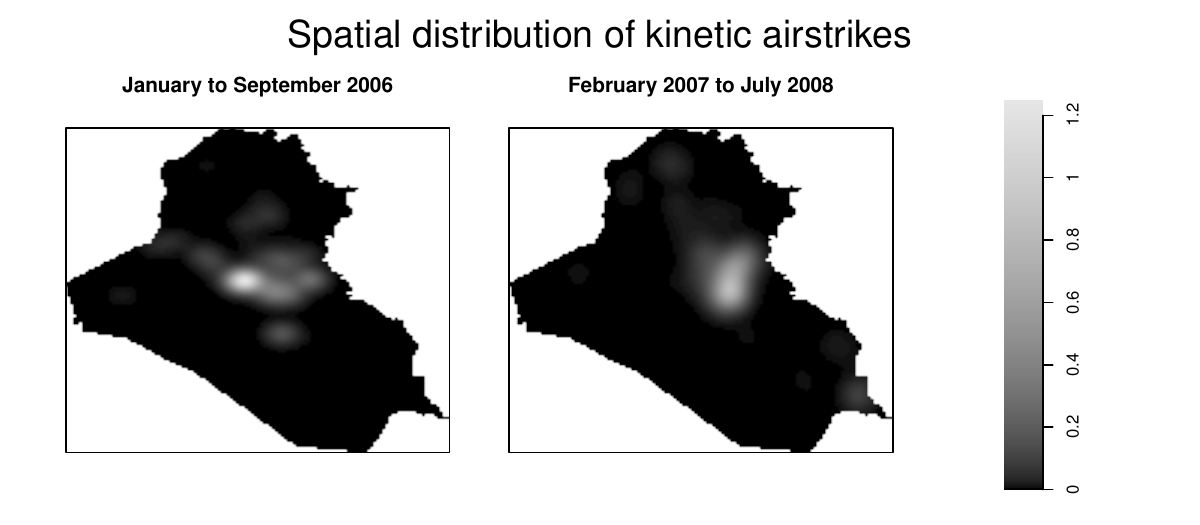}
\caption{Spatial Density Estimate of Airstrike Locations during January 1 -- September 24, 2006 (left) and the Entire Study Period February 2007 -- July 2008 (right).}
\label{fig:air_intens_pre_blackout}
\end{figure}

We consider hypothesized stochastic interventions that generate airstrike locations based on a simple non-homogeneous Poisson point process with finite and non-atomic intensity $\trtintensity: \Omega \rightarrow [0, \infty)$.  We first specify a baseline probability density $\phi_0$ over $\Omega$.  To make this baseline density realistic and increase the credibility of the overlap assumption, we use the airstrike data during January 1 -- September 24, 2006 to define the baseline distribution $\phi_0$ for our stochastic interventions. This subset of the data is not used in the subsequent analysis.  The left plot of \cref{fig:air_intens_pre_blackout} shows the estimated baseline density, using kernel-smoothing of airstrikes with an anisotropic Gaussian kernel and bandwidth specified according to Scott's rule of thumb \citep{scott1992multivariate}.

We consider the following three questions: (1) How does an increase in the number of airstrikes affect insurgent violence? (2) How does the shift in the prioritization of certain locations for airstrikes change the spatial pattern of insurgent attacks? (3) How long does it take for the effects of change in these airstrike strategies to be realized? The last question examines how quickly the insurgents respond to the change in airstrike strategy.

We address the first question by considering stochastic interventions that have the same spatial distribution but vary in the expected number of airstrikes. We represent such strategies using intensities $\trtintensity(\omega) = c\phi_0(\omega)$ with different values of $c > 0$. Since $\int_\Omega \trtintensity(\omega) \mathrm{d}\omega$ represents the expected number of points from a Poisson point process, these interventions have the same spatial distribution $\phi_0$, but the number of airstrikes monotonically increases as a function of $c$. In our analysis, we consider $\{1, 2, \dots, 6\}$ as the range of $c$ which corresponds to the expected number of airstrikes per day, in agreement with the observed data.

For the second question, we fix the expected number of airstrikes but vary their focal locations.  To do this, we specify a distribution over $\Omega$ with power-density $d_\alpha(\omega) = d(\omega)^\alpha / \left(\int_\Omega d(\omega)^\alpha \right)$ and modes located at $s_f \in \Omega$. Based on $d_\alpha$, we specify $\trtintensity_\alpha(\omega) = c_\alpha \phi_0(\omega) d_\alpha(\omega)$ where $c_\alpha$ satisfies the constraint $\int_{\Omega}  \trtintensity_\alpha(\omega) \text{d}\omega = c$, so that the overall expected number of airstrikes remains constant. Locations in $s_f$ are increasingly prioritized under $\trtintensity_\alpha$ for increasing $\alpha$. For our analysis, we choose the center of Baghdad to be the focal point $s_f$ and $d_\alpha$ to be the normal distribution centered at $s_f$ with precision $\alpha$.  We set the expected number of airstrikes per day $c$ to be 3, and vary the precision parameter $\alpha$ from 0 to 3. The visualization of the spatial distributions in $\trtintensity_\alpha$ for the different values of $\alpha$ is shown in \cref{fig:type2_interv}.

As discussed in Section~\ref{sec:estimands}, for both of these questions, we can specify airstrike strategies of interest taking place over a number of time periods, $\lag$, by specifying the stochastic interventions as $\interv[T][] = \interv[F][][\lag]$. In addition, we may also be interested in the lagged effects of airstrike strategies as in the third question.  We specify  lagged intervention to be the one which differs only for the $\lag$ time periods ago, i.e., $\intervdist[T][] =  \intervdist[F][0][\lag-1] \times \intervdist[F][1]$, where $h_0 = \phi_0$ represents the baseline intensity (with $c = 1$), and $h_1 = c\phi_0$ is the increased intensity with different values of $c$ ranging from 1 to 6. We assume that insurgent attacks at day $t$ do not affect airstrikes on the same day, and airstrikes at day $t$ can only affect attacks during subsequent time periods. Thus, causal quantities for interventions taking place over $\lag$ time periods refer to insurgent attacks occurring $\lag$ days later. For our analysis, we consider values of $\lag$ which correspond to 1 day, 3 days, 1 week, and 1 month.

Although full investigation is beyond the scope of this paper, in \cref{app_subsec:adaptive}, we briefly consider an extension to adaptive interventions over a single time period ($M = 1$), and discuss challenges when considering adaptive interventions over multiple time periods ($M > 1$).

\subsection{The Specification and Diagnostics of the Propensity Score Model}
\label{subsec:application_assumptions}

Our propensity score model is a non-homogeneous Poisson point process model with intensity $\lambda_t(\omega) = \exp \{\bm \beta^\top \bm \covs_t(\omega)\}$ where $\bm \covs$ includes an intercept, temporal splines, and 32 spatial surfaces including all the covariates.  The two main drivers of military decisions over airstrikes are the prior number and locations of observed insurgent attacks and airstrikes, which are expected to approximately satisfy unconfoundedness of Assumption~\ref{ass:unmeasured_conf}. Our model includes the observed airstrikes and insurgent attacks during the last day, week, and month (6 spatial surfaces). For example, the airstrike history of time $t$ during the previous week is \( \anyhist[t-1][W]^*(\omega) = \sum_{j = 1}^7 \sum_{s \in \sparseset[W][t-j]} \exp\{ - \mathrm{dist}(s, \omega) \}\), which represents a surface on $\Omega$ with locations closer to the airstrikes in the previous week having greater values than more distant locations.

Our propensity score model also includes additional important covariates that might affect both airstrikes and insurgent attacks. We adjust for shows-of-force (i.e., simulated bombing raids designed to deter insurgents) that occurred one day, one week, and one month before each airstrike (3 spatial surfaces). Patterns of U.S. aid spending might also affect the location and number of insurgent attacks and airstrikes, as we discussed in Section~\ref{sec:iraq}. We therefore include the amount of aid spent (in U.S. dollars) in each Iraqi district in the past month as a time-varying covariate (1 spatial surface). Finally, we also incorporate several time-invariant spatial covariates, including the airstrike's distance from major cities, road networks, rivers, and the population (logged, measured in 2003) of the governorate in which the airstrike took place (4 spatial surfaces). Lastly, we include separate predictors for distances from local settlements in each of the Iraqi districts to incorporate any area specific effects (18 spatial surfaces).

We evaluate the covariate balance by comparing the $p$-values of estimated coefficients in the propensity score model to the $p$-values in the weighted version of the same model, where each time period is inversely weighted by its truncated propensity score estimate (truncated above at the 90$^{th}$ quantile). Although 13 out of 35 estimated coefficients had $p$-values smaller than 0.05 in the fitted propensity score model, all the $p$-values in the weighted propensity score model are close to 1, suggesting that the estimated propensity score adequately balances these confounders (see Figure~\ref{app_fig:sims_balance} of Appendix~\ref{app_subsec:add_sims_iraq_balance}).

\subsection{The Choice of the Bandwidth Parameter for the Spatial Kernel Smoother}
\label{subsec:application_bandwidth}

The kernel smoothing part of our estimator is not necessary for estimating the number of points within any set $B \subset \Omega$ since we can simply use an IPW estimator based on the observed number of points within $B$. However, kernel smoothing is useful for visualizing the estimated intensities of insurgent attacks under an intervention of interest over the entire country.  One can also use it to acquire estimates of the expected number of insurgent attacks under the intervention for any region of Iraq by considering the intensity's integral over the region. \cref{theorem:normality} shows that, for any set $B\subset \Omega$, kernel smoothing does not affect the estimator's asymptotic normality as long as the bandwidth converges to zero.  In practice, the choice of the bandwidth should be partly driven by the size of the sets $B$.

In our analysis, we estimate the causal quantities for the entire country and the Baghdad administrative unit. We choose an adaptive bandwidth separately for each outcome using the \texttt{spatstat} package in R. We consider all observed outcome event locations during our study period, and use Scott's criterion for choosing an optimal, constant bandwidth parameter for isotropic kernel estimation \citep{scott1992multivariate}. Using the estimated density as the pilot density, we calculate the optimal adaptive bandwidth surface according to Abramson's inverse-square-root rule \citep{abramson1982bandwidth}. This procedure yields a value of the bandwidth used for kernel smoothing at each outcome event location.

\subsection{Findings}

\begin{figure}[p]
\centering
\begin{minipage}{0.75\textwidth}
\includegraphics[width =\textwidth]{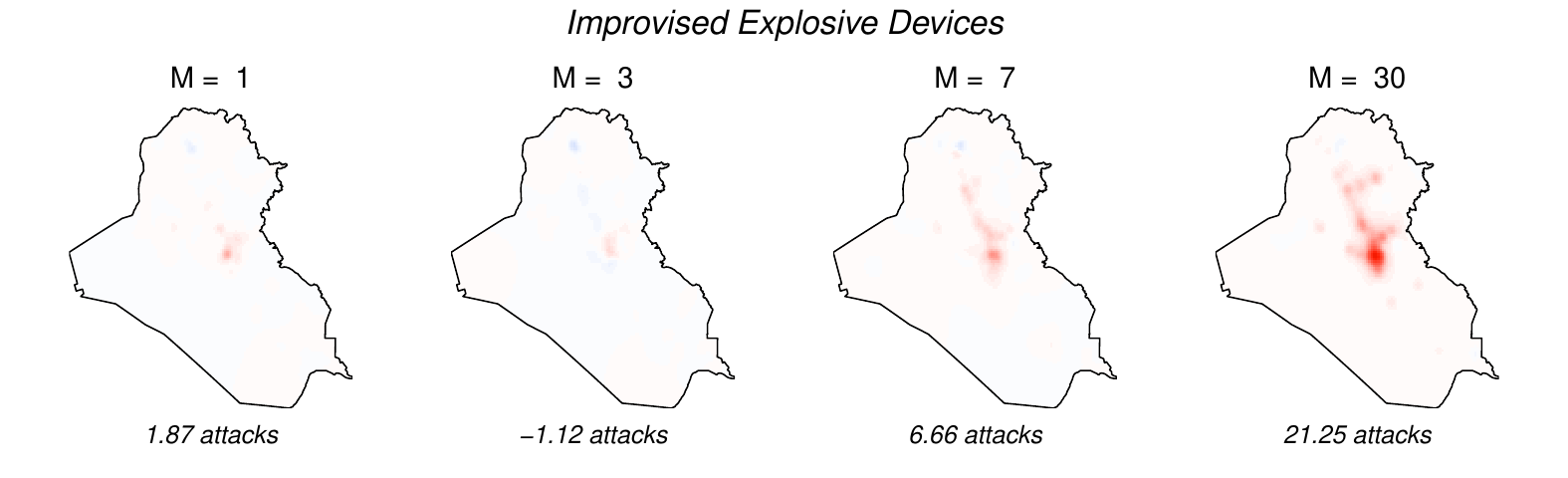} \\
\subfloat[Increasing the expected number of airstrikes from 1 to 6 per day.]{
\includegraphics[width =\textwidth]{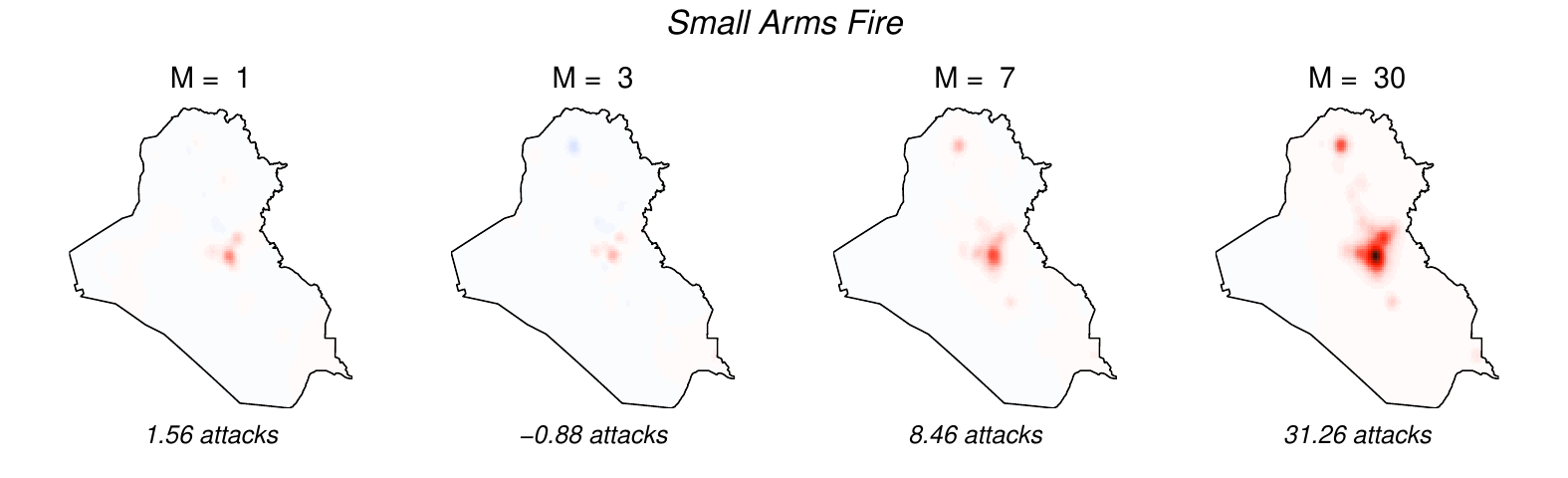}} \\[20pt]
\includegraphics[width =\textwidth]{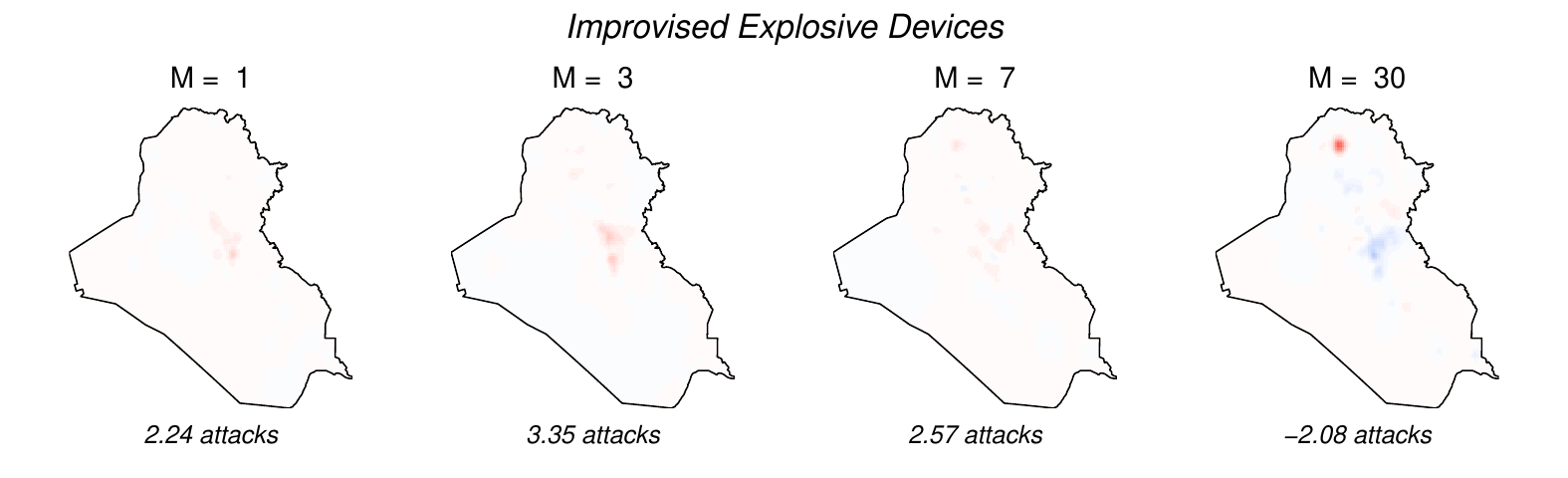} \\
\subfloat[Increasing the priority of Baghdad as focal point of airstrikes from $\alpha = 0$ to $\alpha = 3$.]{
\includegraphics[width =\textwidth]{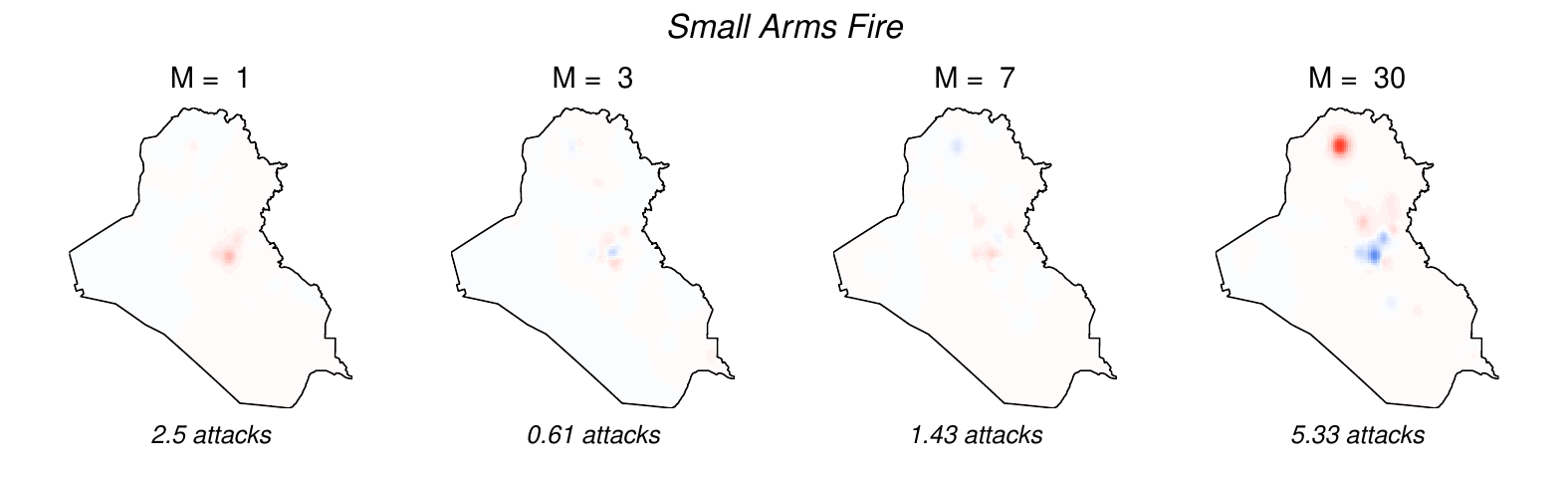}}
\end{minipage}
\begin{minipage}{0.065\textwidth}
\includegraphics[width = \textwidth]{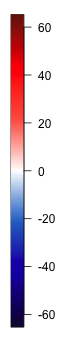}
\end{minipage}
\caption{Changes in Estimated Intensity of Insurgent Attacks when Increasing the Expected Number of Airstrikes (the first two rows) and when Shifting the Focal Point of Airstrikes to Baghdad (the bottom two rows).  Insurgent attacks are measured using Improvised Explosive Devices (IEDs; the first and third rows) and small arms fire (SAFs; second and fourth rows) with the varying number of intervention duration, $\lag = 1, 3, 7, 30$ days (columns). The number shown below each map represents the estimated change in the total number of attacks per day over the entire country, whereas the legend represents the difference in estimated intensities.}
\label{fig:res_IED_SAF}
\end{figure}

\cref{fig:res_IED_SAF} illustrates changes in the estimated intensity surfaces for insurgent attacks (measured using IEDs and SAFs) when increasing the expected number of airstrikes (the first two rows) and when shifting the focal point of airstrikes to Baghdad (the bottom two rows), with the varying duration of interventions, $M=1,3,7,30$ days (columns).  These surfaces can be used to estimate the causal effect of a change in the intervention over any region. Dark blue areas represent areas where the change in the military strategy would reduce insurgent attacks, whereas red areas correspond to those with an increase in insurgent attacks. Statistical significance of these results is shown in \cref{app_tab:res_tau}.

The figure reveals a number of findings.  First, we find no substantial change in insurgent attacks if these interventions last only for one or three days.  When increasing airstrikes for a longer duration, however, a greater number of insurgent attacks are expected to occur.  These changes are concentrated in the Baghdad area and the roads that connect Baghdad and the northern city of Mosul. These patterns apply to both IEDs and SAFs with slightly greater effects estimated for SAFs. Under \cref{ass:unmeasured_conf}, these results suggest that, far from suppressing insurgent attacks, airstrikes actually may {\it increase} them over time.  In this setting, airstrikes can be counterproductive, failing to reduce insurgent violence while also victimizing civilians.  We emphasize that \cref{ass:unmeasured_conf} may be violated and address this issue through our sensitivity analysis.

Under our assumptions, we find that the effect estimates for shifting the focal point of airstrikes to Baghdad for 1, 3, or 7 days are close to null.  However, when the intervention change lasts for 30 days, our analysis suggests that insurgents may shift their attacks to the areas around Mosul while reducing the number of attacks in Baghdad.  This displacement pattern is particularly pronounced for SAFs.  For SAFs, insurgents appear to move their attacks to the Mosul area even with the intervention of 7 days, though the effect size is smaller. In short, the effects of airstrikes may not be localized, but instead can {\it ripple over long distances} as insurgents respond in different parts of the country. Unlike existing approaches which focus on the effect of an intervention in the nearby area \citep[e.g][]{schutte2014matched}, our approach captures this often-considerable displacement of violence.

\begin{figure}[p]
\centering
\subfloat[Estimated effect of increasing the expected number of airstrikes for $M$ days]{\hspace{8pt}
\includegraphics[width=0.93\textwidth]{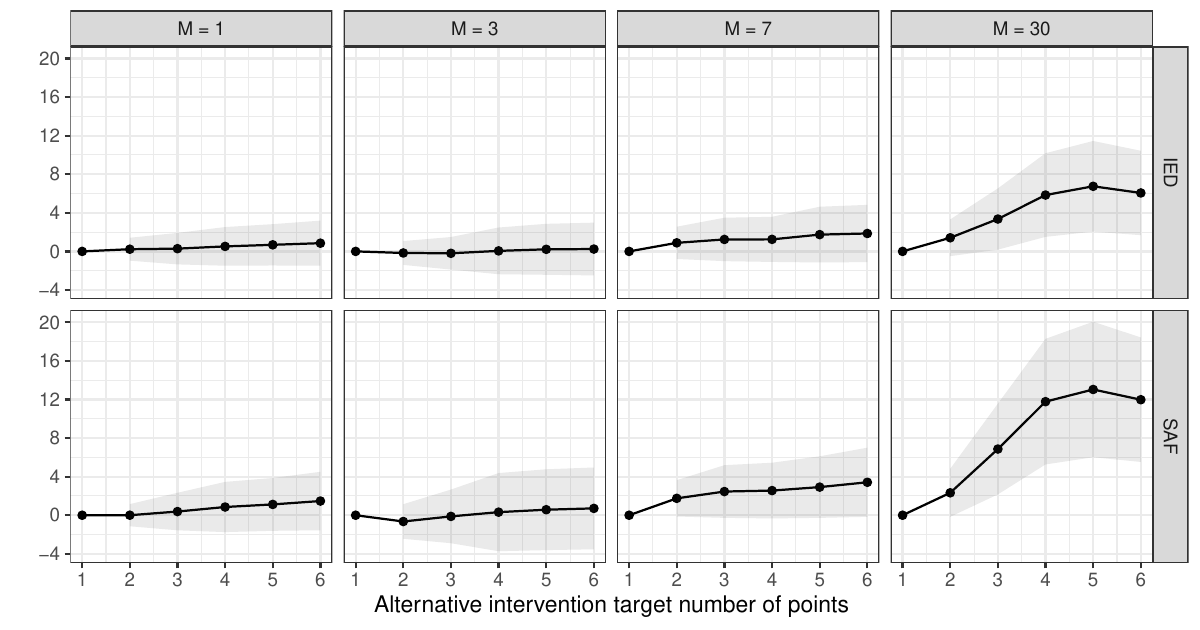}
\label{fig:res_tau_IED_SAF_Type-1}
} \newline
\subfloat[Estimated effect of increasing the expected number of airstrikes $M$ days ago]{
\includegraphics[width=0.93\textwidth]{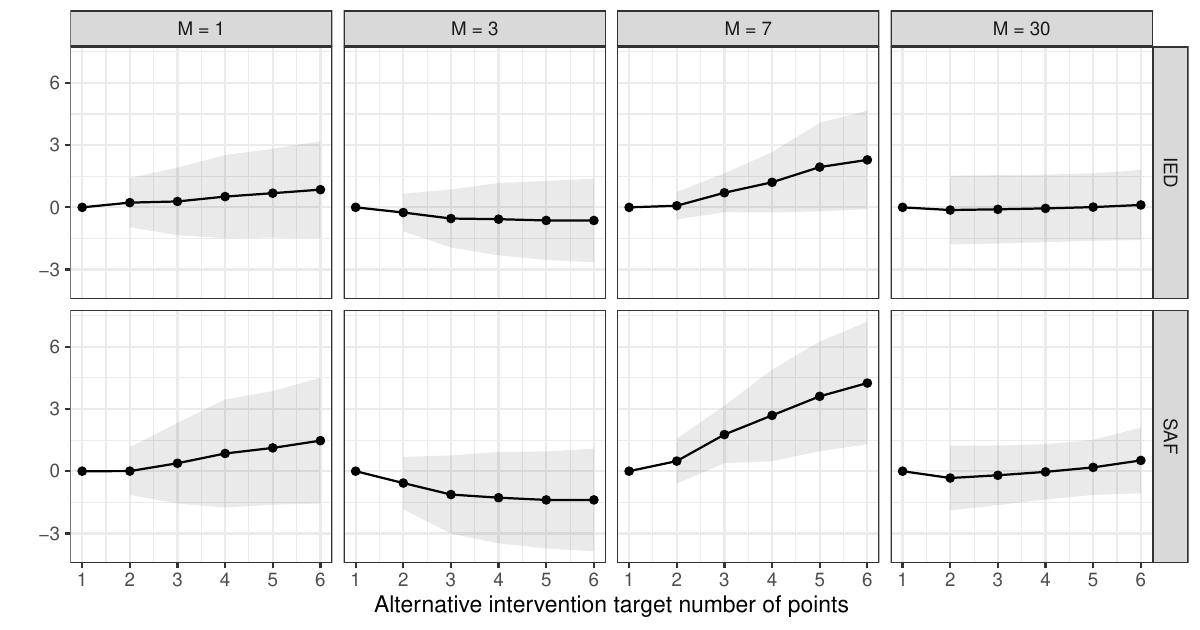}
\label{fig:res_tau_IED_SAF_Lagged}
}
\caption{Changes in the Estimated Number of Insurgency Attacks {\it in Baghdad} when Increasing the Expected Number of Airstrikes (a) for $\lag$ days, and (b) $\lag$ days ago. In each plot, the horizontal axis represents the expected number of airstrike per day under the alternative intervention. The vertical axis represents a change in the estimated average number of insurgency attacks in Baghdad for IEDs (first \& third row) and SAFs (second \& fourth row) when number of airstrikes per day increases from 1 to the value on the horizontal axis. Each column shows different (a) duration or (b) lag length of intervention, $M=1, 3, 7, 30$ days. 95\% confidence intervals are shown as grey bands.}
\end{figure}

\cref{fig:res_tau_IED_SAF_Type-1} shows the changes in the estimated average number of insurgent attacks {\it in Baghdad} as the expected number of airstrikes increases from 1 to $2, 3, \dots, 6$ airstrikes per day in the entire country (horizontal axis).  We also vary the duration of intervention from $M=1$ day to $\lag = 30$ days (columns).  Both the point estimate (solid lines) and 95\% CIs (grey bands) are shown.  Consistent with \cref{fig:res_IED_SAF}, we find that increasing the number of airstrikes leads to a greater number of attacks when the duration of intervention is 7 or 30 days.  These effects appear to be smaller when the intervention is much shorter. The patterns are similar for both IEDs and SAFs.

\cref{fig:res_tau_IED_SAF_Lagged} shows the change in the estimated number of IEDs and SAFs attacks in Baghdad when increasing the number of airstrikes $\lag$ {\it days before}, while the expected number of airstrikes during the following $\lag -1$ days equals one per day.  We find that all estimated lagged effects for $\lag = 3$ are negative, whereas the estimated lagged effects for $\lag = 7$ are positive. This suggests that increasing the number of airstrikes may reduce insurgent violence in a short term while leading to an increase in a longer term. \cref{app_tab:res_tau} presents the effect estimates and 95\% CIs for various interventions and outcomes.

We interpret these localized effects around Baghdad as consistent with prior claims \citep[e.g][]{hashim11} that Sunni insurgents were sufficiently organized to shift their attacks to new fronts in response to American airstrikes. That is, while heavy bombardment in Baghdad might suppress insurgent attacks locally, we find a net increase in overall violence as insurgent commanders displace their violence to new locations such as Mosul that are experiencing less airstrikes. This displacement effect underscores the danger of adopting too-narrow frameworks for casual estimation that miss spillover and other spatial knock-on effects.

We emphasize that the validity of our results hinges on the reliability of our causal assumptions, of which the unconfoundedness assumption \cref{ass:unmeasured_conf} is perhaps the strongest. We evaluate the robustness of our results to violations of this assumption using the sensitivity analysis framework developed in Section~\ref{sec:SA}. We investigate the sensitivity of estimated effects for a change in intervention that corresponds to dosage or increased focus in Baghdad, for all values of $\lag$, for both SAF and IED outcomes, and for effects in the whole country and in Baghdad only. We find that the estimated effects are robust up to the ratio between the misspecified and the true propensity score ($\Gamma$) being bounded by 1.12. The small value of $\Gamma$ indicates that our causal analysis may be sensitive to violations of the unconfoundedness assumption. As discuss before, however, this sensitivity is partially due to the inherently large uncertainty in estimating the point process intensity functions of the propensity scores from sparse data.

\section{Concluding Remarks}
\label{sec:discussion}

In this paper, we provide a framework for causal inference with spatio-temporal point process treatments and outcomes. We illustrate the flexibility of this proposed methodology by applying it to the estimation of airstrike effects on insurgent violence in Iraq. Our central idea is to use a stochastic intervention that represents a distribution of treatments rather than the standard causal inference approach that estimates the average potential outcomes under some fixed treatment values. A key advantage of our approach is its flexibility: it permits unstructured patterns of both spatial spillover and temporal carryover effects. This flexibility is crucial since for many spatio-temporal causal inference problems, including our own application, little is known about how the treatments in one area affect the outcomes in other areas across different time periods.

The estimands and methodology presented in this paper can be applied in a number of settings to estimate the effect of a particular stochastic intervention strategy.  There are several considerations that may be useful when defining a stochastic intervention of interest. First, the choice of intervention should be guided by pressing policy questions or important academic debates where undetected spillover might frustrate traditional methods of causal inference. Second, stochastic interventions should satisfy the overlap assumption (\cref{ass:positivity}). Researchers should not define a stochastic intervention that generates treatment patterns that appear to be far different from those of the observed treatment events. In our application, we achieve this by constructing the stochastic interventions based on the estimated density of point patterns obtained from the past data and the observed number of airstrikes per day.

The proposed framework can also be applied to other high-dimensional, and possibly unstructured, treatments. The standard approach to causal inference, which estimates the causal effects of fixed treatment values, does not perform well in such settings. Indeed, the sparsity of observed treatment patterns alone makes it difficult to satisfy the required overlap assumption \citep{imai:jian:19}. We believe that the stochastic intervention approach proposed here offers an effective solution to a broad class of causal inference problems.  

Future research should further develop the methodology for stochastic interventions.  In particular, it is important to consider an improved weighting method that explicitly targets covariate balance. This might be challenging in the spatiotemporal setting where the notion of covariate balance is not yet well understood. Finally, it is crucial to extend the stochastic intervention framework to adaptive strategies over multiple time periods that might be more reflective of realistic assignments.

\bibliographystyle{natbib}
\bibliography{Iraq,intro}
\newpage

\doparttoc 
\faketableofcontents 
\part{} 

\setcounter{page}{1}

\vspace{-20pt}
\begin{center}
{\sc \LARGE Supplementary Appendix for ``Causal Inference with Spatio-Temporal Data''}
\end{center}

\allowdisplaybreaks
\appendix
\setcounter{equation}{0}
\renewcommand{\theequation}{A.\arabic{equation}}
\setcounter{table}{0}
\renewcommand{\thetable}{A.\arabic{table}}
\setcounter{figure}{0}
\renewcommand{\thefigure}{A.\arabic{figure}}
\setcounter{assumption}{0}
\renewcommand{\theassumption}{A.\arabic{assumption}}
\setcounter{remark}{0}
\renewcommand{\theremark}{A.\arabic{remark}}
\setcounter{lemma}{0}
\renewcommand{\thelemma}{A.\arabic{lemma}}
\setcounter{corollary}{0}
\renewcommand{\thecorollary}{A.\arabic{corollary}}
\setcounter{theorem}{0}
\renewcommand{\thetheorem}{A.\arabic{theorem}}

\vspace{-60pt}
\addcontentsline{toc}{section}{Supplement} 
\part{ } 
\parttoc
\clearpage

\setstretch{1.3}


\section{Notation}
\label{app_sec:notation}

{\renewcommand*{\arraystretch}{1.2}
\begin{table}[H] \spacingset{1.25}
{\small \begin{longtable}{ll L{10.5cm}}
\caption{Notation.}
\label{tab:notation} \\ \hline \hline
Paths & $\Whist$ & Treatments over the time periods $1, \ldots, t$ \\
& $\whist$ & Realized treatment assignments for time periods $1,\ldots, t$ \\
& $\anyhist[t][\allout]$ & Collection of all potential outcomes for time periods $1,\ldots, t$
\\
& $\anyhist[t][\bm Y]$ & Observed outcomes for time periods $1, \ldots, t$ \\
\hline
Intervention & $\lag$ & The number of time periods over which we intervene \\
& $\trtintensity$ & Poisson point process intensity defining the stochastic intervention \\
\hline
Estimands & $\numpoints[t][over]$, $\numpoints[][over]$ &
Expected number of outcome-active locations during time period $t$ for an intervention over $\lag$ time periods, and their average over time \\
& $\tau_t^\lag$, $\tau^\lag $ & Expected change in the number of outcome-active locations comparing two interventions for time period $t$ and their average over time \\
\hline
Estimators & $\widehat{Y}_t^\lag$ & Estimated continuous surface the integral of which is used for calculating $\numpoints[t][hat]$ \\
& $\numpoints[t][hat]$, $\numpoints[][hat]$ & Estimated expected number of points during time period $t$ for an intervention taking place over the preceding $\lag$ time periods, and their average over time \\
& $\widehat{\tau}_t^\lag$, $\widehat{\tau}^\lag $ & Estimated expected change in the number of outcome-active locations for time period $t$ comparing two interventions, and their average over time \\
\hline
Arguments & $B$ & The set over which the number of outcome-active locations are counted \\
\hline \hline
\end{longtable}}
\end{table}}

\section{Theoretical Proofs}
\label{app_sec:proofs}

\subsection{Regularity conditions}

For  $\epsilon > 0$, we use $\mathcal{N}_\epsilon(A)$ to denote the $\epsilon-$neighborhood of a set $A$: $\mathcal{N}_\epsilon(A) = \{\omega \in \Omega: \text{there exists } a \in A \text{ with } \text{dist}(\omega, a) < \epsilon\}$. Also, we use $\boundary$ to denote the  boundary of $B$, formally defined as the set of points for which an open ball of any size centered at them includes points both in and outside $B$, i.e., $\boundary = \{s \in \Omega \text{ such that, for every } \epsilon > 0, \text{ there exist } s_1, s_2 \in \mathcal{N}_\epsilon(s) \text{ for which } s_1 \in B \text{ and } s_2 \not\in B \}$.

\paragraph{Regularity conditions for asymptotic results when using the true or estimated propensity score}

The following assumption includes regularity conditions which are used to show asymptotic normality of the estimator based on the true or estimated propensity score:

\begin{assumption} \spacingset{1.25}
\label{ass:regularity_conditions} The following three conditions hold.
\begin{enumerate}[label=(\alph*)]
\item \label{ass:finite_points}
There exists $\bound[Y] > 0$ such that $|\sparseset[Y]| < \bound[Y]$ for all $t \in \alltimes$ and $\whist \in \alltrt^T$.
\item \label{ass:convergent_variance}
Let $\asymvar_t = \Var\left[ \prod_{j = t - \lag + 1}^t \frac{\intervdistf[][T][j]}{\propscore[j][j][W]} N_B(Y_t) \mid \history[t-\lag]^* \right]$ for $t \geq \lag$. Then, there exists $\asymvar \in \mathbb{R}^+$ such that $(T-\lag+1)^{-1} \sum_{t = \lag}^T \asymvar_t \overset{p}{\rightarrow} \asymvar $ as $T \rightarrow \infty$.
\item \label{ass:neighborhood_boundary}
There exists $\bound[B] > 0$ and $Q^* \in (1/2, 1)$ such that
\[ P \left( \sum_{t = \lag}^T I\Big( \exists s \in \sparseset[{}Y] \cap \mathcal{N}_{\bound[B]}(\boundary) \Big) > T^{1-Q^*} \right) \rightarrow 0, \text{ as } T \rightarrow \infty. \]
\end{enumerate}
\end{assumption}
\noindent
\cref{ass:regularity_conditions}\ref{ass:finite_points} states that there is an upper limit on the number of outcome-active locations at any time period and under any treatment path. In our application, it is reasonable to assume that the number of insurgent attacks occurring during any day is bounded. In \cref{ass:regularity_conditions}\ref{ass:convergent_variance}, $\history[t]^*$ represents the expanded history preceding $W_{t+1}$, including previous treatments, all potential outcomes, and all potential confounders. Given the assumptions of bounded relative positivity and bounded number of outcome-active locations, \cref{ass:regularity_conditions}\ref{ass:convergent_variance} is a weak condition, as it states that the average of bounded quantities converges. Lastly, \cref{ass:regularity_conditions}\ref{ass:neighborhood_boundary} states that the probability that we observe more than $T^{1-Q^*}$ time periods with outcome-active locations within a $\bound[B]-$neighborhood of $B$'s boundary goes to zero as the number of observed time periods increases. Since the size of the boundary's neighborhood can be arbitrarily small, this assumption is also reasonable. Informally, \cref{ass:regularity_conditions}\ref{ass:neighborhood_boundary} would be violated in our study if insurgent attacks occurred at the {\it boundary} of region $B$ more often than during $\sqrt T$ time periods. As long as the regions $B$ are decided upon substantive interest, we would expect this assumption to be satisfied. Alternatively, regions $B$ can be defined by avoiding setting the region's boundary at observed outcome-active locations.

\paragraph{Regularity conditions for asymptotic results when using the estimated propensity score}

Next, we formalize the regularity conditions on the propensity score model.  These conditions are used for establishing the asymptotic normality of the estimator based on the estimated propensity score.

\begin{assumption}
Assume that the parametric form of the propensity score indexed by $\pspar$, $f(W_t = w_t \mid \history ; \pspar)$, is correctly specified and differentiable with respect to $\pspar$, and let
$\scorefun = \frac \partial {\partial \pspar} \log f(W_t = w_t \mid \history = \anyhist[t-1][h] ; \pspar) $ be twice continuously differentiable score functions. Let $\pspar_0$ denote the true values of the parameters, where $\pspar_0$ is in an open subset of the Euclidean space. Denote $\filtration[t] = \history[t-\lag+1]^* = \{\Whist[t-\lag+1], \anyhist[T][\allout], \anyhist[T][\allcovs] \}$, as in the proof of \cref{theorem:normality}. We assume that the following conditions hold:
\begin{enumerate}
\item
\begin{enumerate}
\item $E_{\pspar_0} \Big[\| \scorefun[T][T][0] \|^2 \Big] < \infty$,

\item There exists a positive definite matrix $V_{ps}$ such that
$$\frac1T \sum_{t = 1}^T E_{\pspar_0} \Big( \scorefun[T][T][0] \scorefun[T][T][0]^\top \mid \filtration \Big) \overset{p}{\rightarrow} V_{ps}$$
\label{app_cond:ps_information_matrix}

\item \( \displaystyle \frac1T \sum_{t = 1}^T E_{\theta_0} \Big[ \| \scorefun[T][T][0] \|^2 I \Big( \| \scorefun[T][T][0] \| > \epsilon \sqrt{T} \Big) \mid \filtration \Big] \overset{p}{\rightarrow} 0 \), for all $\epsilon > 0$,
\end{enumerate}

\item
\label{ps_ass:conv_zero}
For all $k,j$, if we denote the $k^{th}$ element of the $\scorefun$ vector by $\scorefun[F][F][][k]$ and
$P_{kjt} = \frac{\partial}{\partial \pspar[F]_j} \scorefun[T][T][][k] \big|_{\pspar_0}$, then $E_{\pspar_0}\left[ \left| P_{kjt} \right| \right] < \infty$ and
there exists $0 < r_{kj} \leq 2$ such that
\( \displaystyle
\sum_{t = 1}^T \frac1{t^{r_{kj}}}  E_{\pspar_0} \left( \left| P_{kjt} - E_{\theta_0}(P_{kjt} \mid \filtration) \right|^{r_{kj}} \mid \filtration \right) \overset{p}{\rightarrow} 0 \)

\item There exists an integrable function $\overset{\bigcdot\bigcdot}{\psi}(w_t, \anyhist[t-1][h])$ such that $\overset{\bigcdot\bigcdot}{\psi}(w_t, \anyhist[t-1][h])$ dominates the second partial derivatives of $\scorefun$ in a neighborhood of $\pspar_0$ for all $(w_t, \anyhist[t-1][h])$.
\label{ass:ps_dominated}

\end{enumerate}
\label{ass:propensity_score}
\end{assumption}

\begin{assumption}
Suppose that $\scorefun$ are the score functions of a propensity score model that satisfies \cref{ass:propensity_score}
with true parameters $\pspar_0$, and
$$\displaystyle
s(\anyhist[t-1][h], w_t, y_t; \pspar) = \Bigg[ \prod_{j = t - \lag + 1}^t \frac{\intervdistf[][F](w_j)}{\parpropscore[j][j][w]} \Bigg] N_B(y_t) - \avgout[F][][\lag].$$ Then, the following conditions hold.

\begin{enumerate}
\item There exists $u \in \mathbb{R}^K$ such that
\[
\frac1{T - \lag + 1} \sum_{t = \lag}^T E_{\theta_0} \big[s(\history[t-1], W_t, Y_t; \pspar_0) \scorefun[T][T][0] \mid \filtration \big] \overset{p}{\rightarrow} u,
\]
\item If \( \displaystyle
P_{jt} = \frac \partial{\partial \pspar_j} s(\history[t-1], W_t, Y_t; \pspar) \Big|_{\pspar_0} \), where $\pspar_j$ is the $j^{th}$ entry of $\pspar$, then there exists $r_j \in (0, 2]$ such that
\[
\sum_{t = 1}^T \frac1{t^{r_j}}  E_{\pspar_0} \left( \left| P_{jt} - E_{\pspar_0}(P_{jt} \mid \filtration) \right|^{r_j} \mid \filtration \right) \overset{p}{\rightarrow} 0.
\]
\end{enumerate}

\label{app_ass:estimator_ps}
\end{assumption}

\begin{remark}
Given the previous assumptions, \cref{app_ass:estimator_ps} is quite weak. We look at the two parts separately:
\begin{enumerate}
\item
For the $k^{th}$ entry, we can write:
\begin{align*}
& \frac1{T - \lag + 1} \sum_{t = \lag}^T \left| E_{\theta_0} \big[s(\history[t-1], W_t, Y_t; \pspar_0) \scorefun[T][T][0][k] \mid \filtration \big] \right| \leq \\
& \leq \frac1{T - \lag + 1} \sum_{t = \lag}^T \sqrt{E_{\theta_0} \big[s(\history[t-1], W_t, Y_t; \pspar_0)^2 \mid \filtration \big] } \ \sqrt{ E_{\theta_0} \big[ \scorefun[T][T][0][k]^2 \mid \filtration \big]}
\tag{Cauchy-Schwarz} \\
& \leq \frac1{2(T - \lag + 1)} \sum_{t = \lag}^T \left( E_{\theta_0} \big[s(\history[t-1], W_t, Y_t; \pspar_0)^2 \mid \filtration \big] + E_{\theta_0} \big[ \scorefun[T][T][0][k]^2 \mid \filtration \big] \right) \tag{$2ab \leq a^2 + b^2$} \\
& \overset{p}{\rightarrow} \frac12 \left(\asymvar + [V_{ps}]_{kk} \right).
\end{align*}
The proof that the first part converges to $v$ will be shown in Equation~\cref{proof_eq:asym_var}, and the second part is based on \cref{ass:propensity_score}, where $[V_{ps}]_{kk}$ denotes the $k^{th}$ diagonal entry of $V_{ps}$. Since the expression is already bounded at the limit, the assumption that it converges is reasonable. Furthermore, we have that
$|u_k| \leq \frac12 (\asymvar + [V_{ps}]_{kk})$, where $u_k$ is
the $k^{th}$ entry of $u$.
\label{remark:regularity_condition}

\item This assumption limits how much the derivative of \( \displaystyle
s(\anyhist[t-1][h], w_t, y_t; \pspar) \) can vary around its conditional expectation. As we will see in \cref{app_lemma:a_psi}, this derivative can be re-written as a sum that involves three terms: the number of outcome active locations, the inverse probability ratios, and the score functions. The first two of these terms are bounded, and \cref{ass:propensity_score} already controls how variable the score functions can be.  Thus, this assumption is also reasonable.

\end{enumerate}
\end{remark}

\subsection{Proofs: The propensity score as a balancing score}

\renewcommand*{\proofname}{\textbf{Proof of \cref{theorem:balancing_score}}}
\begin{proof}
Note that $f(W_t = w \mid \propscore,\history) = f(W_t = w \mid \history) = \propscore$ since $\propscore$ is a function of $\history$.
Therefore, it suffices to show that $f(W_t = w \mid \propscore) = \propscore$:
\begin{align}
f(W_t = w \mid \propscore) &= E[f(W_t = w \mid \history) \mid \propscore] = E[\propscore \mid \propscore] = \propscore.
\label{app_eq:in_proof_balance}
\end{align}
\end{proof}

\renewcommand*{\proofname}{\textbf{Proof of \cref{theorem:ps_unconfoundedness}}}
\begin{proof}
\begin{align*}
f \big(W_t = w & \mid \Whist[t - 1], \anyhist[T][\allout] , \anyhist[T][\allcovs] \big) \\
&= f \big(W_t = w \mid \history, \Whist[t - 1], \anyhist[T][\allout] , \anyhist[T][\allcovs] \big)
\tag{Since $\history \subset \{\Whist[t-1], \anyhist[T][\allout] , \anyhist[T][\allcovs] \} $}\\
&= f \big(W_t = w \mid \history \big)
\tag{From \cref{ass:unmeasured_conf}} \\
&= \propscore \\
&= f(W_t = w \mid \propscore) \tag{From \cref{app_eq:in_proof_balance}}
\end{align*}
%
%
\end{proof}

\subsection{Proofs: Asymptotic normality based on the true propensity score}
\label{subsec:proofs_truePS}

\renewcommand*{\proofname}{\textbf{Proof of \cref{theorem:normality}}}

\begin{proof}

Note that the collection of variables temporally precedent to treatment at time period $t$ is the expanded history $\history^*$, defined in Assumption \ref{ass:regularity_conditions}. The expanded history $\history^*$ is a filtration generated by the collection of potential confounders $\anyhist[T][\allcovs]$, the collection of potential outcomes $\anyhist[T][\allout]$, and the previous treatments, and satisfies $\history^* \subset \history[t]^*$.

Let $\error = \estimatorNt - \avgout$ be the estimation error for time period $t$ and lag $\lag$. We will decompose $\error$ in two components, one corresponding to the error due to the treatment assignment ($A_{1t}$), and the other corresponding to the error due to spatial smoothing ($A_{2t}$). Since the bandwidth parameter of the kernel depends on $T$, we write $K_{b_T}$ instead of $K_b$. Specifically,
\begin{equation}
\begin{aligned}
\error & \ = \ \Bigg[ \prod_{j = t - \lag + 1}^t \frac{\intervdistf[][F](W_j)}{\propscore[j][j][W]} \Bigg]
\int_B \sum_{s \in \sparseset[{}Y]} K_{b_T}(\omega, s) \mathrm{d}\omega
- \avgout[F][][\lag] \\
& \ = \ \underbrace{\Bigg[ \prod_{j = t - \lag + 1}^t \frac{\intervdistf[][F](W_j)}{\propscore[j][j][W]} \Bigg] N_B(Y_t) - \avgout[F][][\lag]}_{\quant[1]} + \\
& \hspace{40pt}
   \underbrace{\Bigg[ \prod_{j = t - \lag + 1}^t \frac{\intervdistf[][F](W_j)}{\propscore[j][j][W]} \Bigg] \Bigg[
    \int_B \sum_{s \in \sparseset[{}Y]} K_{b_T}(\omega, s) \mathrm{d}\omega -
    N_B(Y_t) \Bigg]}_{\quant[2]}.
\end{aligned}
\label{app_eq:est_error_split}
\end{equation}
We show that
\begin{enumerate}
    \item $\sqrt{T}\big(\frac1{T-\lag+1} \sum_{t = \lag}^T \quant[1] \big)$ is asymptotically normal, and
    \item $\sqrt{T}\big(\frac1{T-\lag+1} \sum_{t = \lag}^T \quant[2] \big)$ converges to zero in probability.
\end{enumerate}

\paragraph{Asymptotic normality of the first error.}
$\ $\\[10pt]
We use the central limit theorem for martingale difference series (Theorem 4.16 of \cite{VanDerVaart2010timeseries}) to establish the asymptotic normality of $(T-\lag+1)^{-1}\sum_{t = \lag}^T \quant[1]$.

\begin{claim*} \normalfont
$\quant$ is a martingale difference series with respect to the filtration $\filtration[t] = \history[t-\lag+1]^*$.
\end{claim*}
\noindent
To prove this, we show that $E(|\quant|) < \infty$ and $E(\quant \mid \filtration) = E(\quant[1][t] \mid \history[t - \lag]^*) = 0$. For the first part, Assumptions~\ref{ass:positivity}~and~\cref{ass:regularity_conditions}\ref{ass:finite_points} imply that $\quant$ is bounded and hence $E[|\quant|] < \infty$:
\begin{equation}
|\quant| \  \leq \  \Bigg| \prod_{j = t - \lag + 1}^t \frac{\intervdistf[][F](W_j)}{\propscore[j][j][W]} N_B(Y_t) \Bigg| + \Big| \avgout[F][][\lag] \ \Big| \ \leq \ \bound[W]^M \bound[Y] + \bound[Y]
\label{proof_eq:bounded_A1t}
\end{equation}
For the second part, it suffices to show
$$ E\Bigg\{
\Bigg[ \prod_{j = t - \lag + 1}^t \frac{\intervdistf[][F](W_j)}{\propscore[j][j][W]} \Bigg] N_B(Y_t) \ \Big| \ \history[t - \lag]^* \Bigg \} \ = \ \avgout[F][][\lag], $$
where the expectation is taken with respect to the assignment of treatments $\bW_{(t - \lag + 1):t}$.
\begin{align*}
 & \quad E\Bigg\{  \Bigg[ \prod_{j = t - \lag + 1}^t \frac{\intervdistf[][F](W_j)}{\propscore[j][j][W]}\Bigg]
N_B(Y_t) \ \Big | \  \history[t - \lag]^* \Bigg \} \\
&= \int \Bigg[ \prod_{j = t - \lag + 1}^t \frac{\intervdistf[][F](w_j)}{\propscore[j][j]}\Bigg]
N_B \Big(Y_t \big(\Whist[t-\lag], \underbrace{w_{t - \lag + 1}, \dots, w_t}_{\bw_{(t-M+1):t}} \big) \Big) \times \\
& \hspace{0.4in}
f(w_{t - \lag + 1} \mid \history[t-\lag]^*)
f(w_{t - \lag + 2} \mid \history[t-\lag]^*, W_{t - \lag + 1}) \cdots \times  \\
& \hspace{0.4in}
f(w_t \mid \history[t-\lag]^*, \bW_{(t - \lag + 1):(t - 1)}) \ \mathrm{d}\bw_{(t-M+1):t} \\
&= \int \Bigg[ \prod_{j = t - \lag + 1}^t \frac{\intervdistf[][F](w_j)}{\propscore[j][j]}\Bigg]
N_B\Big(Y_t\big(\Whist[t-\lag], w_{t - \lag + 1}, \dots, w_t\big)\Big) \times \\
& \hspace{0.4in}
f(w_{t - \lag + 1} \mid \history[t-\lag]^*)
f(w_{t - \lag + 2} \mid \history[t-\lag+1]^*) \cdots
f(w_t \mid \history[t-1]^*) \ \mathrm{d}\bw_{(t-M+1):t}
\tag{because $\history[t' +1]^* = \history[t']^* \cup \{W_{t' + 1}\}$} \\
&= \int N_B \Big(Y_t\big(\Whist[t-\lag], w_{t - \lag + 1}, \dots, w_t\big)\Big)
\left[ \prod_{j = t - \lag + 1}^t \intervdistf[][F](w_j) \right] \ \mathrm{d}\bw_{(t-M+1):t}
\tag{By \cref{ass:unmeasured_conf}} \\
&= \avgout[F][][\lag]. \numberthis \label{app_eq:expectation_A1t}
\end{align*}
This proves that $\quant$ is a martingale difference series with respect to filtration $\filtration$.


\begin{claim*} \normalfont
$(T - \lag + 1)^{-1} \sum_{t = \lag}^T E\{\quant^2 I(|\quant| > \epsilon \sqrt{T - \lag + 1}) \mid \filtration\} \overset{p}{\rightarrow} 0$ for every $\epsilon > 0$.
\end{claim*}
\noindent
Let $\epsilon > 0$. Note that $\quant$ is bounded by $\bound[Y](\bound[W]^\lag +1)$ (see Equation~\cref{proof_eq:bounded_A1t}). Choose $T_0$ as
\begin{align*}
T_0 & \ = \ \underset{t \in \mathbb{N}^+}{\text{argmin}} \{ \epsilon\sqrt{t - \lag + 1} > \bound[Y](\bound[W]^\lag + 1) \} \\
& \ = \ \underset{t \in \mathbb{N}^+}{\text{argmin}} \Big\{ t > \lag - 1 + \Big[ \frac{\bound[Y](\bound[W]^\lag + 1)}{\epsilon} \Big]^2 \Big\} \\
& \ = \ \ceil[\Bigg]{\lag - 1 + \Big[ \frac{\bound[Y](\bound[W]^\lag + 1)}{\epsilon} \Big]^2}.
\end{align*}
Then, for $T > T_0$, we have that $\epsilon \sqrt{T - \lag + 1} > \epsilon \sqrt{T_0 - \lag + 1} > \bound[Y](\bound[W]^\lag +1)$ which leads to $I(|\quant| > \epsilon \sqrt{T+\lag+1}) = 0$ and
$ E(\quant^2 I(|\quant| > \epsilon \sqrt{T - \lag + 1}) \mid \filtration) = 0. $
This proves the claim.\\

We now combine the above claims to establish the asymptotic normality of the first error.  Since $\quant$ has mean zero, $E(\quant^2 \mid \filtration) = \Var(\quant \mid \filtration)$, and since $\avgout$ is fixed, \[ \Var(\quant \mid \filtration) =
\Var \left( \prod_{j = t - \lag + 1}^t \frac{\intervdistf[][F](W_j)}{\propscore[j][j][W]} N_B(Y_t) \mid \history[t-\lag]^* \right) \]
which yields
\begin{equation}
\frac1{T-\lag +1} \sum_{t = \lag}^T E(\quant^2 \mid \filtration)  \overset{p}{\rightarrow} \asymvar,
\label{proof_eq:asym_var}
\end{equation}
from Assumption~\ref{ass:regularity_conditions}\ref{ass:convergent_variance}.
Combining these results, using that $\sqrt{T} / \sqrt{T - \lag + 1} \rightarrow 1$ and Theorem 4.16 of \cite{VanDerVaart2010timeseries}, we have the desired result,
$$
\sqrt{T} \left( \frac{1}{T - \lag + 1} \sum_{t = \lag}^T \quant \right) \overset{d}{\rightarrow} N(0, v).
$$

\paragraph{Convergence to zero of the second error.}
$\ $\\[10pt]
The second error compares the integral of the kernel-smoothed outcome surface over the region of interest $B$ with the actual number of points within the set $B$. We show that as $T$ goes to infinity, and since the bandwidth of the kernel converges to 0, the error due to kernel smoothing also goes to zero.
Specifically, we will show that
$$ \sqrt{T} \left(\frac{1}{T-\lag+1}\sum_{t = \lag}^T \quant[2] \right) \overset{p}{\rightarrow} 0.$$

\noindent
Let $c_t = \prod_{j = t - \lag + 1}^t \intervdistf[][F](W_j) / \propscore[j][j][W]$, and write
\begin{align*}
\left| \frac1{T-\lag+1} \sum_{t = \lag}^T \quant[2] \right|
& \ = \ \left| \frac1{T-\lag+1} \sum_{t = \lag}^T c_t
\left[ \int_B \sum_{s \in \sparseset[{}Y]} K_{b_T}(\omega;s) \mathrm{d}\omega - N_B(Y_t) \right] \right|.
\end{align*}
Then,
\begin{align*}
& \int_B \sum_{s \in \sparseset[{}Y]} K_{b_T}(\omega;s) \mathrm{d}\omega - N_B(Y_t) \\
= \ & \sum_{s \in \sparseset[{}Y]\cap B} \int_B K_{b_T}(\omega;s) \mathrm{d}\omega +
\sum_{s \in \sparseset[{}Y] \cap B\complement} \int_B K_{b_T}(\omega;s) \mathrm{d}\omega -
N_B(Y_t) \\
= \ & \sum_{s \in \sparseset[{}Y]\cap B} \Big[1 - \int_{B\complement} K_{b_T}(\omega;s) \mathrm{d}\omega \Big] +
\sum_{s \in \sparseset[{}Y] \cap B\complement} \int_B K_{b_T}(\omega;s) \mathrm{d}\omega -
N_B(Y_t) \\
=\ &  \sum_{s \in \sparseset[{}Y] \cap B\complement} \int_B K_{b_T}(\omega;s) \mathrm{d}\omega -
\sum_{s \in \sparseset[{}Y]\cap B} \int_{B\complement} K_{b_T}(\omega;s) \mathrm{d}\omega.
\end{align*}
This shows that the error from smoothing the outcome surface at time $t$ comes from (1) the kernel weight from points outside of $B$ that falls within $B$, and (2) the kernel weight from points inside $B$ that falls outside $B$.
Using this, we write:
\begin{align*}
& \left| \frac1{T-\lag+1} \sum_{t = \lag}^T \quant[2] \right| \ = \\
& \hspace{20pt} \left| \frac1{T-\lag+1} \sum_{t = \lag}^T c_t \Bigg[ \sum_{s \in \sparseset[{}Y] \cap B\complement} \int_B K_{b_T}(\omega;s) \mathrm{d}\omega -
\sum_{s \in \sparseset[{}Y]\cap B} \int_{B\complement} K_{b_T}(\omega;s) \mathrm{d}\omega \Bigg] \right|.
\end{align*}

Take $\epsilon > 0$, and $Q \in (1/2, Q^*)$ where $Q^*$ is the one in \cref{ass:regularity_conditions}\ref{ass:neighborhood_boundary}. Then, we will show that $P ( T^Q \{ | \frac1{T-\lag+1} \sum_{t = \lag}^T \quant[2] |
\} > \epsilon) \rightarrow 0$ as $T \rightarrow \infty$, which implies that the second error converges to zero faster than $\sqrt{T}$ (since $Q  > 1/2$).
{\small
\begin{align*}
& P \Bigg( T^Q \Bigg\{ \Bigg| \frac1{T-\lag+1} \sum_{t = \lag}^T \quant[2] \Bigg|
\Bigg\} > \epsilon \Bigg) \\
= \ & P \Bigg( \Bigg| \frac1{T-\lag+1} \sum_{t = \lag}^T c_t \Bigg[ \sum_{s \in \sparseset[{}Y] \cap B\complement} \int_B K_{b_T}(\omega;s) \mathrm{d}\omega -
\sum_{s \in \sparseset[{}Y]\cap B} \int_{B\complement} K_{b_T}(\omega;s) \mathrm{d}\omega \Bigg] \Bigg| > \frac{\epsilon}{T^Q} \Bigg) \\
 \leq \ & P \Bigg( \frac1{T-\lag+1} \sum_{t = \lag}^T c_t \sum_{s \in \sparseset[{}Y] \cap B\complement} \int_B K_{b_T}(\omega;s) \mathrm{d}\omega > \frac{\epsilon}{2 T^Q} \Bigg) + \\
& \hspace{40pt}
P \Bigg( \frac1{T-\lag+1} \sum_{t = \lag}^T c_t
\sum_{s \in \sparseset[{}Y]\cap B} \int_{B\complement} K_{b_T}(\omega;s) \mathrm{d}\omega > \frac{\epsilon}{2T^Q} \Bigg),
\end{align*}
where the last equation holds because $|A - B| > \epsilon$ implies that at least one of $|A|, |B| > \epsilon / 2$. Also, since  all quantities are positive, we can drop the absolute value. Then, since $c_t \leq \bound[W]^\lag$ from Assumption~\ref{ass:positivity},
\begin{align*}
& P \Bigg( T^Q \Bigg\{ \Bigg| \frac1{T-\lag+1} \sum_{t = \lag}^T \quant[2] \Bigg|
\Bigg\} > \epsilon \Bigg) \\
\leq \ & P \Bigg( \frac1{T-\lag+1} \sum_{t = \lag}^T \sum_{s \in \sparseset[{}Y] \cap B\complement} \int_B K_{b_T}(\omega;s) \mathrm{d}\omega > \frac{\epsilon}{2 T^Q \bound[W]^\lag} \Bigg) + \\
& \hspace{60pt}
P \Bigg(\frac1{T-\lag+1} \sum_{t = \lag}^T
\sum_{s \in \sparseset[{}Y]\cap B} \int_{B\complement} K_{b_T}(\omega;s) \mathrm{d}\omega > \frac{\epsilon}{2 T^Q \bound[W]^\lag} \Bigg).
\end{align*}}

Use $\spoint[out]$ to denote the point in $\sparseset[{}Y]$ that lies outside B and is the closest to $B$:
\(\displaystyle \spoint[out] = \{s \in \sparseset[{}Y] \cap B\complement: \text{dist}(s, B) = \min_{s' \in \sparseset[{}Y] \cap B\complement} \text{dist}(s', B) \} \).
Similarly, $\spoint$ is the point in $\sparseset[{}Y] \cap B$ that is closest to $B\complement$. These points are shown graphically in \cref{app_fig:proof_outline}.
\begin{figure}[!t]
\centering
\includegraphics[width=0.4\textwidth]{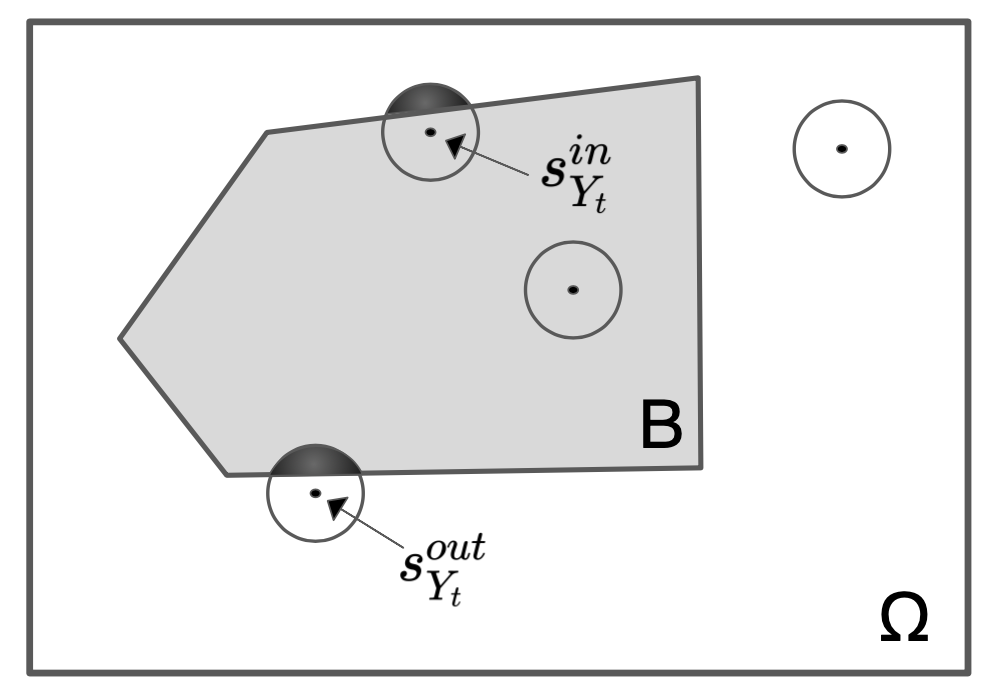}
\caption{Kernel-smoothed outcome surface, and points $\spoint[in], \spoint[out]$ as the points closest to the boundary of $B$ that lie within and outside $B$ respectively. The amount of kernel weight falling within $B$ from points outside of $B$ is necessarily less or equal to the kernel weight from $\spoint[out]$ (shaded), and similarly for $\spoint[in]$.}
\label{app_fig:proof_outline}
\end{figure}
Because there are at most $\bound[Y]$ outcome-active locations, from the definition of $\spoint[in], \spoint[out]$, and because kernels are defined to be decreasing in distance, we have that
\begin{align*}
P \Bigg( T^Q \Bigg\{ \Bigg| & \frac1{T-\lag+1} \sum_{t = \lag}^T \quant[2] \Bigg|
\Bigg\} > \epsilon \Bigg) \\
 \leq \ & P \Bigg( \frac1{T-\lag+1} \sum_{t = \lag}^T \int_B K_{b_T}(\omega; \spoint[out]) \mathrm{d}\omega > \frac{\epsilon}{2 T^Q \bound[W]^\lag \bound[Y]} \Bigg) \\
& \hspace{40pt} +
P \Bigg( \frac1{T-\lag+1} \sum_{t = \lag}^T
\int_{B\complement} K_{b_T}(\omega; \spoint[in]) \mathrm{d}\omega > \frac{\epsilon}{2 T^Q \bound[W]^\lag \bound[Y]} \Bigg) \\
= \ & \underbrace{P \Bigg( \sum_{t = \lag}^T \int_B K_{b_T}(\omega; \spoint[out]) \mathrm{d}\omega > \frac{\epsilon (T-\lag+1)}{2 T^Q \bound[W]^\lag \bound[Y]} \Bigg)}_{\quantt} \\
& \hspace{40pt} +
\underbrace{P \Bigg( \sum_{t = \lag}^T
\int_{B\complement} K_{b_T}(\omega; \spoint[in]) \mathrm{d}\omega > \frac{\epsilon (T-\lag+1)}{2 T^Q \bound[W]^\lag \bound[Y]} \Bigg)}_{\quantt[2]}.
\end{align*}
%
We show that $\quantt, \quantt[2]$ converge to zero separately. Take $\quantt$:
{\small
\begin{align*}
\quantt \ & = \ P \Bigg( \sum_{t = \lag}^T \int_B K_{b_T}(\omega; \spoint[out]) \mathrm{d}\omega > \frac{\epsilon (T-\lag+1)}{2 T^Q \bound[W]^\lag \bound[Y]} \ \Bigg\vert \ \sum_{t=\lag}^T I(\spoint[out] \in \mathcal{N}_{\bound[B]}(\boundary) ) > T^{1-Q^*} \Bigg) \\
& \hspace{6cm}
\times \ P \Bigg( \sum_{t=\lag}^T I(\spoint[out] \in \mathcal{N}_{\bound[B]}(\boundary)) > T^{1-Q^*} \Bigg) \\
& \hspace{0.5cm} + P \Bigg( \sum_{t = \lag}^T \int_B K_{b_T}(\omega; \spoint[out]) \mathrm{d}\omega > \frac{\epsilon (T-\lag+1)}{2 T^Q \bound[W]^\lag \bound[Y]} \ \Bigg\vert \ \sum_{t=\lag}^T I(\spoint[out] \in \mathcal{N}_{\bound[B]}(\boundary)) \leq T^{1-Q^*} \Bigg)  \\
& \hspace{6cm}
\times \ P \Bigg( \sum_{t=\lag}^T I(\spoint[out] \in \mathcal{N}_{\bound[B]}(\boundary)) \leq T^{1-Q^*} \Bigg)
\end{align*}}
From Assumption~\ref{ass:neighborhood_boundary} we have that
\begin{align*}
P \Bigg( \sum_{t=\lag}^T & I(\spoint[out] \in \mathcal{N}_{\bound[B]}(\boundary)) > T^{1-Q^*} \Bigg)
\\
& \leq \ P \left( \sum_{t = \lag}^T I\Big( \exists s \in \sparseset[{}Y] \cap \mathcal{N}_{\bound[B]}(\boundary) \Big) > T^{1-Q^*} \right)
\rightarrow 0,
\end{align*}
and $\lim_{T \rightarrow \infty} \quantt$ is equal to
\begin{align*}
\lim_{T \rightarrow \infty}
P \Bigg( \sum_{t = \lag}^T \int_B K_{b_T}(\omega; \spoint[out]) \mathrm{d}\omega > \frac{\epsilon (T-\lag+1)}{2 T^Q \bound[W]^\lag \bound[Y]} \ \Bigg\vert \ \sum_{t=\lag}^T I(\spoint[out] \in \mathcal{N}_{\bound[B]}(\boundary)) \leq T^{1-Q^*} \Bigg).
\end{align*}
Studying the latter quantity, we have that
\begin{align*}
& \quad P \Bigg( \sum_{t = \lag}^T \int_B K_{b_T}(\omega; \spoint[out]) \mathrm{d}\omega > \frac{\epsilon (T-\lag+1)}{2 T^Q \bound[W]^\lag \bound[Y]} \Bigg\vert \sum_{t=\lag}^T I(\spoint[out] \in \mathcal{N}_{\bound[B]}(\boundary)) \leq T^{1-Q^*} \Bigg) \\
& \leq P \Bigg( \hspace{20pt} \sum_{\mathclap{\substack{t = \lag \\ \spoint[out] \not\in \mathcal{N}_{\bound[B]}(\boundary)}}}^T \hspace{20pt}
\int_B K_{b_T}(\omega; \spoint[out]) \mathrm{d}\omega > \frac{\epsilon (T-\lag+1)}{2 T^Q \bound[W]^\lag \bound[Y]} - T^{1-Q^*} \Bigg) \\
&
\leq P \Bigg( \hspace{20pt} \sum_{\mathclap{\substack{t = \lag \\ \spoint[out] \not\in \mathcal{N}_{\bound[B]}(\boundary)}}}^T \hspace{20pt}
\int_{\omega: \|\omega\| > \bound[B]} K_{b_T}(\omega; \bm 0) \mathrm{d}\omega > \frac{\epsilon (T-\lag+1)}{2 T^Q \bound[W]^\lag \bound[Y]} - T^{1-Q^*} \Bigg) \\
& \leq P \Bigg( (T-\lag+1)
\int_{\omega: \|\omega\| > \bound[B]} K_{b_T}(\omega; \bm 0) \mathrm{d}\omega > \frac{\epsilon (T-\lag+1)}{2 T^Q \bound[W]^\lag \bound[Y]} - T^{1-Q^*} \Bigg)\\
&= I \Bigg( (T-\lag+1)
\int_{\omega: \|\omega\| > \bound[B]} K_{b_T}(\omega; \bm 0) \mathrm{d}\omega > \frac{\epsilon (T-\lag+1)}{2 T^Q \bound[W]^\lag \bound[Y]} - T^{1-Q^*} \Bigg)
\numberthis \label{proof_eq:indicator}
\end{align*}
where the first inequality follows from the fact that at most $T^{1-Q^*}$ time periods had $\spoint[out]$ within $\bound[B]$ of set's $B$ boundary, and $\int_B K_{b_T}(\omega; \spoint[out]) \leq 1$ for those time periods. The second inequality follows from the fact that during the remaining time periods $\spoint[out]$ was further than $\bound[B]$ from $B$ and $\int_B K_{b_T}(\omega;\spoint[out]) \leq \int_{\omega: \|\omega - \spoint[out]\| > \bound[B]} K_{b_T}(\omega; \spoint[out]) = \int_{\omega: \|\omega\| > \bound[B]} K_{b_T}(\omega; \bm 0)$. The third inequality follows from not excluding the time periods with $\spoint[out] \in \mathcal{N}_{\bound[B]}(\boundary)$. Finally, the last equality holds because there is no uncertainty in the statement so the probability turns to an indicator.

Since $b_T \rightarrow 0$ as $T \rightarrow 0$, there exists $T_1 \in \mathbb{N}$ such that $b_T < \bound[B]$ and $\int_{\omega: \|\omega\| > \bound[B]} K_{b_T}(\omega; \bm 0) \mathrm{d}\omega = 0$ for all $T \geq T_1$.
Also, since $\frac{\epsilon (T-\lag+1)}{2 T^Q \bound[W]^\lag \bound[Y]} - T^{1-Q^*} \rightarrow \infty$, there exists $T_2 \in \mathbb{N}$ such that $\frac{\epsilon (T-\lag+1)}{2 T^Q \bound[W]^\lag \bound[Y]} - T^{1-Q^*} > 1$ for all $T \geq T_2$. Then, for all $T \geq T_0 = \max\{T_1, T_2\}$ we have that the quantity in Equation~\cref{proof_eq:indicator} is equal to 0, showing that
$\lim_{T\rightarrow \infty} B_1 = 0$. Similarly, we can show that $\lim_{T\rightarrow \infty} B_2 = 0$.

Combining all of these results we have that
$$P \Bigg( T^Q \Bigg\{ \Bigg| \frac1{T-\lag+1} \sum_{t = \lag}^T \quant[2] \Bigg|
\Bigg\} > \epsilon \Bigg) \rightarrow 0,$$ as $T \rightarrow \infty$,
establishing that the second error converges to zero faster than $1/\sqrt{T}$.
\end{proof}

\renewcommand*{\proofname}{\textbf{Proof of \cref{lemma:consistent_variance}}}
\begin{proof}
Define $\Psi_t = \big[ \estimatorNt[F][][\lag] \big]^2 - \asymvar_t^*$. Then, $\Psi_t$ is a martingale difference  series with respect to $\filtration[t] = \history[t - \lag +  1]$ since the following two hold:
\begin{enumerate*}[label=(\arabic*)]
\item $E(|\Psi_t|) < \infty$ since $\Psi_t$ is bounded, and
\item $ \text{E}(\Psi_t \mid \filtration) = \text{E} \Big\{ \big[ \estimatorNt[F][][\lag] \big]^2 \mid \history[t-\lag]^* \Big\} - \asymvar_t^* = 0. $
\end{enumerate*}
Also, since  $\estimatorNt[F][][\lag]$ is bounded we have that
$\sum_{t= \lag}^\infty t^{-2} \text{E}\big(\Psi_t^2\big) < \infty.$
From Theorem 1 in \cite{Csorgo1969} we have that
$$
\frac1{T-\lag +1} \sum_{t = \lag}^T \Psi_t = \frac1{T-\lag +1} \sum_{t = \lag}^T \big[ \estimatorNt[F][][\lag] \big]^2 - \frac1{T-\lag +1} \sum_{t = \lag}^T \asymvar_t^* \overset{p}{\rightarrow} 0.
$$
\end{proof}

We use the results above to acquire asymptotic normality of the estimator for the causal effect, $\tempeffect[T][F][1][2]$:
\begin{theorem} \spacingset{1.25} \label{theorem:normality_tau}
Suppose that Assumptions~\ref{ass:unmeasured_conf}~and~\ref{ass:positivity} as well as the regularity conditions (Assumption~\ref{ass:regularity_conditions}) hold. If the bandwidth $b_T \rightarrow 0$, then we have that
$$\sqrt{T}(\tempeffect[T][F][1][2] - \tempeffect[F][F][1][2]) \overset{d}{\rightarrow} N(0, \eta),$$ as $T \rightarrow \infty,$ for some $\eta > 0$. Finally, an upper bound of the asymptotic variance $\eta$ can be consistently estimated by $$\frac{1}{T - \lag + 1} \sum_{t = \lag}^T \left[\effect[T][F][1][2]\right]^2 \overset{p}{\rightarrow} \eta^\ast \ge \eta.$$
\end{theorem}

\renewcommand*{\proofname}{\textbf{Proof}}
\begin{proof}
In order to prove the asymptotic normality of $\tempeffect[T][F][1][2]$ we will rely on results in the proof of \cref{theorem:normality} above. Take
{\small
\begin{align*}
& \effect[T][F][1][2] -  \effect[F][F][1][2] = \\
& \Bigg\{ \prod_{j = t - \lag + 1}^t \frac{\intervdistf[2][F](W_j)}{\propscore[j][j][W]} -
  \prod_{j = t - \lag + 1}^t \frac{\intervdistf[1][F](W_j)}{\propscore[j][j][W]}  \Bigg\}
\int_B \sum_{s \in \sparseset[{}Y]} K_{b_T}(\omega, s) \mathrm{d}\omega
- \effect[F][F][1][2] = \\
& \underbrace{
\Bigg\{  \prod_{j = t - \lag + 1}^t \frac{\intervdistf[2][F](W_j)}{\propscore[j][j][W]}  -
  \prod_{j = t - \lag + 1}^t \frac{\intervdistf[1][F](W_j)}{\propscore[j][j][W]}  \Bigg\}
N_B(Y_t) - \effect[F][F][1][2]}_{\quanttt[1][]} + \\
& \hspace{40pt}
   \underbrace{\Bigg[ \prod_{j = t - \lag + 1}^t \frac{\intervdistf[2][F](W_j)}{\propscore[j][j][W]} \Bigg] \Bigg[ \int_B \sum_{s \in \sparseset[{}Y]} K_{b_T}(\omega, s) \mathrm{d}\omega -
    N_B(Y_t) \Bigg]}_{\quanttt[2][2]} - \\
& \hspace{40pt}
   \underbrace{\Bigg[ \prod_{j = t - \lag + 1}^t \frac{\intervdistf[1][F](W_j)}{\propscore[j][j][W]} \Bigg] \Bigg[\int_B \sum_{s \in \sparseset[{}Y]} K_{b_T}(\omega, s) \mathrm{d}\omega -
    N_B(Y_t) \Bigg]}_{\quanttt[2][1]}
\end{align*}}
Following steps identical to showing  $\sqrt{T} \Big[(T-\lag+1)^{-1}\sum_{t = \lag}^T \quant[2] \Big] \overset{p}{\rightarrow} 0$ in the proof of \cref{theorem:normality}, we can equivalently show that
$ \sqrt{T} \Big[(T-\lag+1)^{-1}\sum_{t = \lag}^T \quanttt[2][1] \Big] \overset{p}{\rightarrow} 0$
and
$ \sqrt{T} \Big[(T-\lag+1)^{-1}\sum_{t = \lag}^T \quanttt[2][2] \Big] \overset{p}{\rightarrow} 0$.

Therefore, all we need to show is that
$ \sqrt{T} \Big[(T-\lag+1)^{-1}\sum_{t = \lag}^T \quanttt[1][] \Big] \overset{d}{\rightarrow} N(0, \eta).$ We will do so by showing again that $\quanttt[1][]$ is a martingale difference series with respect to the filtration $\filtration$:
\begin{enumerate}
    \item Since $E(|\quant|) < \infty$, from the triangular inequality we straightforwardly have that $E(|\quanttt[1][]|) < \infty$.
    \item Since $E(\quant | \filtration[t - 1]) = 0$, we also have that $E(\quanttt[1][] | \filtration[t - 1]) = 0$, from linearity of expectation.
\end{enumerate}
Then, using the triangular inequality and Equation~\cref{proof_eq:bounded_A1t}, we have that $\quanttt[1][]$ is bounded by $2 \bound[Y](\bound[W]^\lag +1)$. Then, for $\epsilon > 0$, choosing
$T_0 = \underset{t \in \mathbb{N}^+}{\text{argmin}} \{ \epsilon\sqrt{t - \lag + 1} > 2 \bound[Y](\bound[W]^\lag + 1) \}$ satisfies that, for $T > T_0$,
$ E(\quanttt[1][2] I(|\quanttt[1][]| > \epsilon \sqrt{T - \lag + 1}) | \filtration) = 0. $
Combining these results, we have that $\sqrt{T}\big[\effect[T][F][1][2] - \effect[F][F][1][2] \big] \rightarrow N(0, \eta)$.

To show $(T - \lag + 1)^{-1} \sum_{t = \lag}^T \Big\{ \big[\effect[T][F][1][2] \big] ^2 - E \big\{ \big[\effect[F][F][1][2] \big]^2 | \history[t - \lag]^*\big\}  \Big\} \overset{p}{\rightarrow} 0$, the proof  follows exactly the same way as the proof of \cref{lemma:consistent_variance} and is omitted here.

\end{proof}

\subsection{Proofs: Asymptotic normality based on the estimated propensity score}
\label{app_subsec:asym_estps}

\renewcommand*{\proofname}{\textbf{Proof}}

We will prove the asymptotic normality of the proposed estimators when the propensity score is estimated using a correctly specified parametric model. We extend Theorem~4.16 of \cite{VanDerVaart2010timeseries} to multivariate martingale difference series. To our knowledge, this result is new even though the related results exist in the continuous time setting \citep{kuchler1999note, crimaldi2005convergence}. Under some additional assumptions on the martingale series, we show that the solution to the empirical estimating equation is also asymptotically normal. This result will be crucial in establishing the asymptotic normality of the maximum likelihood estimator for the propensity score model parameters. Finally, we combine these results and apply them to our specific context.

\begin{theorem}[Central limit theorem for multivariate martingale difference series]
Let $X_t = (X_{1t}, X_{2t}, \dots, X_{Kt})^\top$ be a multivariate martingale difference series with respect to the filtration $\filtration[t]$ in that $E[X_t \mid \filtration] = 0$ and $E[\|X_t\|] < \infty$, where $\| X_t \| = \sqrt{X_t^\top X_t} = \sqrt{ \sum_{k = 1}^K X_{kt}^2 }$. Suppose that the following conditions hold.
\begin{enumerate}
\item There exists positive definite matrix $V \in \mathbb{R}^{K \times K}$ such that
\( \displaystyle \frac1T \sum_{t = 1}^T E \Big( X_t X_t^\top \mid \filtration \Big) \overset{p}{\rightarrow} V \),
\item \( \displaystyle \frac1T \sum_{t = 1}^T E \Big[ \| X_t \|^2 I \Big( \| X_t \| > \epsilon \sqrt{T} \Big) \mid \filtration \Big] \overset{p}{\rightarrow} 0 \), for all $\epsilon > 0$.
\end{enumerate}
Then, we have,  $$\frac1{\sqrt T} \sum_{t = 1}^T X_t \overset{d}{\rightarrow} N(0, V).$$
\label{theorem:CLT_multi_mds}
\end{theorem}

\begin{proof}
We will use the Cramer-Wold device. We show that for every $\alpha = (\alpha_1, \alpha_2, \dots, \alpha_K) \in \mathbb{R}^K$, it holds that
\( \displaystyle
\alpha^\top \frac1{\sqrt T} \sum_{t = 1}^T X_t \overset{d}{\rightarrow} \alpha^\top N(0, V). \)
If this is true, it is implied that
\[
\frac1{\sqrt T} \sum_{t = 1}^T X_t \overset{d}{\rightarrow} N(0, V).
\]
Clearly, if $\alpha$ is the zero-vector, the result is trivial. So we focus on vectors $\alpha$ such that $\| \alpha \| \neq 0$.

Define $Y_t = \alpha^\top X_t.$ First we show that $Y_t$ is a martingale difference series with respect to $\filtration[t]$:
\begin{align*}
E(|Y_t|) & \leq \sum_{k = 1}^K |\alpha_k| E(|X_{kt}|) \leq \sum_{k = 1}^K |\alpha_k| E(\|X_t\|) < \infty, \quad \text{and} \\
E(Y_t \mid \filtration) & = \sum_{k = 1}^K \alpha_k E(X_{kt} \mid \filtration) = 0,
\end{align*}
since $X_t$ is a martingale difference series with respect to $\filtration[t]$. So $Y_t$ is also a martingale difference series with respect to $\filtration[t]$.
Next we will show that the conditions of Theorem 4.16 of \cite{VanDerVaart2010timeseries} hold for $Y_t$.
For Condition 2 we will use the fact that $\alpha \alpha^\top$ is a rank 1 symmetric matrix of dimension $K$ with only non-zero eigenvalue equal to $\| \alpha \|^2$, and for that reason $X_t^\top \alpha \alpha^\top X_t = \| \alpha \| ^ 2 X_t^\top X_t = \| \alpha \|^2 \| X_t \|^2 $.
\begin{align*}
\text{Condition 1} \quad\quad & \frac1T \sum_{t = 1}^T E \big(Y_t^2 \mid \filtration \big)
= \frac1T \sum_{t = 1}^T E \Big(\alpha^\top X_t X_t^\top \alpha \mid \filtration \Big) \\
&= \alpha^\top \frac1T \sum_{t = 1}^T E \Big(X_t X_t^\top \mid \filtration \Big) \alpha
\overset{p}{\rightarrow} \alpha^\top V \alpha  \tag{from the first assumption of \cref{theorem:CLT_multi_mds}} \\[10pt]
%
\text{Condition 2} \quad\quad &
E \Big[ Y_t^2 I \big( |Y_t| > \epsilon \sqrt{n} \big) \mid \filtration \Big] = \\
&= E \Big[ X_t^\top \alpha \alpha^\top X_t \ I \big( X_t^\top \alpha \alpha^\top X_t > \epsilon^2 n \big) \mid \filtration \Big] \\
&= E \Big[ \| \alpha \| ^ 2 \| X_t \|^2 \ I \big( \| \alpha \| ^ 2 \| X_t \|^2 > \epsilon ^ 2 n \big) \mid \filtration \Big] \\
&= \| \alpha \| ^ 2 E \left[ \| X_t \|^2 \ I \left( \| X_t \| > \frac{\epsilon}{\| \alpha \|} \sqrt{n} \right) \mid \filtration \right] \overset{p}{\rightarrow} 0 \tag{From the second condition of the Theorem for $\epsilon' = \epsilon / \| \alpha \|$}
\end{align*}
Using Theorem 4.16 from \cite{VanDerVaart2010timeseries}:
\[
\alpha^\top \frac1{\sqrt T} \sum_{t = 1}^T X_t =
\sqrt{T} \frac1T \sum_{t = 1}^T Y_t
\overset{d}{\rightarrow} N(0, \alpha^\top V \alpha) \overset{d}{=} \alpha^\top N(0, V).
\]
\end{proof}

Now that a multivariate central limit theorem (CLT) for martingale difference series is established, we prove the next result which will be crucial in obtaining the asymptotic normality of estimators for the propensity score parameters. To our knowledge, this result is also new in martingale theory, but a related result in the iid setting is given as Theorem~5.21 of \cite{van1998asymptotic}.
\begin{theorem}[Asymptotic normality of the solution to the estimating equation]
Let $\theta \rightarrow s(x, \theta) = (s_1(x, \theta), s_2(\theta), \dots, s_K(x, \theta))^\top \in \mathbb{R}^K$ be twice continuously differentiable with respect to $\theta = (\theta_1, \theta_2, \dots, \theta_K)^\top \in \Theta$, open subset of $\mathbb{R}^K$. Suppose that the following conditions hold.
\begin{enumerate}
\item
$s(X_t, \theta_0)$ satisfies the conditions of \cref{theorem:CLT_multi_mds} under $\theta_0$, in that there exists filtration $\filtration[t]$ such that
\begin{enumerate}
\item $E_{\theta_0}[s(X_t, \theta_0) \mid \filtration] = 0$ and $E_{\theta_0}[\|s(X_t, \theta_0)\|] < \infty$ (and therefore it is a martingale difference series),
\label{app_cond:mds}
\item $\exists \ V \in \mathbb{R}^{K \times K}$ positive definite such that
\( \displaystyle \frac1T \sum_{t = 1}^T E_{\theta_0} \Big( s(X_t, \theta_0) s(X_t, \theta_0)^\top \mid \filtration \Big) \overset{p}{\rightarrow} V \), and
\label{app_cond:asym_var}
\item \( \displaystyle \frac1T \sum_{t = 1}^T E_{\theta_0} \Big[ \| s(X_t, \theta_0) \|^2 I \Big( \| s(X_t, \theta_0) \| > \epsilon \sqrt{T} \Big) \mid \filtration \Big] \overset{p}{\rightarrow} 0 \), for all $\epsilon > 0$,
\label{app_cond:not_exploding}
\end{enumerate}
\item \( \displaystyle \frac1T \sum_{t = 1}^T E_{\theta_0} \left( \frac \partial{\partial \theta^T} s(X_t, \theta) \Big|_{\theta_0} \mid \filtration \right) \overset{p}{\rightarrow} V_d, \) for $V_d \in \mathbb{R}^{K \times K}$ invertible,
\label{app_cond:expectation_derivative}
\item for all $k,j$, if we denote \( \displaystyle P_{kjt} = \frac \partial{\partial \theta_j} s_k(X_t, \theta) \Big|_{\theta_0}$,
we have that
\( E_{\theta_0} [ | P_{kjt} | ] < \infty \), and there exists $0 < r_{kj} \leq 2$ such that \( \displaystyle \sum_{t = 1}^T \frac1{t^{r_{kj}}} E_{\theta_0} \left( \left| P_{kjt} - E_{\theta_0} \left[P_{kjt} \mid \filtration \right] \right|^{r_{kj}} \mid \filtration \right) \overset{p}{\rightarrow} 0 \),
\label{app_cond:squared_derivative}
\item there exists an integrable function $\overset{\bigcdot\bigcdot}{\psi}(x)$ such that $\overset{\bigcdot\bigcdot}{\psi}(x)$ dominates the second partial derivatives of $s_k(x, \theta)$ in a neighborhood of $\theta_0$ for all $x$, and $k =1, 2, \dots, K$.
\label{app_cond:dominated}
\end{enumerate}
If $\Psi_T(\theta) = \frac1T \sum_{t = 1}^T s(X_t, \theta)$, and the solution to $\Psi_T(\theta) = 0$, $\widehat\theta_T$, is consistent for $\theta_0$, then
\[
\sqrt{T}\left(\widehat \theta_T - \theta_0 \right) \overset{d}{\rightarrow} N \Big(0, V_d^{-1} V (V_d^{-1})^\top \Big).
\]
\label{theorem:CLT_solution_multi_mds}
\end{theorem}

\begin{proof}
We extend the proof of Theorem 5.41 of \cite{van1998asymptotic} from the iid to the time series setting. Since the conditions of \cref{theorem:CLT_multi_mds} are satisfied under $\theta_0$, we have that
\[
\frac1{\sqrt{T}} \sum_{t = 1}^T s(X_t, \theta_0) = \sqrt{T} \Psi_T(\theta_0) \rightarrow N(0, V).
\]
We will use the Taylor expansion for the vector valued $\Psi_T(\widehat \theta)$ around $\theta_0 = (\theta_{01}, \theta_{02}, \dots, \theta_{0K})^T$. To do so, we define the matrix $\overset{\bigcdot}{\Psi_T} (\theta) \in \mathbb{R}^{K \times K}$ and array $\overset{\bigcdot\bigcdot}{\Psi_T}(\theta) \in \mathbb{R}^{K \times K \times K}$ of first and second derivatives as
\begin{align*}
\Big[ \overset{\bigcdot}{\Psi_T} (\theta) \Big]_{kj}
&= \frac {\partial}{\partial\theta_j} \Psi_{kT}(\theta) \Big|_{\theta}
= \frac1T \sum_{t = 1}^T \frac {\partial}{\partial\theta_j} s_k(X_t, \theta) \Big|_{\theta}
\qquad \qquad \text{and} \\
\Big[ \overset{\bigcdot \bigcdot}{\Psi_T} (\theta) \Big]_{kji}
&= \frac {\rm \partial^2}{\partial\theta_j \partial\theta_i} \Psi_{kT}(\theta) \Big|_{\theta}
= \frac1T \sum_{t = 1}^T \frac {\rm \partial^2}{\partial\theta_j \partial\theta_i} s_k(X_t, \theta) \Big|_{\theta},
\end{align*}
for $i,j,k = 1, 2, \dots, q$, where $\Psi_{kT}$ is the $k^{th}$ element of the $\Psi_T$ vector. Then, we can write the Taylor expansion as
\begin{equation}
\Psi_T(\widehat \theta) = \Psi_T(\theta_0) + \overset{\bigcdot}{\Psi_T} (\theta_0) (\widehat \theta_T - \theta_0) + \overset{\bigcdot\bigcdot}{\Psi_T} (\theta^*)(\widehat \theta_T - \theta_0, \widehat \theta_T - \theta_0),
\label{app_eq:vector_taylor}
\end{equation}
where $\theta^*$ is between $\widehat \theta_T$ and $\theta_0$, and
\( \displaystyle \overset{\bigcdot \bigcdot}{\Psi_T} (\theta^*) (\widehat \theta_T - \theta_0, \widehat \theta_T - \theta_0) \)
is a vector of length $K$ with $k^{th}$ entry
\[
\sum_{j,i=1}^K \Big[ \overset{\bigcdot \bigcdot}{\Psi_T} (\theta^*) \Big]_{kji} (\widehat\theta_{Tj} - \theta_{0j})(\widehat \theta_{Ti} - \theta_{0i}),
\]
and $\widehat\theta_{Ti}$ is the $i^{th}$ entry of $\widehat\theta$.
Therefore, we can write \( \displaystyle \overset{\bigcdot\bigcdot}{\Psi_T} (\theta^*)(\widehat \theta_T - \theta_0, \widehat \theta_T - \theta_0) \) as \( A_T (\widehat\theta_T - \theta_0) \) where $A_T$ is the $K \times K$ matrix which is the result of multiplying the tensor
\( \displaystyle \overset{\bigcdot\bigcdot}{\Psi_T} (\theta^*) \)
with the vector $\widehat \theta_T - \theta_0$ along the second mode, and it has $(k,i)$ entry equal to
\[
[A_T]_{ki} = \sum_{j=1}^K \Big[ \overset{\bigcdot \bigcdot}{\Psi_T} (\theta^*) \Big]_{kji} (\widehat\theta_{Tj} - \theta_{0j})
\]
For notational simplicity, we do not include $\theta^*$ and $\widehat \theta_T - \theta_0$ in the notation of $A_T$. Since $\widehat \theta$ is the solution to $\Psi_T(\theta) = 0,$ and based on the above, we can re-write Equation~\cref{app_eq:vector_taylor} as
\begin{align*}
& 0 = \Psi_T(\theta_0) + \overset{\bigcdot}{\Psi_T} (\theta_0) (\widehat \theta_T - \theta_0) + A_T (\widehat \theta_T - \theta_0) \\
\implies & - \sqrt{T} \Psi_T(\theta_0) = \sqrt{T} \Big[ \overset{\bigcdot}{\Psi_T} (\theta_0) + A_T \Big] (\widehat \theta_T - \theta_0) \\
\implies & - \sqrt{T} \Psi_T(\theta_0) = \sqrt{T} \Big[ \underbrace{\overset{\bigcdot}{\Psi_T} (\theta_0) - \frac1T \sum_{t = 1}^T E_{\theta_0} \left( \frac \partial{\partial \theta^T} s(X_t, \theta) \Big|_{\theta_0} \mid \filtration \right)}_{(*)} + \\
& \hspace{5cm} + \frac1T \sum_{t = 1}^T E_{\theta_0} \left( \frac \partial{\partial \theta^T} s(X_t, \theta) \Big|_{\theta_0} \mid \filtration \right) + \underbrace{A_T}_{(**)} \Big] (\widehat \theta_T - \theta_0)
\end{align*}
We will show that the under-braced terms ($K\times K$ matrices) are $o_P(1)$.
For the first term $(*)$, note that it involves the average over $t$ of the $P_{kjt}$ terms defined in  Condition \ref{app_cond:squared_derivative} of the theorem. Clearly, we have that $E_{\theta_0}[P_{kjt} - E_{\theta_0} \left[P_{kjt} \mid \filtration \right] \mid \filtration] = 0$, and we also have that
\begin{align*}
&E_{\theta_0}[|P_{kjt} - E_{\theta_0} \left[P_{kjt} \mid \filtration \right]|] \\
=\ & E_{\theta_0} \left\{ \left| \frac \partial{\partial \theta_j} s_k(X_t, \theta) \Big|_{\theta_0} - E_{\theta_0} \left[ \frac \partial{\partial \theta_j} s_k(X_t, \theta) \Big|_{\theta_0} \mid \filtration \right]  \right| \right\} \\
\leq \ &  E_{\theta_0} \left\{ \left| \frac \partial{\partial \theta_j} s_k(X_t, \theta) \Big|_{\theta_0} \right| \right\} + E_{\theta_0} \left\{ \left| E_{\theta_0} \left[ \frac \partial{\partial \theta_j} s_k(X_t, \theta) \Big|_{\theta_0} \mid \filtration \right]  \right| \right\} \tag{Triangle inequality} \\
\leq \ &
E_{\theta_0} \left\{ \left| \frac \partial{\partial \theta_j} s_k(X_t, \theta) \Big|_{\theta_0} \right| \right\} +
E_{\theta_0} \left\{ E_{\theta_0} \left[ \left| \frac \partial{\partial \theta_j} s_k(X_t, \theta) \Big|_{\theta_0} \right| \mid \filtration \right]  \right\} \tag{Jensen's inequality} \\
= \ & 2 E_{\theta_0} \left\{ \left| \frac \partial{\partial \theta_j} s_k(X_t, \theta) \Big|_{\theta_0} \right| \right\} < \infty \tag{Condition \ref{app_cond:squared_derivative}}
\end{align*}
So the assumptions of \cite{chow1965local} (Theorem~5), which is also stated in \cite{stout1974almost} (Theorem~3.3.1), are satisfied and we have that
\begin{align*}
& \overset{\bigcdot}{\Psi_T} (\theta_0) - \frac1T \sum_{t = 1}^T E_{\theta_0} \left( \frac \partial{\partial \theta^\top} s(X_t, \theta) \Big|_{\theta_0} \mid \filtration \right) \\
= &  \frac1T \sum_{t = 1}^T \left[ \frac {\partial}{\partial\theta^\top} X_{t,i}(\theta) \Big|_{\theta_0} -
E_{\theta_0} \left( \frac \partial{\partial \theta^\top} s(X_t, \theta) \Big|_{\theta_0} \mid \filtration \right)
\right] \overset{p}{\rightarrow} 0.
\end{align*}
Then for $A_T$ we notice that
\[
\left| \Big[ \overset{\bigcdot \bigcdot}{\Psi_T} (\theta^*) \Big]_{kji} \right|
\leq
\frac1T \sum_{t = 1}^T \left| \frac {\rm \partial^2}{\partial\theta_j \partial\theta_i} s_k(X_t, \theta) \Big|_{\theta^*} \right| \leq \frac1T \sum_{t = 1}^T \overset{\bigcdot\bigcdot}{\psi}(X_t) ,
\]
where the last inequality holds for large $T$ because $\widehat\theta_T$ is consistent for $\theta_0$ and the parameter space $\Theta$ is an open subset of $\mathbb{R}^n$ which imply that $\widehat\theta_T$ is within the neighborhood of $\theta_0$ that satisfies Condition \ref{app_cond:dominated} of the theorem with probability that tends to 1, and therefore so will $\theta^*$. Since $\overset{\bigcdot\bigcdot}{\psi}(x)$ is integrable, the right hand side above is bounded with probability 1 from the law of large numbers. Then, using Cauchy-Schwarz on $[A_T]_{ki}$ and since $\widehat \theta_T$ is consistent for $\theta_0$, we have that $[A_T]_{ki} \overset{p}{\rightarrow} 0$ for all $k,i$. Therefore, using Condition \ref{app_cond:expectation_derivative} of the theorem
$$
- \sqrt{T} \Psi_T(\theta_0) = \sqrt{T} [ V_d + o_P(1) ] (\widehat \theta_T - \theta_0)
$$
which, since $V_d$ is invertible, implies asymptotically that
\[
\sqrt{T}\left(\widehat \theta_T - \theta_0 \right) \overset{d}{\rightarrow} N \Big(0, V_d^{-1} V (V_d^{-1})^\top \Big).
\]
\end{proof}

\cref{theorem:CLT_solution_multi_mds} will be the basis for showing asymptotic normality of our estimators when the propensity score is estimated using a correctly specified parametric propensity score.

\begin{lemma}[Properties of the time series score functions.]
If \cref{ass:unmeasured_conf} holds, and $\scorefun$ are score functions that satisfy \cref{ass:propensity_score}, then
\begin{enumerate}
\item $E_{\pspar_0}[ \scorefun[T][T][0] \mid \filtration] = 0$, $E_{\pspar_0} \Big[\| \scorefun[T][T][0] \| \Big] < \infty$, and
\item \( \displaystyle
E_{\pspar} \left( - \frac \partial{\partial \pspar^\top} \scorefun[T][T] \mid \filtration \right) = E_{\pspar} \Big( \scorefun[T][T] \scorefun[T][T]^\top \mid \filtration \Big) \)
which in turn implies that
\( \displaystyle
\frac1T \sum_{t = 1}^T E_{\pspar_0} \left( - \frac \partial{\partial \pspar^\top} \scorefun[T][T] \Big|_{\pspar_0} \mid \filtration \right) \overset{p}{\rightarrow} V_{ps}, \)
for $V_{ps}$ positive definite, symmetric and therefore invertible.
\end{enumerate}
\label{lemma:ps_score_functions}
\end{lemma}

\begin{proof}
First, we show that $E_{\pspar_0} \Big[\| \scorefun[T][T][0] \| \Big] < \infty$. From Jensen's inequality we have that
\[
E_{\pspar_0}^2 \Big[\| \scorefun[T][T][0] \| \Big] \leq E_{\pspar_0} \Big[\| \scorefun[T][T][0] \|^2 \Big] < \infty,
\]
so this part is shown.
The remaining of the proof follows steps similar to the ones in the iid setting while conditioning on the corresponding filtration. Since $\scorefun$ are the score functions, we have that
\begin{equation}
\begin{aligned}
&\scorefun f(W_t = w_t \mid \history = \anyhist[t-1][h] ; \pspar) \\
=& \left[ \frac\partial{\partial \pspar} \log f(W_t = w_t \mid \history = \anyhist[t-1][h] ; \pspar) \right] f(W_t = w_t \mid \history = \anyhist[t-1][h] ; \pspar) \\
=& \frac\partial{\partial \pspar} f(W_t = w_t \mid \history = \anyhist[t-1][h] ; \pspar).
\end{aligned}
\label{app_eq:score_times_density}
\end{equation}
Then,
\begin{align*}
& E_{\pspar}[ \scorefun[T][T] \mid \filtration]
= E_{\pspar} \Big[ \scorefun[T][T] \mid \history[t - \lag]^* \Big] \\
&= E_{\pspar} \Big\{ E_{\pspar} \Big[ \scorefun[T][T] \mid \history[t - 1]^* \Big] \mid \history[t - \lag]^* \Big\} \tag{Since $\history[t-1]^* \supseteq \history[t-\lag]^*$} \\
&= E_{\pspar} \Big\{ \Big[ \int \scorefun[F][T] f(W_t = w_t \mid \history[t - 1]^*) \mathrm{d}w_t \Big] \mid \history[t - \lag]^* \Big\} \\
&= E_{\pspar} \Big\{ \Big[ \int \scorefun[F][T] f(W_t = w_t \mid \history[t - 1]) \mathrm{d}w_t \Big] \mid \history[t - \lag]^* \Big\} \tag{\cref{ass:unmeasured_conf}} \\
&= E_{\pspar} \Big\{ \Big[ \int \scorefun[F][T] f(W_t = w_t \mid \history[t - 1]; \pspar) \mathrm{d}w_t \Big] \mid \history[t - \lag]^* \Big\} \\
&= E_{\pspar} \Big\{ \Big[ \int \frac\partial{\partial \pspar^\top} f(W_t = w_t \mid \history; \pspar) \mathrm{d}w_t \Big] \mid \history[t - \lag]^* \Big\} \tag{Equation~\ref{app_eq:score_times_density}} \\
&= E_{\pspar} \Big\{ \frac\partial{\partial \pspar^\top} \Big[ \int f(W_t = w_t \mid \history; \pspar) \mathrm{d}w_t \Big] \mid \history[t - \lag]^* \Big\} = 0,
\end{align*}
where reversing the integral and derivative is valid using the Leibniz's rule which requires mild regularity conditions (continuity of the propensity score and its partial derivatives with respect to $\pspar$). The last equation is equal to zero since the integral of the propensity score over its support is equal to 1, and the derivative to 1 is equal to 0.

To show the second part, we differentiate $E_{\pspar}[ \scorefun[T][T] \mid \filtration] = 0$ with respect to $\pspar$:
\begin{align*}
0 &= \frac{\partial}{\partial \pspar^\top} E_{\pspar} \Big[ \scorefun[T][T] \mid \filtration \Big]  \\
&= \frac{\partial}{\partial \pspar^\top} E_{\pspar} \Big\{ E_{\pspar} \Big[ \scorefun[T][T] \mid \history[t - 1]^* \Big] \mid \history[t - \lag]^* \Big\} \\
&= \frac{\partial}{\partial \pspar^\top} E_{\pspar} \Big\{  \int \scorefun[F][T] f(W_t = w_t \mid \history[t - 1]; \pspar) \mathrm{d}w_t \mid \history[t - \lag]^* \Big\} \tag{\cref{ass:unmeasured_conf}} \\
&= E_{\pspar} \Big\{ \int \frac{\partial}{\partial \pspar^\top} \Big[ \scorefun[F][T] f(W_t = w_t \mid \history[t - 1]; \pspar) \Big] \mathrm{d}w_t \mid \history[t - \lag]^* \Big\} \tag{Leibniz's rule} \\
&= E_{\pspar} \Big\{ \int \scorefun[F][T] \frac{\partial}{\partial \pspar^\top} \Big[ f(W_t = w_t \mid \history[t - 1]; \pspar) \Big] \mathrm{d}w_t \mid \history[t - \lag]^* \Big\} \\
& \hspace{3cm}
+ E_{\pspar} \Big\{ \int \Big[  \frac{\partial}{\partial \pspar^\top} \scorefun[F][T] \Big] f(W_t = w_t \mid \history[t - 1]; \pspar) \mathrm{d}w_t \mid \history[t - \lag]^* \Big\} \\
&= E_{\pspar} \Big\{ \int \scorefun[F][T] \scorefun[F][T]^\top f(W_t = w_t \mid \history; \pspar) \mathrm{d}w_t \mid \history[t - \lag]^* \Big\} \tag{Equation~\cref{app_eq:score_times_density}} \\
& \hspace{3cm}
+ E_{\pspar} \Big\{ \int \Big[  \frac{\partial}{\partial \pspar^\top} \scorefun[F][T] \Big] f(W_t = w_t \mid \history[t - 1]; \pspar) \mathrm{d}w_t \mid \history[t - \lag]^* \Big\} \\
&= E_{\pspar} \Big\{ \int \scorefun[F][T] \scorefun[F][T]^\top f(W_t = w_t \mid \history^*; \pspar) \mathrm{d}w_t \mid \history[t - \lag]^* \Big\} \\
& \hspace{3cm}
+ E_{\pspar} \Big\{ \int \Big[  \frac{\partial}{\partial \pspar^\top} \scorefun[F][T] \Big] f(W_t = w_t \mid \history[t - 1]^*; \pspar) \mathrm{d}w_t \mid \history[t - \lag]^* \Big\} \tag{\cref{ass:unmeasured_conf}} \\
&= E_{\pspar} \Big\{ E_{\pspar} \Big[ \scorefun[T][T] \scorefun[T][T]^\top \mid \history^* \Big] \mid \history[t - \lag]^* \Big\} \\
& \hspace{3cm}
+ E_{\pspar} \Big\{ E_{\pspar} \Big[ \frac{\partial}{\partial \pspar^\top} \scorefun[T][T] \mid \history^* \Big] \mid \history[t - \lag]^* \Big\} \\
&= E_{\pspar} \Big[ \scorefun[T][T] \scorefun[T][T]^\top \mid \history[t - \lag]^* \Big] +
E_{\pspar} \Big[ \frac{\partial}{\partial \pspar^\top} \scorefun[T][T] \mid \history[t - \lag]^* \Big] \\
\implies &
E_{\pspar} \Big[ \scorefun[T][T] \scorefun[T][T]^\top \mid \history[t - \lag]^* \Big] =
E_{\pspar} \Big[ -  \frac{\partial}{\partial \pspar^\top} \scorefun[T][T] \mid \history[t - \lag]^* \Big].
\end{align*}
From Condition \ref{app_cond:ps_information_matrix} of \cref{ass:propensity_score} we have the last result.
\end{proof}

\begin{corollary}[Asymptotic normality of spatio-temporal propensity score parameters]
Consider a propensity score model that satisfies \cref{ass:propensity_score} and therefore the results of \cref{lemma:ps_score_functions} hold. \cref{theorem:CLT_solution_multi_mds} implies that the MLE of the propensity score parameters are asymptotically normal centered at the true value and with asymptotic variance $V_{ps}^{-1}$, as in the iid setting.
\end{corollary}

Before we state our main theorem we establish a useful Lemma.

\begin{lemma}
Aassume that \cref{ass:unmeasured_conf} holds. Let $\scorefun$ be the score functions of a propensity score model that satisfies \cref{ass:propensity_score} as in \cref{lemma:ps_score_functions} and $\filtration$ be as above. For
\[
s(\history[t-1], W_t, Y_t; \pspar) = \Bigg[ \prod_{j = t - \lag + 1}^t \frac{\intervdistf[][F](W_j)}{\parpropscore[j][j][W]} \Bigg] N_B(Y_t) - \avgout[F][][\lag],
\]
it holds that
\begin{enumerate}
\item
\( \displaystyle E_{\pspar_0} \left[ s(\history[t-1], W_t, Y_t; \pspar_0)  \scorefun[T][T][0] \mid \filtration \right] = - E_{\pspar_0} \left[ \frac\partial{\partial \pspar} s(\history[t-1], W_t, Y_t; \pspar) \ \Big|_{\pspar_0} \mid \filtration \right], \)
\label{lemma:spsi_equal_ders}
\item
\( \displaystyle
\frac \partial{\partial \pspar_l} s(\anyhist[t-1][h], w_t, y_t; \pspar) = - N_B(y_t) \left[ \prod_{j = t - \lag + 1}^t \frac{\intervdistf[][F](w_j)}{\parpropscore[j][j][w]} \right]
\sum_{j = t - \lag + 1}^t \scorefun[F][F][][l][j]
\),
where we use $\scorefun[F][F][][l][j]$ to denote the $l^{th}$ element of the $\scorefun$ vector, and
\item similarly \( \displaystyle \frac \partial{\partial \pspar_m} \frac \partial{\partial \pspar_l} s(\anyhist[t-1][h], w_t, y_t; \pspar) \) is equal to
\begin{align*}
& - N_B(y_t) \left[ \prod_{j = t - \lag + 1}^t \frac{\intervdistf[][F](w_j)}{\parpropscore[j][j][w]} \right] \left\{ \left[ \sum_{j = t - \lag + 1}^t  \frac \partial{\partial \pspar_m}
\scorefun[F][F][][l][j] \right] - \right. \\
& \hspace{120pt} - \left. \left[ \sum_{j = t - \lag + 1}^t \scorefun[F][F][][m][j] \right]
\left[ \sum_{j = t - \lag + 1}^t \scorefun[F][F][][l][j] \right] \right\}
\end{align*}
\end{enumerate}
Note: \( \displaystyle s(\history[t-1], W_t, Y_t; \pspar_0) \) is the term $\quant[1]$ in the proof of \cref{theorem:normality}.
\label{app_lemma:a_psi}
\end{lemma}

\begin{proof} $\ $
\begin{enumerate}
\item We will show it for $\lag = 1$, and the proof for $\lag > 1$ is similar. For $\lag = 1$, $\filtration[t-1] = \history^* = \{\Whist[t-1], \anyhist[T][\allout], \anyhist[T][\allcovs] \}$, we consider
\begin{align*}
& E_{\pspar_0} \left[ s(\history[t-1], W_t, Y_t; \pspar_0) \scorefun[T][T][0] \mid \filtration \right] \\
& = \int s(\history[t-1], w_t, Y_t; \pspar_0) \scorefun[F][T][0] f(W_t = w_t \mid \filtration[t-1] ; \pspar_0) \mathrm{d}w_t \\
& = \int s(\history[t-1], w_t, Y_t; \pspar_0) \scorefun[F][T][0] f(W_t = w_t \mid \history = \anyhist[t-1][h] ; \pspar_0) \mathrm{d}w_t
\tag{\cref{ass:unmeasured_conf}}
\\
&= \int s(\history[t-1], w_t, Y_t; \pspar_0) \frac{\partial}{\partial \pspar} f(W_t = w_t \mid \history = \anyhist[t-1][h] ; \pspar) \Big|_{\pspar_0} \mathrm{d}w_t
\tag{Equation~\cref{app_eq:score_times_density}}
\\
&= \int \frac{\partial}{\partial \pspar} \left[ s(\history[t-1], w_t, Y_t; \pspar_0) f(W_t = w_t \mid \history = \anyhist[t-1][h] ; \pspar_0) \right] \mathrm{d}w_t - \\
& \hspace{20pt} - \int \frac{\partial}{\partial \pspar} s(\history[t-1], w_t, Y_t; \pspar) \Big|_{\pspar_0} f(W_t = w_t \mid \history = \anyhist[t-1][h] ; \pspar_0) \mathrm{d}w_t  \\
&= \int \frac{\partial}{\partial \pspar} \left[ s(\history[t-1], w_t, Y_t; \pspar_0) f(W_t = w_t \mid \filtration[t-1]  ; \pspar_0) \right] \mathrm{d}w_t - \\
& \hspace{20pt} - \int \frac{\partial}{\partial \pspar} s(\history[t-1], w_t, Y_t; \pspar) \Big|_{\pspar_0} f(W_t = w_t \mid \filtration[t-1] ; \pspar_0) \mathrm{d}w_t \tag{\cref{ass:unmeasured_conf}} \\
&= \frac{\partial}{\partial \pspar} E_{\pspar} \left[s(\history[t-1], W_t, Y_t; \pspar) \mid \filtration[t-1]  \right] \Big|_{\pspar_0} -
E_{\pspar_0} \left[ \frac{\partial}{\partial \pspar} s(\history[t-1], w_t, Y_t; \pspar) \Big|_{\pspar_0} \mid \filtration[t-1] \right] \\
&= - E_{\pspar_0} \left[ \frac{\partial}{\partial \pspar} s(\history[t-1], w_t, Y_t; \pspar) \Big|_{\pspar_0} \mid \filtration[t-1] \right]
\end{align*}
where the last equation holds from Equation~\cref{app_eq:expectation_A1t}.  This shows that the expectation is 0, so the derivative is also 0.

Note that at the second line of the proof, we would also need the distribution of $Y_t$ given the filtration $\filtration[t-1]$ and the treatment at time period $t$, $W_t = w_t$. However, given both $\filtration[t-1]$ and $W_t$, the variable $Y_t$ is no longer random, and it is equal to its potential value $Y_t(\Whist[t-1], w_t)$, where $\Whist[t-1]$ is specified in $\filtration[t-1]$. We refrain from explicitly including this in the proof for simplicity.

\item
\begin{align*}
& \frac \partial{\partial \pspar_l} s(\anyhist[t-1][h], w_t, y_t; \pspar) \\
&= \frac \partial{\partial \pspar_l} \left[ \prod_{j = t - \lag + 1}^t \frac{\intervdistf[][F](w_j)}{\parpropscore[j][j]} N_B(y_t) \right] \\
&= N_B(y_t) \left[ \prod_{j = t - \lag + 1}^t \intervdistf[][F](w_j) \right]
\left[ \frac \partial{\partial \pspar[F]_l} \frac1{\prod_{j = t - \lag + 1}^t \parpropscore[j][j][w]} \right] \\
&= - N_B(y_t) \left[ \prod_{j = t - \lag + 1}^t \intervdistf[][F](w_j) \right]
\frac{\frac{\partial}{\partial \pspar[F]_l} \prod_{j = t - \lag + 1}^t \parpropscore[j][j][w] }{\left[\prod_{j = t - \lag + 1}^t \parpropscore[j][j][w] \right]^2} \\
&= - N_B(y_t) \left[ \prod_{j = t - \lag + 1}^t \frac{\intervdistf[][F](w_j)}{\parpropscore[j][j][w]} \right]
\frac{\frac{\partial}{\partial \pspar[F]_l} \prod_{j = t - \lag + 1}^t \parpropscore[j][j][w] }{\prod_{j = t - \lag + 1}^t \parpropscore[j][j][w]} \\
&= - N_B(y_t) \left[ \prod_{j = t - \lag + 1}^t \frac{\intervdistf[][F](w_j)}{\parpropscore[j][j][w]} \right]
\sum_{j = t - \lag + 1}^t \frac{\frac{\partial}{\partial \pspar[F]_l} \parpropscore[j][j][w]}{\parpropscore[j][j][w]} \\
&= - N_B(y_t) \left[ \prod_{j = t - \lag + 1}^t \frac{\intervdistf[][F](w_j)}{\parpropscore[j][j][w]} \right]
\sum_{j = t - \lag + 1}^t \scorefun[F][F][][l-1][j].
\tag{Equation~\cref{app_eq:score_times_density}}
\end{align*}

\item Following a similar procedure we have that
\begin{align*}
& \frac \partial{\partial \pspar_m} \frac \partial{\partial \pspar_l} s(\anyhist[t-1][h], w_t, y_t; \pspar) \\
&= - N_B(y_t) \left\{  \left[ \frac \partial{\partial \pspar_m} \prod_{j = t - \lag + 1}^t \frac{\intervdistf[][F](w_j)}{\parpropscore[j][j][w]} \right]  \sum_{j = t - \lag + 1}^t \scorefun[F][F][][l][j] \right. \\
& \hspace{80pt} \left.
+\left[ \prod_{j = t - \lag + 1}^t \frac{\intervdistf[][F](w_j)}{\parpropscore[j][j][w]} \right]
\left[ \frac \partial{\partial \pspar_m} \sum_{j = t - \lag + 1}^t \scorefun[F][F][][l][j] \right] \right\} \\
&= - N_B(y_t) \left\{ -
\left[ \prod_{j = t - \lag + 1}^t \frac{\intervdistf[][F](w_j)}{\parpropscore[j][j][w]} \right]
\left[ \sum_{j = t - \lag + 1}^t \scorefun[F][F][][m][j] \right]
\left[ \sum_{j = t - \lag + 1}^t \scorefun[F][F][][l][j] \right] \right. \\
& \hspace{80pt} \left. +
\left[ \prod_{j = t - \lag + 1}^t \frac{\intervdistf[][F](w_j)}{\parpropscore[j][j][w]} \right]
\left[ \sum_{j = t - \lag + 1}^t  \frac \partial{\partial \pspar_m}
\scorefun[F][F][][l][j] \right] \right\} \\
&= - N_B(y_t) \left[ \prod_{j = t - \lag + 1}^t \frac{\intervdistf[][F](w_j)}{\parpropscore[j][j][w]} \right] \left\{ \left[ \sum_{j = t - \lag + 1}^t  \frac \partial{\partial \pspar_m}
\scorefun[F][F][][l][j] \right] \right. \\
& \hspace{80pt} - \left. \left[ \sum_{j = t - \lag + 1}^t \scorefun[F][F][][m][j] \right]
\left[ \sum_{j = t - \lag + 1}^t \scorefun[F][F][][l][j] \right] \right\}
\end{align*}

\end{enumerate}
\end{proof}

\begin{corollary}
Part \ref{lemma:spsi_equal_ders} of \cref{lemma:ps_score_functions} holds for any function $s(\history, W_t, Y_t; \gamma)$ for which
\[
E_{\pspar} \left[s(\history[t-1], W_t, Y_t; \pspar) \mid \filtration[t-1]  \right] = 0.
\]
(The proof is identical, hence it is omitted.)
\end{corollary}

We remind one last result from real analysis which we will use in our theorem. We state it here to avoid unnecessarily complicated notation in the proof of the main theorem. The result extends to multivariate functions.
\begin{remark}
For a function $f: \mathbb{R} \rightarrow \mathbb{R}$ differentiable, if $|f'(x)| \leq \alpha$ for $x \in (x_0 - \epsilon, x_0 + \epsilon)$ and some $\alpha in \mathbb{R}^+$, then $|f(x)|$ is also bounded on $(x_0 - \epsilon, x_0 + \epsilon)$.
\label{remark:derivative_bounded}
\end{remark}
\begin{proof}
The proof is straightforward using Taylor expansion:
\[
f(x) = f(x_0) + f'(x^*)(x - x_0)
\rightarrow |f(x)| \leq |f(x_0)| + \alpha \epsilon.
\]
\end{proof}

\hspace{10pt}
Now we can prove our theorem on asymptotic normality of the causal estimators using propensity scores that are estimated based on a correctly specified propensity score model.

\renewcommand*{\proofname}{\textbf{Proof of \cref{theorem:normality_estps}}}
\begin{proof}
We will use \cref{theorem:CLT_solution_multi_mds} to show asymptotic normality for the causal estimator based on the estimated propensity score model.

Remember that $\history[t] = \{\anyhist[t][\bW], \anyhist[t][\bm Y], \anyhist[t+1][\covs] \} $. Then $\{ \history[t-1], W_t, Y_t\} = \history[t] \setminus \{\covs_{t-1}\}$ is the set of observed variables until (and including) the $t^{th}$ outcome. Let $\mu \in \mathbb{R}$ and $\pspar \in \mathbb{R}^K$ be the parameters of the propensity score model with score functions $\scorefun$, and define $\allpars^\top = (\mu, \pspar^\top)$.
Again based on Equation~\cref{app_eq:est_error_split}, we will show the asymptotic normality of the estimator that excludes spatial smoothing.  We will them prove that the spatial smoothing does not affect estimation asymptotically because it converges to zero faster than $T^{-1/2}$. Focusing on the first part of the error, define the $K + 1$ vector
\[
\estimeq = \left( \begin{array}{c}
\Bigg[ \prod_{j = t - \lag + 1}^t \frac{\intervdistf[][F](W_j)}{\parpropscore[j][j][W]} \Bigg] N_B(Y_t) - \avgout[F][][\lag] - \mu \\ \scorefun[T][T]
\end{array} \right)
=  \left( \begin{array}{c}
\quant[1] - \mu \\ \scorefun[T][T]
\end{array} \right),
\]
where $\quant[1]$ is defined in the proof of \cref{theorem:normality}.
We again work with the filtration $\filtration[t] = \history[t-\lag+1]^* = \{\anyhist[t - \lag + 1][\bW], \anyhist[T][\allout], \anyhist[T][\allcovs]\}$. We will show that the conditions of \cref{theorem:CLT_solution_multi_mds} hold.

\paragraph{Condition \ref{app_cond:mds}}
We wish to show the expectation of $s$ conditional on the filtration is 0.  Since we showed in the proof of \cref{theorem:normality} that
\[
{\rm E} \left\{ \Bigg[ \prod_{j = t - \lag + 1}^t \frac{\intervdistf[][F](W_j)}{\parpropscore[j][j][W]} \Bigg] N_B(Y_t) \mid \filtration \right\} = \avgout[F][][\lag],
\]
we have $\allpars_0^\top = (\mu_0, \pspar_0^\top) = (0, \pspar_0^\top)$, where $\pspar_0$ represents the true value for the parametric propensity score. Then, based on \cref{lemma:ps_score_functions}, we have that
\( \displaystyle
{\rm E}_{\allpars_0} \left[ \estimeq[T] \mid \filtration \right] = 0.
\)
Also, from Jensen's inequality we have that
\begin{align*}
{\rm E}_{\allpars_0}^2 \left[ \| \estimeq[T] \| \right]
& \leq {\rm E}_{\allpars_0} \left[ \| \estimeq[T] \|^2 \right] \\
&= {\rm E}_{\allpars_0} (\quant[1]^2) + {\rm E}_{\allpars_0} \left\{ \|\scorefun[T][T][0] \|^2 \right\}
<\infty
\end{align*}
where the first term is finite because $\quant[1]$ is bounded as shown in Equation~\cref{proof_eq:bounded_A1t}, and the second term is finite based on \cref{ass:propensity_score}.

\paragraph{Condition \ref{app_cond:asym_var}}

Since all terms are under the $\allpars_0$-law, we work with $\mu = \mu_0 = 0$. We have that
\begin{align*}
& E_{\allpars_0} \Big( \estimeq[T] \estimeq[T]^\top \mid \filtration \Big) \\
&= \left[ \begin{array}{cc}
E_{\allpars_0} \big[\quant[1]^2 \mid \filtration \big]  & E_{\allpars_0} \big[\quant[1] \scorefun[T][T][0]^\top \mid \filtration \big] \\
 E_{\allpars_0} \big[\quant[1] \scorefun[T][T][0] \mid \filtration \big] & E_{\theta_0} \big[ \scorefun[T][T][0] \scorefun[T][T][0]^\top \mid \filtration \big].
\end{array} \right]
\end{align*}
Equation~\cref{proof_eq:asym_var} implies that
\( \displaystyle
(T - \lag + 1)^{-1} \sum_{t = \lag}^T E_{\allpars_0} \big[\quant[1]^2 \mid \filtration \big] \overset{p}{\rightarrow} v \). In addition, due to \cref{ass:propensity_score}(\ref{app_cond:ps_information_matrix}), we also know that
\( \displaystyle (T - \lag + 1)^{-1} \sum_{t = \lag}^T E_{\pspar_0} \Big( \scorefun[T][T][0] \scorefun[T][T][0]^\top \mid \filtration \Big) \overset{p}{\rightarrow} V_{ps} \).  Lastly, \cref{app_ass:estimator_ps} implies that
\( \displaystyle
(T - \lag + 1)^{-1} \sum_{t = \lag}^T E_{\allpars_0} \big[\quant[1] \scorefun[T][T][0] \mid \filtration \big]
\overset{p}{\rightarrow} u.
\)
Since all the entries of the matrix converge, we are left to show that the resulting matrix
is positive definite. However, since
\[
M = \left[ \begin{array}{cc} \quant[1]^2 & \quant[1] \scorefun[T][T][0]^\top \\ \quant[1]\scorefun[T][T][0] & \scorefun[T][T][0]^2 \end{array} \right]
\]
is positive definite (easy to check by taking vector $\bm x \in \mathbb{R}^k$, not all zero, and showing that $\bm x^\top M \bm x > 0$), we have that
\[
\left[ \begin{array}{cc} v & u^\top \\ u & V_{ps} \end{array} \right]
\]
will also be positive definite.

\paragraph{Condition \ref{app_cond:not_exploding}}
Take $\epsilon > 0$ and write
\begin{align*}
& \frac1{T - \lag +1} \sum_{t = \lag}^T E_{\theta_0} \Big[ \| \estimeq[T] \|^2 I \Big( \| \estimeq[T]  \| > \epsilon \sqrt{T} \Big) \mid \filtration \Big] \\
&= \frac1{T - \lag +1} \sum_{t = \lag}^T
E_{\theta_0} \Big[ \big( \quant[1]^2 + \| \scorefun[T][T][0] \|^2 \big) I \Big( \quant[1]^2 + \| \scorefun[T][T][0] \|^2 > \epsilon^2 T \Big) \mid \filtration \Big] \\
&= \frac1{T - \lag +1} \sum_{t = \lag}^T
E_{\theta_0} \Big[ \quant[1]^2 \ I \Big( \quant[1]^2 + \| \scorefun[T][T][0] \|^2  > \epsilon^2 T\Big) \mid \filtration \Big] \\
& \hspace{30pt}
+ \frac1{T - \lag +1} \sum_{t = \lag}^T
E_{\theta_0} \Big[ \| \scorefun[T][T][0] \|^2 \ I \Big( \| \scorefun[T][T][0] \|^2 > \epsilon^2 T - \quant[1]^2 \Big) \mid \filtration \Big]
\end{align*}

We start with the second term:
Since $\quant[1]^2$ cannot exceed $(\bound[W]^M \bound[Y] + \bound[Y])^2$ based on Equation~\cref{proof_eq:bounded_A1t}, we have that
{\small
\begin{align*}
& \frac1{T - \lag +1} \sum_{t = \lag}^T
E_{\theta_0} \Big[ \| \scorefun[T][T][0] \|^2 \ I \Big( \| \scorefun[T][T][0] \|^2 > \epsilon^2 T - \quant[1]^2 \Big) \mid \filtration \Big] \\
& \leq
\frac1{T - \lag +1} \sum_{t = \lag}^T
E_{\theta_0} \Big[ \| \scorefun[T][T][0] \|^2 \ I \Big( \| \scorefun[T][T][0] \| > \sqrt{\epsilon^2 T - (\bound[W]^M \bound[Y] + \bound[Y])^2} \Big) \mid \filtration \Big] \overset{p}{\rightarrow} 0,
\end{align*}}
based on \cref{ass:propensity_score} and since $\bound[W]^M \bound[Y] + \bound[Y]$ is fixed.

For the first term, since $I(\quant[1]^2 + \|\scorefun[T][T][0]\|^2) > \epsilon^2 T$ implies that at least one of
$\quant[1]^2$ and $\|\scorefun[T][T][0]\|^2$ is greater than $\epsilon^2 T / 2$, we have that
\[
I \big(\quant[1]^2 + \|\scorefun[T][T][0]\|^2) > \epsilon^2 T \big) \leq
I \big(\quant[1]^2 > \epsilon^2 T/2 \big) + I \big( \|\scorefun[T][T][0]\|^2 > \epsilon^2 T/2 \big).
\]
This leads to
\begin{align*}
&
E_{\theta_0} \Big[ \quant[1]^2 \ I \Big( \quant[1]^2 + \| \scorefun[T][T][0] \|^2  > \epsilon^2 T\Big) \mid \filtration \Big] \\
&\leq E_{\theta_0} \Big[ \quant[1]^2 \ I \Big( \quant[1]^2 > \epsilon^2 T/ 2\Big) \mid \filtration \Big] +
E_{\theta_0} \Big[ \quant[1]^2 \ I \Big(\| \scorefun[T][T][0] \|^2  > \epsilon^2 T / 2\Big) \mid \filtration \Big].
\end{align*}
In the proof of \cref{theorem:normality} we have already shown that because $\quant[1]$ is bounded we have that
\[
\frac1{T - \lag +1} \sum_{t = \lag}^T E_{\theta_0} \Big[ \quant[1]^2 \ I \Big( |\quant[1]| > \frac{\epsilon}{\sqrt 2} \sqrt{T} \Big) \mid \filtration \Big] \overset{p}{\rightarrow} 0,
\]
and we want to show that
\[
\frac1{T - \lag +1} \sum_{t = \lag}^T E_{\theta_0} \Big[ \quant[1]^2 \ I \Big(\| \scorefun[T][T][0] \|^2  > \epsilon^2 T / 2\Big) \mid \filtration \Big] \overset{p}{\rightarrow} 0.
\]
We write
\begin{align*}
& E_{\theta_0} \Big[ \quant[1]^2 \ I \Big(\| \scorefun[T][T][0] \|^2  > \epsilon^2 T / 2\Big) \mid \filtration \Big] \\
&= E_{\theta_0} \Big[ \quant[1]^2 \ I \Big(\| \scorefun[T][T][0] \|^2  > \epsilon^2 T / 2\Big) \mid \quant[1]^2 \leq \|\scorefun\|^2, \filtration \Big] \times \\
& \hspace{100pt} \times P(\quant[1]^2 \leq \|\scorefun\|^2 \mid \filtration) + \\
& \hspace{30pt} + E_{\theta_0} \Big[ \quant[1]^2 \ I \Big(\| \scorefun[T][T][0] \|^2  > \epsilon^2 T / 2\Big) \mid \quant[1]^2 \geq \|\scorefun\|^2, \filtration \Big] \times \\
& \hspace{130pt} \times P(\quant[1]^2 \geq \|\scorefun\|^2 \mid \filtration) \\
& \leq E_{\theta_0} \Big[ \| \scorefun[T][T][0] \|^2 \ I \Big(\| \scorefun[T][T][0] \|^2  > \epsilon^2 T / 2\Big) \mid \quant[1]^2 \leq \|\scorefun\|^2, \filtration \Big] \times \\
& \hspace{100pt} \times P(\quant[1]^2 \leq \|\scorefun\|^2 \mid \filtration) + \\
& \hspace{30pt} +
E_{\theta_0} \Big[ \quant[1]^2 \ I \Big(\| \quant[1]^2  > \epsilon^2 T / 2\Big) \mid \quant[1]^2 \geq \|\scorefun\|^2, \filtration \Big],
\end{align*}
where again the average over time of the last term will be converging to zero in probability since $\quant[1]$ is bounded and using similar arguments. Using the law of total expectation we can write
\begin{align*}
& E_{\theta_0} \Big[ \| \scorefun[T][T][0] \|^2 \ I \Big(\| \scorefun[T][T][0] \|^2  > \epsilon^2 T / 2\Big) \mid  \filtration \Big] \\
&= E_{\theta_0} \Big[ \| \scorefun[T][T][0] \|^2 \ I \Big(\| \scorefun[T][T][0] \|^2  > \epsilon^2 T / 2\Big) \mid \quant[1]^2 \leq \|\scorefun\|^2, \filtration \Big] \times \\
& \hspace{100pt} \times P(\quant[1]^2 \leq \|\scorefun\|^2 \mid \filtration) + \\
& \hspace{30pt} E_{\theta_0} \Big[ \| \scorefun[T][T][0] \|^2 \ I \Big(\| \scorefun[T][T][0] \|^2  > \epsilon^2 T / 2\Big) \mid \quant[1]^2 \geq \|\scorefun\|^2, \filtration \Big] \times \\
& \hspace{130pt} \times P(\quant[1]^2 \geq \|\scorefun\|^2 \mid \filtration) \\
& \geq E_{\theta_0} \Big[ \| \scorefun[T][T][0] \|^2 \ I \Big(\| \scorefun[T][T][0] \|^2  > \epsilon^2 T / 2\Big) \mid \quant[1]^2 \leq \|\scorefun\|^2, \filtration \Big] \times \\
& \hspace{100pt} \times P(\quant[1]^2 \leq \|\scorefun\|^2 \mid \filtration).
\end{align*}
Since all the terms in the expectations are positive and since (from \cref{ass:propensity_score}) we have that
\[
\frac1{T - \lag +1} \sum_{t = \lag}^T E_{\theta_0} \Big[ \| \scorefun[T][T][0] \|^2 \ I \Big(\| \scorefun[T][T][0] \|^2  > \epsilon^2 T / 2\Big) \mid  \filtration \Big] \overset{p}{\rightarrow} 0
\]
we also have that
{\small
\begin{align*}
& \frac1{T - \lag +1} \sum_{t = \lag}^T \Bigg\{ \\
& \hspace{20pt}
E_{\theta_0} \Big[ \| \scorefun[T][T][0] \|^2 \ I \Big(\| \scorefun[T][T][0] \|^2  > \epsilon^2 T / 2\Big) \mid \quant[1]^2 \leq \|\scorefun\|^2, \filtration \Big] \times \\
& \hspace{100pt} \times P(\quant[1]^2 \leq \|\scorefun\|^2 \mid \filtration)
\Bigg\} \overset{p}{\rightarrow} 0
\end{align*}}
which completes the proof that Condition \ref{app_cond:not_exploding} holds.

\paragraph{Condition \ref{app_cond:expectation_derivative}}

We denote $\allpars^\top = (\allpars[F]_1, \allpars[F]_2, \dots, \allpars[F]_{K+1}) = (\mu, \pspar^\top)$ and use $\estimeq[F][k]$ to denote the $k^{th}$ entry of the $\estimeq$ vector.
We note that
\[
\frac \partial{\partial \allpars^T} \estimeq =
\left[ \begin{array}{cc}
 -1  & \frac{\partial}{\partial \pspar^T} \estimeq[F][1]  \\
\bm 0  & \frac{\partial}{\partial \pspar^T} \scorefun[T][T]
\end{array} \right]
\]
\cref{lemma:ps_score_functions} implies that $(T  - \lag + 1)^{-1} \sum_{t = \lag}^T E_{\pspar_0} \left[ \frac{\partial}{\partial \pspar^T} \scorefun[T][T] \Big|_{\pspar_0} \mid \filtration \right] \rightarrow - V_{ps}$ (invertible).
\cref{app_ass:estimator_ps} and \cref{app_lemma:a_psi} imply that
\begin{align*}
& (T - \lag + 1)^{-1} \sum_{t = \lag}^T E_{\allpars_0} \left( \frac \partial{\partial \pspar^T} \estimeq[F][1] \Big|_{\allpars_0} \mid \filtration \right) \\
&= - (T - \lag + 1)^{-1} \sum_{t = \lag}^T E_{\allpars_0}
\left[ \estimeq[T][1]  \scorefun[T][T][0] \mid \filtration \right] \\
& \overset{p}{\rightarrow} - u^T.
\end{align*}
Putting these together we have that
\[
(T - \lag + 1)^{-1} \sum_{t = \lag}^T E_{\allpars_0} \left[ \frac \partial{\partial \allpars^T} \estimeq \Big|_{\allpars_0} \mid \filtration \right] \overset{p}{\rightarrow}
\left[ \begin{array}{cc} -1 & -u \\ 0 & -V_{ps} \end{array} \right].
\]
Since $V_{ps}$ is invertible and the first row is the only one to have a non-zero first element we have that this limit matrix is invertible.

\paragraph{Condition \ref{app_cond:squared_derivative}}

We want to show that for all $k, j = 1, 2, \dots, K + 1$, if we use $P_{kjt}$ to denote
\[
P_{kjt} = \frac \partial{\partial \allpars[F]_j} \estimeq[F][k] \Big|_{\allpars_0},
\]
then $E_{\allpars_0}|P_{kjt}| < \infty$, and
there exists $0 < r_{kj} \leq 2$ such that
\[ \sum_{t = \lag}^T \frac1{t^{r_{kj}}}  E_{\allpars_0} \big[ \left| P_{kjt} - E_{\allpars_0} \left( P_{kjt} \mid \filtration \right) \right|^{r_{kj}} \mid \filtration \big] \overset{p}{\rightarrow} 0.\]
For $k, j \geq 2$, this is given by Condition \ref{ps_ass:conv_zero} of \cref{ass:propensity_score}. For $j = 1$ and $k \geq 2$, we have that
\[ \frac \partial{\partial \theta_1} \estimeq[F][k] = 0, \]
so the result holds for any $r_{k1}$. Similarly, for $k = j = 1$, we have that
\[ \frac \partial{\partial \theta_1} \estimeq[F][1] = -1, \] so the result holds for a value $r_{11} \in (1, 2]$. Therefore, it is left to show that it holds for $k = 1$ and $j \geq 2$.
For $k = 1$ and $j \geq 2$, the condition that there exists $0 < r_{1j} \leq 2$ such that
\[ \sum_{t = \lag}^T \frac1{t^{r_{1j}}}  E_{\allpars_0} \big[ \left| P_{1jt} - E_{\allpars_0} \left( P_{1jt} \mid \filtration \right) \right|^{r_{1j}} \mid \filtration \big] \overset{p}{\rightarrow} 0 \]
is given by \cref{app_ass:estimator_ps}. So we are left to show that $E_{\allpars[0]} (|P_{1jt}|) < \infty$. \cref{app_lemma:a_psi} implies that
\begin{align*}
P_{1jt} &= \frac \partial{\partial \pspar_{j-1}} \estimeq[F][1] \Big|_{\allpars_0} \\
&= - N_B(Y_t) \left[ \prod_{t' = t - \lag + 1}^t \frac{\intervdistf[][F](W_{t'})}{\parpropscore[t'][t'][W][T]} \right] \sum_{t' = t - \lag + 1}^t \scorefun[T][T][0][j-1][t'] \\
\implies |P_{1jt}| &\leq \bound[Y] \bound[W]^M \sum_{t' = t - \lag + 1}^t \Big| \scorefun[T][T][0][j-1][t'] \Big| \\
\implies E_{\allpars_0}|P_{1jt}| & \leq \bound[Y] \bound[W]^M \sum_{t' = t - \lag + 1}^t E_{\pspar_0} \Big| \scorefun[T][T][0][j-1][t'] \Big|.
\end{align*}
Since
\begin{align*}
E_{\pspar_0}^2 \Big| \scorefun[T][T][0][j-1][t'] \Big| & \leq
E_{\pspar_0} \left[ \scorefun[T][T][0][j-1][t'] ^2 \right]
\tag{Jensen's inequality}
\\
&\leq E_{\pspar_0} \left[ \| \scorefun[T][T][0][j-1][t'] \|^2 \right] < \infty, \tag{\cref{ass:propensity_score}}
\end{align*}
we have that $E_{\allpars[0]}|P_{1jt}| < \infty$.

\paragraph{Condition \ref{app_cond:dominated}}

We want to show that there exists
integrable function $\overset{\bigcdot\bigcdot}{\psi}(x)$ which dominates the second partial derivatives of $\estimeqsmall[F][]$ in a neighborhood of $\allpars_0$ for all $(w_t, \anyhist[t-1][h], y_t)$.
We consider derivatives of $\estimeqsmall$ with respect to $\allpars[F]_m, \allpars[F]_l$. For $k, m, l \geq 2$,
\[
\frac{\partial}{\allpars[F]_m}
\frac{\partial}{\allpars[F]_l}
\estimeqsmall[F][k] =
\frac{\partial}{\pspar[F]_{m-1}}
\frac{\partial}{\pspar[F]_{l-1}}
\scorefun[F][F][][k-1]
\]
where $\scorefun[T][T][][k-1]$ is the $k-1$ entry of the $\scorefun[T][T]$ vector.
From Condition \ref{ass:ps_dominated} of \cref{ass:propensity_score}, we know that the above is dominated by an integrable function. For $k \geq 2$ and if $l=1$ or $m = 1$ we have that the second partial derivative is equal to 0, since
\[
\frac{\partial}{\partial \allpars[F]_1} \estimeqsmall[F][k]
= \frac{\partial}{\partial \mu} \scorefun[F][F][][k-1] = 0.
\]
So for $k \geq 2$, all second partial derivatives are dominated by the function in Condition \ref{ass:ps_dominated} of \cref{ass:propensity_score}. Then, for $k = 1$, if at least one of $l = 1$ or $m = 1$ we have that the second partial derivative is also zero, since
\[
\frac{\partial}{\partial \allpars[F]_1} \estimeqsmall[F][1] = \frac{\partial}{\partial \mu} \estimeqsmall[F][1] = -1.
\]
So we need to show it only for $k = 1$, and $l, m \geq 2$. From Lemma \ref{app_lemma:a_psi} we have that
\begin{align*}
& \frac {\partial^2}{\partial \allpars[F]_m \partial \allpars[F]_l} \estimeqsmall[F][1] \\
&= - N_B(y_t) \left[ \prod_{j = t - \lag + 1}^t \frac{\intervdistf[][F](w_j)}{\parpropscore[j][j][w]} \right] \left\{ \left[ \sum_{j = t - \lag + 1}^t  \frac \partial{\partial \pspar_{m-1}}
\scorefun[F][F][][l-1][j] \right] \right. \\
& \hspace{120pt} - \left. \left[ \sum_{j = t - \lag + 1}^t \scorefun[F][F][][m-1][j] \right]
\left[ \sum_{j = t - \lag + 1}^t \scorefun[F][F][][l-1][j] \right] \right\}
\end{align*}
Because of \cref{ass:regularity_conditions}\ref{ass:finite_points} and \cref{ass:positivity} we have that
\[
|N_B(y_t)| \leq \bound[Y] \quad \text{and} \quad 0 \leq \prod_{j = t - \lag + 1}^t \frac{\intervdistf[][F](w_j)}{\parpropscore[j][j][w]} \leq \bound[W]^\lag,
\]
which implies that
\begin{align*}
\left| \frac {\partial^2}{\partial \allpars[F]_m \partial \allpars[F]_l} \estimeqsmall[F][1] \right| & \leq
\bound[Y] \bound[W]^\lag \left| \sum_{j = t - \lag + 1}^t  \frac \partial{\partial \pspar_{m-1}}
\scorefun[F][F][][l-1][j] \right| + \\
& \hspace{10pt} + \bound[Y] \bound[W]^\lag
\left| \sum_{j, j' = t - \lag + 1}^t \scorefun[F][F][][m-1][j] \scorefun[F][F][][l-1][j'] \right|  \\
& \leq \sum_{j = t - \lag + 1}^t \bound[Y] \bound[W]^\lag \left| \frac \partial{\partial \pspar_{m-1}} \scorefun[F][F][][l-1][j]  \right| + \\
& \hspace{10pt} + \sum_{j,j'=t - \lag + 1}^t \bound[Y] \bound[W]^\lag \left| \scorefun[F][F][][m-1][j] \scorefun[F][F][][l-1][j'] \right|
\end{align*}
We work first with the first term. Since the summation is over $\lag$ terms with $\lag$ finite, we only need to study the quantity in the absolute value.
We know from \cref{ass:propensity_score} that the second partial derivatives of $\scorefun$ are dominated by $\overset{\bigcdot\bigcdot}{\psi}(w_t, \anyhist[t-1][h])$  in a neighborhood of $\pspar_0$. Assume that this neighborhood is the $\epsilon-$ball around $\pspar_0$ (this always exists since a neighborhood is an open set around $\pspar_0$). Then, from \cref{remark:derivative_bounded} we know that
\[
\left| \frac \partial{\partial \pspar_{m-1}} \scorefun[F][F][][l-1][j] \right|
\leq
\left| \frac \partial{\partial \pspar_{m-1}} \scorefun[F][F][][l-1][j] \Big|_{\pspar_0} \right|
+ \epsilon \ K \ \overset{\bigcdot\bigcdot}{\psi}(w_t, \anyhist[t-1][h]),
\]
where the $K$ appears because we consider all $K$ second partial derivatives which are all bounded by $\overset{\bigcdot\bigcdot}{\psi}$. From \cref{ass:propensity_score}(\ref{ps_ass:conv_zero}), we have that the quantity on the right has finite expectation and is fixed in $\pspar$. Therefore, it is an integrable function that dominates the first partial derivatives of $\scorefun$ in a neighborhood of $\pspar_0$ for all $l, m$. Denote the maximum of these functions over $l, m$ by $\overset{\bigcdot\bigcdot}{\psi}_1(w_t, \anyhist[t-1][h])$.

We now turn our attention to the second term. Since (using again \cref{remark:derivative_bounded})
\[
\left| \scorefun[F][F][][m-1][j] \right|
\leq
\left| \scorefun[F][F][0][m-1][j] \right|
+ \epsilon \ K \ \overset{\bigcdot\bigcdot}{\psi_1}(w_t, \anyhist[t-1][h]),
\]
and $E_{\pspar_0} \left[ \left| \scorefun[F][F][0][m-1][j] \right| \right] < \infty $ from \cref{ass:propensity_score}, we have that this quantity is also dominated by an integrable function that is constant in $\pspar$. Denote the maximum of these functions over $m$ as $\overset{\bigcdot\bigcdot}{\psi_2}(w_t, \anyhist[t-1][h])$.

Putting these together we have that
\begin{align*}
\left| \frac {\partial^2}{\partial \allpars[F]_m \partial \allpars[F]_l} \estimeqsmall[F][1] \right| & \leq
M \bound[Y] \bound[W]^\lag \overset{\bigcdot\bigcdot}{\psi_1}(w_t, \anyhist[t-1][h]) + M^2 \bound[Y] \bound[W]^\lag \left[ \overset{\bigcdot\bigcdot}{\psi_2}(w_t, \anyhist[t-1][h]) \right]^2,
\end{align*}
where the right hand side is integrable. By defining taking the maximum of the quantity on the right hand side and $\overset{\bigcdot\bigcdot}{\psi}(w_t, \anyhist[t-1][h])$ for each $(w_t, \anyhist[t-1][h])$ we have that the condition holds using this new integrable function.

\paragraph{Consistency of the solution}

The last condition of \cref{theorem:CLT_solution_multi_mds} that we need to show is that the solution to \( \sum_{t = \lag}^T \estimeq = 0 \) is consistent for $\allpars_0$. Since the estimator of the propensity score parameters based on the score functions are consistent, we only need to show that the solution to \( \sum_{t = \lag}^T \estimeq[F][1] = 0 \) is consistent for $\mu_0 = 0$.

Since the estimator based on the true propensity score was shown to be consistent in \cref{theorem:normality}, the propensity score estimators $\widehat \pspar$ are consistent for $\pspar[T]$, $\estimeq[F][1]$ is a continuous function of the propensity score which is itself continuous in $\pspar$, using Slutsky's theorem we have that the solution to \( \sum_{t = \lag}^T \estimeq[F][1] = 0 \) using the estimated propensity score parameters is also consistent.

\paragraph{Asymptotic normality of the estimator without spatial smoothing} Since the conditions of \cref{theorem:CLT_solution_multi_mds} are satisfied, we have that the solution $\widehat \allpars_T$ to $\sum_{t = \lag}^T \estimeq = 0$ are asymptotically normal with
\[
\sqrt{T} \left(\widehat \allpars_T - \allpars_0 \right) \overset{d}{\rightarrow} N\left(0, V_{\allpars} \right),
\]
where $V_{\allpars} = A^{-1} B \left( A^{-1} \right)^T$ for
\begin{align}
A = \left[ \begin{array}{cc} -1 & -u^T \\ \bm 0_K & - V_{ps} \end{array} \right]
\quad \text{and} \quad
B = \left[ \begin{array}{cc} \asymvar & u^T \\ u & V_{ps} \end{array} \right].
\label{app_eq:asymvar_matrices}
\end{align}
As a result, focusing on the first entry of $\widehat \theta$ and since $\mu_0 = 0$, we have that
\[
\sqrt{T} \Big\{ \underbrace{(T - \lag + 1)^{-1} \sum_{t = \lag}^T \Bigg[ \prod_{j = t - \lag + 1}^t \frac{\intervdistf[][F](W_j)}{\parpropscore[j][j][W]} \Bigg] N_B(Y_t)}_{\text{estimator without spatial smoothing}} - \tempavgout \Big\} \rightarrow N \left( 0, \asymvar^e \right),
\]
where $\asymvar^e = [V_{\allpars}]_{11}$ is the $(1, 1)$ entry of $V_{\allpars}$.

\paragraph{Asymptotic normality of the estimator {\it with} spatial smoothing}

To prove the asymptotic normality of the estimator with spatial smoothing (our proposed estimator in Equation~\cref{eq:estimatorN}), we again decompose the estimation error in two components like in Equation~\cref{app_eq:est_error_split} for the proof of \cref{theorem:normality}. We write
\begin{align*}
\error & \ = \ \underbrace{\Bigg[ \prod_{j = t - \lag + 1}^t \frac{\intervdistf[][F](W_j)}{\parpropscore[j][j][W]} \Bigg] N_B(Y_t) - \avgout[F][][\lag]}_{\quant[1]} \\
& \hspace{40pt}
   + \underbrace{\Bigg[ \prod_{j = t - \lag + 1}^t \frac{\intervdistf[][F](W_j)}{\parpropscore[j][j][W]} \Bigg] \Bigg[
    \int_B \sum_{s \in \sparseset[{}Y]} K_{b_T}(\omega, s) \mathrm{d}\omega -
    N_B(Y_t) \Bigg]}_{\quant[2]},
\end{align*}
where we use the parametric propensity score. We showed the asymptotic normality based on $\quant[1]$, so we are left to show that
\(
\sqrt{T} \left( (T - \lag + 1)^{-1} \sum_{t = \lag}^T \quant[2] \right) \overset{p}{\rightarrow} 0.
\)
In the proof of \cref{theorem:normality} we already showed that the above result holds. The proof there can be directly used here also if the known propensity score is used (instead of the estimated one). By re-defining the terms $c_t$ defined there to use the estimated propensity score as
\[
c_t = \prod_{j = t - \lag + 1}^t \frac{\intervdistf[][F](W_j)}{\parpropscore[j][j][W]},
\]
it suffices to show that $c_t$ is bounded, and the steps of the proof with the known propensity score will follow identically. But since the propensity score $\parpropscore$ is continuous in $\pspar$ (since it is differentiable), the function $1 / x$ is continuous for $x > 0$,
and
\( \intervdistf[][F](w_j) / \parpropscore[j][j][w][T] \leq \bound[W] \)
then $c_t$ will be bounded in a neighborhood of $\pspar_0$. And since $\widehat \pspar \overset{p}{\rightarrow} \pspar_0$, $\widehat \pspar$ will be in the neighborhood of $\pspar_0$ with probability 1 as $T$ increases, so $c_t$ will be bounded.

Putting these results together we have asymptotic normality of the spatially smoothed estimator and
\[
\sqrt{T} \left( \estimatorN - \tempavgout \right) \overset{d}{\rightarrow} N \left( 0, v^e \right).
\]
\end{proof}

\renewcommand*{\proofname}{\textbf{Proof of \cref{theorem:smaller_asymvar}}}
\begin{proof}
The asymptotic variance $\asymvar^e$ corresponds to the (1, 1) entry of the matrix $A^{-1} B (A^{-1})^\top$, where $A, B$ are defined in Equation~\cref{app_eq:asymvar_matrices}.
\begin{align*}
A^{-1} B (A^{-1})^\top
&=
\left[ \begin{array}{cc} 1 & u^T \\ \bm 0_K & V_{ps} \end{array} \right]^{-1}
\left[ \begin{array}{cc} \asymvar & u^\top \\ u & V_{ps} \end{array} \right]
\left\{ \left[ \begin{array}{cc} 1 & u^\top \\ \bm 0_K & V_{ps} \end{array} \right]^{-1}
\right\}^\top \\
&=
\left[ \begin{array}{cc} 1 & - u^\top V_{ps}^{-1} \\ \bm 0_K & V_{ps}^{-1} \end{array} \right]
\left[ \begin{array}{cc} \asymvar & u^\top \\ u & V_{ps} \end{array} \right]
\left[ \begin{array}{cc} 1 & \bm 0_K \\ - V_{ps}^{-1} u & V_{ps}^{-1} \end{array} \right] \\
&=
\left[ \begin{array}{cc}
v - u^\top V_{ps}^{-1} u & \bm 0_K^\top \\ \ldots & \ldots
\end{array} \right]
\left[ \begin{array}{cc} 1 & \ldots \\ - V_{ps}^{-1}u & \ldots \end{array} \right] \\
&=
\left[ \begin{array}{cc}
v - u^\top V_{ps}^{-1} u & \ldots \\
\ldots & \ldots
\end{array} \right]
\end{align*}
so $\asymvar^e = \asymvar - u^\top V_{ps}^{-1} u$, and since $V_{ps}$ is positive definite and therefore $V_{ps}^{-1}$ is positive definite we have that $u^\top V_{ps}^{-1} u \geq 0$ and
$\asymvar^e \leq \asymvar$.
\end{proof}
\renewcommand*{\proofname}{\textbf{Proof}}

\subsection{Asymptotics for an increasing number of independent regions}
\label{app_subsec:sprawl-asymptotics}

All the asymptotic results that have been discussed up to now correspond to the scenario where 1 region is observed repeatedly over time, and the asymptotic properties are derived when the number of time periods $T$ increases to infinity. However, there might also be interest in situations where the number of time periods is fixed, but there exist an increasing number of independent-acting regions. 

Here we consider this related but separate scenario. We start by defining relevant estimands in this setting, ensuring that these new estimands are as closely comparable to the estimands in the manuscript. We propose similar estimators, and derive the asymptotic properties of the new estimators when the number of regions $R$ goes to infinity.

\subsubsection{Estimands for independently-acting regions}

For this scenario, we decompose the treatments, potential outcomes, outcomes, and history over all the regions to region-specific components and write
$w_t = (w_{1t}, w_{2t}, \dots, w_{Rt})$, $\whist = (\whist[1t], \whist[2t], \dots, \whist[Rt])$,
$Y_t(\whist) = (Y_{1t}(\whist), Y_{2t}(\whist), \dots, Y_{Rt}(\whist))$, $Y_t = (Y_{1t}, Y_{2t}, \dots, Y_{Rt})$, and
$\history[t] = (\history[1t], \history[2t], \dots, \history[Rt])$,
where $\history[rt] = \{\Whist[rt], \anyhist[rt][\bm Y], \anyhist[r(t+1)][\covs] \}$.
We make the following assumption that describes that the regions do not interfere spatially, and that treatment assignment is local within regions:
\begin{assumption}[Independently acting spatial regions] We assume the following:
\begin{enumerate}
\item For $\whist, \whist'$ such that $\whist[rt]= \whist[rt]'$, we have that $Y_{rt}(\whist) = Y_{rt}(\whist')$ (and a similar assumption for the time-varying covariates), and
\item the treatment assignment of region $r$ at time $t$ does not depend on unobserved potential outcomes or potential time varying covariates, nor on any information from other regions, denoted as $W_{rt} \independent \history[t-1], \anyhist[T][\allout], \anyhist[T][\allcovs] \mid \history[r(t-1)].$
\end{enumerate}
\label{app_ass:independent_regions}
\end{assumption}
\noindent
This assumption allows us to denote potential outcomes using their region-specific treatments {\it only}, and write
$Y_t(\whist) = (Y_{1t}(\whist[1t]), Y_{2t}(\whist[2t]), \dots, Y_{Rt}(\whist[Rt]))$. It also allows us to think of the $R$ regions as completely separately acting regions, as outcomes, covariates and treatments of one region do not depend on any information of any other region. Based on this assumption, we can use $\anyhist[rT][\allout]$ to denote the collection of potential outcomes for region $r$ over all time periods and for any regional treatment path (and similarly for covariates).

For the purpose of this section {\it only}, we also assume the temporal carryover effect is limited to up to some lag $\lag_Y$. Specifically, we assume that the outcome at time $t$ can only depend on treatments during the preceding $\lag_Y$ time periods, formalized as
\begin{assumption}[Limited temporal carryover effect]
There exists positive integer $\lag_Y$ such that
for $\whist[rt], \whist[rt]'$ for which
$w_{r\tau} = w_{r\tau}'$ for all $\tau = t - \lag_Y + 1, \dots, t - 1, t$,
it holds that
$Y_{rt}(\whist[rt]) = Y_{rt}(\whist[rt]')$.
\label{app_ass:no_carryover_regions}
\end{assumption}

We start by defining region and time specific estimands that are as closely related to the estimands defined in Section~\ref{sec:estimands}. We again focus on point pattern treatments and outcomes and on estimands that represent the number of outcome active locations in each region. For simplicity we focus on the scenario where the temporal carryover lag $\lag_Y$ and the intervention length $\lag$ are both equal to 1, but we note that the results would also follow in all scenarios where $\lag \geq \lag_Y$.

Let $\intervdist$ be a stochastic treatment assignment that is constant across regions. The stochastic intervention can depend on baseline covariates of the regions, but we refrain for explicitly denoting that for simplicity. We define the expected number of outcome active locations at region $r$ at time $t$ as
\[
\avgout[F][][][t][r] =
\int_{w_{rt}} N_r \Big( Y_{rt}(\Whist[r(t-1)], w_{rt}) \Big) d\intervdist(w_{rt}) =
\int_{w_{rt}} N_r \big( Y_{rt}(w_{rt}) \big) d\intervdist(w_{rt}),
\]
where we define the estimand as in the first equation to be more closely related to the estimands in \cref{sec:estimands}, and the second equation holds because of \cref{app_ass:no_carryover_regions} for $\lag_Y = 1$. We specify region-specific estimands, averaged over time, as
\[
\tempavgout[F][][][r] = \frac1T \sum_{t = 1}^T \avgout[F][][][t][r],
\]
and estimands averaged over region and time as
\[
\tempavgout[F][][][] = \frac1R \sum_{r = 1}^R \tempavgout[F][][][r] = \frac1R \sum_{r = 1}^R \frac1T \sum_{t = 1}^T \avgout[F][{}][{}][t][r].
\]

\subsubsection{Estimators for independently-acting regions}

Like in Section~\ref{sec:estimation}, assume that $\intervdist$ admits density $\intervdistf[][F]$.
Based on \cref{app_ass:independent_regions}, we can separate the treatment assignment over all regions to the treatment assignment of each region separately, as
\[
\propscore[t][t] = \prod_{r = 1}^R \propscore[rt][rt]
\]
where $\propscore[rt][rt] = f(W_{rt} = w_{rt} \mid \history[r(t-1)])$ is the region-specific propensity score.
We propose corresponding region and time-specific estimator
\[
\estimatorNt[F][][][r] = \frac{\intervdistf[][T][rt]}{\propscore[rt][rt][W]} N_r(Y_{rt}),
\]
where we use $N_r(Y_{rt})$ to denote the number of outcome active locations in the observed outcome for region $r$ at time $t$. We also propose the corresponding estimators averaged over time and over regions as
\[
\estimatorN[F][][][r] = \frac1T \sum_{t = 1}^T \estimatorNt[F][][][r]
\quad \text{and} \quad
\estimatorN[F][][][] = \frac1R \sum_{r = 1}^T \estimatorN[F][][][r].
\]

\subsubsection{Consistency and asymptotic normality for independent-acting regions}

We will show the consistency and asymptotic normality of these estimators when the propensity score is known for an increasing number of independently acting regions. The proof here follows closely the proof in \cite{papadogeorgou2019causal} for weighting estimators under a known propensity score and for stochastic interventions. We do not show the asymptotic properties of an estimator based on a correctly specified parametric propensity score, since, once the baseline conditions for the known propensity score are established, the proof for the estimated propensity score would resemble the corresponding proof in \cite{papadogeorgou2019causal}.

To establish the asymptotic properties for an increasing number of regions, we first assume that our observed regions are a random sample from some super-population of regions. Let $(\anyhist[rT][\allout], \anyhist[rT][\allcovs], \Whist[rT])$ be a draw from a super-population distribution $F^{sp}$.
We assume that \cref{app_ass:independent_regions} holds over $F^{sp}$ and we make the following super-population positivity assumption for the independent regions (which resembles the one in \cref{ass:positivity}):
\begin{assumption}[Positivity of treatment assignment in the super-population]
There exists $\bound[W]$ such that
$\propscore[rt][rt] > \bound[W] \cdot \intervdistf[][F](w_{rt})$  for all treatment point patterns $w_{rt}$.
\label{app_ass:positivity_regions}
\end{assumption}
\noindent
We also assume that there is a bounded number of outcome active locations within each region, similarly to \cref{ass:regularity_conditions}\ref{ass:finite_points}, but region-specific:
\begin{assumption}
There exists $\bound[Y] > 0$ such that $N_r(Y_{rt}(\whist[rt])) < \bound[Y]$ with probability 1 over $F^{sp}$, where $\whist[rt]$ is any possible treatment path.
\label{app_ass:boundedY_regions}
\end{assumption}

\begin{theorem}
If Assumptions \ref{app_ass:independent_regions}, \ref{app_ass:no_carryover_regions} and \ref{app_ass:positivity_regions} hold, then, for $R \rightarrow \infty$,
$\estimatorN[F][][][]$ is consistent for $\tempavgout[F][][][]$ and
\( \displaystyle
\sqrt{R} \left( \estimatorN[F][][][] - \tempavgout[F][][][] \right) \rightarrow N(0, \sigma^2),
\)
for some $\sigma^2 > 0$, where $\tempavgout[F][][][]$ is the super-population estimand defined as
$\tempavgout[F][][][] = E_{F^{sp}} \left[ \tempavgout[F][][][r] \right]$.

\begin{proof}
Let $\D[r] = (\Whist[T], \anyhist[T][\bm Y], \anyhist[T][\covs])$ denote all observed data for region $r$, and $\D = (\D[1], \D[2], \dots, \D[R])$. We define
\[
\psi_r(\D[r]; \mu) = \left(\frac1T \sum_{t = 1}^T \frac{\intervdistf[][T][rt]}{\propscore[rt][rt][W]} N_r(Y_{rt}) \right) - \mu
\]
and $\Psi_R(\D; \mu) = \sum_{r = 1}^T \psi_r (\D[r]; \mu)$. Then obviously the estimator $\widehat \mu = \estimatorN[F][][][]$ is the solution to $\Psi_R(\D; \mu) = 0$. Then we calculate the solution to $\Psi^{sp}(\mu) = E_{F^{sp}}(\psi_r(\D[r]; \mu)) = 0$ which is equal to
\begin{align*}
\mu_0 &= E_{F^{sp}} \left[ \frac1T \sum_{t = 1}^T \frac{\intervdistf[][T][rt]}{\propscore[rt][rt][W]} N_r(Y_{rt}) \right] \\
&= \frac1T \sum_{t = 1}^T E_{F^{sp}} \left[ \frac{\intervdistf[][T][rt]}{\propscore[rt][rt][W]} N_r(Y_{rt}) \right] \\
&= \frac1T \sum_{t = 1}^T \int_{\anyhist[rT][\allout], \anyhist[rT][\allcovs]} \int_{w_{r1}} \int_{w_{r2}} \dots \int_{w_{rt}} \frac{\intervdistf[][F](w_{rt})}{\propscore[rt][rt][w]} N_r(Y_{rt}(\whist[rt])) \ \mathrm{d}F^{sp}(\anyhist[rT][\allout], \anyhist[rT][\allcovs], \whist[rt])  \\
&= \frac1T \sum_{t = 1}^T \int_{\anyhist[rT][\allout], \anyhist[rT][\allcovs]} \int_{w_{r1}} \int_{w_{r2}} \dots \int_{w_{r(t-1)}}\\
& \hspace{4cm}
\left[ \int_{w_{rt}} \frac{\intervdistf[][F](w_{rt})}{\propscore[rt][rt][w]} N_r(Y_{rt}(\whist[rt])) f_{W_{rt}}(w_{rt} \mid \Whist[r(t-1)] = \whist[r(t-1)],  \anyhist[rT][\allout], \anyhist[rT][\allcovs]) \ \mathrm{d}w_{rt} \right] \\
& \hspace{8cm}
 \ \mathrm{d}F^{sp}(\anyhist[rT][\allout], \anyhist[rT][\allcovs], \whist[r(t-1)])  \\
&= \frac1T \sum_{t = 1}^T \int_{\anyhist[rT][\allout], \anyhist[rT][\allcovs]} \int_{w_{r1}} \int_{w_{r2}} \dots \int_{w_{r(t-1)}}
\left[ \int_{w_{rt}} \frac{\intervdistf[][F](w_{rt})}{\propscore[rt][rt][w]} N_r(Y_{rt}(\whist[rt])) f_{W_{rt}}(w_{rt} \mid \history[r(t-1)]) \ \mathrm{d}w_{rt} \right] \\
& \hspace{8cm}
 \ \mathrm{d}F^{sp}(\anyhist[rT][\allout], \anyhist[rT][\allcovs], \whist[r(t-1)])
 \tag{From \cref{app_ass:independent_regions}}  \\
&= \frac1T \sum_{t = 1}^T \int_{\anyhist[rT][\allout], \anyhist[rT][\allcovs]} \int_{w_{r1}} \int_{w_{r2}} \dots \int_{w_{r(t-1)}}
\left[ \int_{w_{rt}} \intervdistf[][F](w_{rt}) N_r(Y_{rt}(\whist[rt])) \ \mathrm{d}w_{rt} \right]
\ \mathrm{d}F^{sp}(\anyhist[rT][\allout], \anyhist[rT][\allcovs], \whist[r(t-1)])
\tag{From the definition of the region-specific propensity score} \\
&= \frac1T \sum_{t = 1}^T E_{F^{sp}} \left[ \avgout[F][][][t][r]  \right] \\
&= E_{F^{sp}} \left[ \tempavgout[F][][][r] \right]
\end{align*}

\paragraph{Consistency} We use an alteration of Lemma~A in Section~7.2.1 of \cite{Serfling1980}.
Since $\psi_r(\D[r]; \mu)$ is monotone in $\mu$ with
$ \partial \psi_r(\D[r]; \mu)/\partial \mu = - 1 < 0 $, we have that
$\Psi_R(\D; \mu)$ and $\Psi^{sp}(\mu)$
are also monotone which implies uniqueness of their roots, $\widehat \mu$ and $\mu_0$.
From the strong law of large numbers we have that
$\Psi_R(\D; \mu) \overset{a.s.}{\rightarrow} \Psi^{sp}(\mu)$, and
$$
|\Psi^{sp}(\widehat\mu) - \Psi^{sp}(\mu_0)| =
|\Psi^{sp}(\widehat\mu) - \Psi_R(\widehat\mu)| \leq
\sup_\mu |\Psi^{sp}(\mu) - \Psi_R(\mu)| \rightarrow 0,
$$
which, by the uniqueness of the roots for $\Psi^{sp}$ and $\Psi_R$ implies that
$\estimatorN[F][][][] \overset{as}{\rightarrow} \tempavgout[F][][][]$ and
$\estimatorN[F][][][]$ is consistent for $\tempavgout[F][][][]$.

\paragraph{Asymptotic normality}
For asymptotic normality we will use Theorem~A in Section~7.2.2 of \cite{Serfling1980}. We have already shown that $\mu_0$ is an isolated root of $\Psi^{sp}(\mu) = 0$ (since it is unique) and that $\psi_r(\D[r];\mu)$ is monotone in $\mu$. We also have that $\Psi^{sp}(\mu)$ is differentiable in $\mu$ with $\frac{\partial}{\partial \mu} \Psi^{sp}(\mu) = -1 \neq 0$. Lastly we will show that $E_{F^{sp}} \left[ \psi_r^2(\D[r]; \mu) \right]$ is finite in a neighborhood of $\mu_0$. To do so, consider $\mu$ in an $\epsilon$-neighborhood of $\mu_0$, $\mu \in (\mu_0 - \epsilon, \mu_0 + \epsilon)$. Then
\begin{align*}
E_{F^{sp}} \left[ \psi_r^2(\D[r]; \mu) \right] &=
E_{F^{sp}} \left\{ \left[ \frac1T \sum_{t = 1}^T \frac{\intervdistf[][T][rt]}{\propscore[rt][rt][W]} N_r(Y_{rt}) - \mu \right]^2 \right\} \\
&=
E_{F^{sp}} \left\{ \left| \frac1T \sum_{t = 1}^T \frac{\intervdistf[][T][rt]}{\propscore[rt][rt][W]} N_r(Y_{rt}) - \mu \right|^2 \right\} \\
& \leq E_{F^{sp}} \left\{ \left[ \frac1T \sum_{t = 1}^T \frac{\intervdistf[][T][rt]}{\propscore[rt][rt][W]} N_r(Y_{rt}) + |\mu| \right]^2 \right\} \tag{Triangle inequality} \\
&= E_{F^{sp}} \left\{ \left[ \frac1T \sum_{t = 1}^T \frac{\intervdistf[][T][rt]}{\propscore[rt][rt][W]} N_r(Y_{rt}) \right]^2 \right\} + 2|\mu| E_{F^{sp}} \left[ \frac1T \sum_{t = 1}^T \frac{\intervdistf[][T][rt]}{\propscore[rt][rt][W]} N_r(Y_{rt}) \right] +
\mu^2 \\
& \leq E_{F^{sp}} \left\{ \left[ \frac1T \sum_{t = 1}^T \bound[W] \bound[Y] \right]^2 \right\} +
2 |\mu| E_{F^{sp}} \left[ \frac1T \sum_{t = 1}^T \bound[W] \bound[Y] \right] + |\mu|^2
\\
&= (\bound[W]\bound[Y])^2 + 2 |\mu| \bound[W]\bound[Y] + \mu^2
\end{align*}
where we used that all terms in the summation are positive along with Assumptions \ref{app_ass:positivity_regions} and \ref{app_ass:boundedY_regions}. Since $\mu \in (\mu_0 + \epsilon, \mu_0 + \epsilon)$ it is bounded, so the expectation above exists.

Then, since all the conditions of the theorem are satisfied we have that
\[
\sqrt{R} \left( \estimatorN[F][][][] - \tempavgout[F][][][] \right) \rightarrow N(0, \sigma^2),
\]
where $\sigma^2 = E_{F^{sp}} \left[ \psi_r^2(\D[r]; \mu_0) \right]$.
\end{proof}

\end{theorem}

\section{The H\'ajek Estimator}
\label{app_sec:hajek}

The standardization of weights used in the H\'ajek estimator is known to be effective in the settings where the weights are extreme.  Its sample boundedness property guarantees that the resulting estimate is always within the range of the observed outcome.  In our case, the H\'ajek estimator replaces the division by $T - \lag + 1$ with that by $\sum_{t = \lag}^T w_t$ where $w_t$ is the product of fractions in Equation~\cref{eq:estimatort}. For example,
\begin{align*}
\estimatorN_{\text{H\'ajek}} & \ = \ \frac{1}{\sum_{t = \lag}^T w_t} \sum_{t = \lag}^T \estimatorNt[F][][\lag] 
\end{align*}
%
The new martingale theorem stated in \cref{theorem:CLT_solution_multi_mds} can be used in future research to show that the H\'ajek estimator is consistent and asymptotically normal, and derive the functional form of its asymptotic variance. However, for now, we use a heuristic approach to estimating the variance bound of the H\'ajek estimator. Since the H\'ajek estimator simply rescales the corresponding IPW estimator by
$(T-\lag+1)/\sum_{t = \lag}^T w_t$, we scale the variance bound derived for the estimator by $[(T-\lag+1)/(\sum_{t = \lag}^T w_t)]^2$.

\section{Sensitivity analysis}
\label{app_sec:SA}

In this section we discuss sensitivity analysis for the IPW estimators. In the main text of the manuscript we discuss sensitivity analysis for the H\'ajek estimator, which is admittedly a much harder problem due to the standardization of weights performed in the H\'ajek correction.

In this section we discuss sensitivity analysis based on the IPW estimator. We quickly see that bounding the estimator for different amounts of propensity score misspecification $\Gamma$ can be directly achieved by solving a linear program. We can similarly bound the causal effect estimator exactly. In contrast, in the main text, bounding the value of the H\'ajek estimator requires additional tools to {\it transform} the problem to a linear program. This transformation forbids us from bounding the effect estimator exactly and forces us to acquire possibly conservative bounds for the effect estimator (see \cref{theorem:SA_bound_both}).

\subsection{For the IPW estimator}

We focus again on bounding the estimators for intervention over a single time period, though extensions to multiple time periods are direct, and discussed in more detail for the H\'ajek estimator in \cref{app_subsec:SA_hajek}. The IPW estimators that use the correct propensity score can be written as:
\begin{align*}
\estimatorN[F][][][\bm \rho] &= \frac1T \sum_{t = 1}^T \rho_t \ w_t(\interv) \ \widetilde N_B(Y_t), \quad \text{and} \\
\tempeffect[T][F][1][2][1][\bm \rho] &=\frac1T \sum_{t = 1}^T \rho_t \ w_t(\interv[F][2]) \ \widetilde N_B(Y_t) -
\frac1T \sum_{t = 1}^T \rho_t \ w_t(\interv[F][1]) \ \widetilde N_B(Y_t) \\
&= \frac1T \sum_{t = 1}^T \rho_t \left[ w_t(\interv[F][2]) - w_t(\interv[F][1])  \right] \widetilde N_B(Y_t)
\end{align*}
where
\begin{equation*}
w_t(\interv) = \frac{\intervdistf[][T]}{\propscore[t][t][W]} \quad \text{and} \quad
\widetilde N_B(Y_t) =
\int_B \sum_{s \in \sparseset[{}Y][t]} K_b(\|\omega - s\|) \mathrm{d} \omega.
\end{equation*}
Both of the IPW estimators $\estimatorN[F][][][\bm \rho]$ and $\tempeffect[T][F][1][2][1][\bm \rho]$ are linear in $\bm \rho$, so finding their maximum/minimum
over $\rho_t \in [\Gamma^{-1}, \Gamma]^T$ for each $t$ is a linear problem and can be easily solved.

\subsection{For the H\'ajek estimator}
\label{app_subsec:SA_hajek}

The standardization of the weights in the H\'ajek estimator implies that maximizing/minimizing the value of the estimator is no longer linear in $\rho_t$. This is evident in the form of the H\'ajek estimator for the number of points and the effect in a region, defined respectively as
\[
\frac {\sum_{t = 1}^T \rho_t \ w_t(\interv) \ \widetilde N_B(Y_t) }{ \sum_{t = 1}^T \rho_t \ w_t(\interv) } \quad \text{and} \quad
\frac {\sum_{t = 1}^T \rho_t \ w_t(\interv[F][2]) \ \widetilde N_B(Y_t) }{ \sum_{t = 1}^T \rho_t \ w_t(\interv[F][2]) } -
\frac {\sum_{t = 1}^T \rho_t \ w_t(\interv[F][1]) \ \widetilde N_B(Y_t) }{ \sum_{t = 1}^T \rho_t \ w_t(\interv[F][1]) }
\]
where
\[
w_t(\interv) = \frac{\intervdistf[][T]}{\propscore[t][t][W]} \quad \text{and} \quad
\widetilde N_B(Y_t) =
\int_B \sum_{s \in \sparseset[{}Y][t]} K_b(\|\omega - s\|) \mathrm{d} \omega.
\]

\cref{theorem:SA_bound_both} states that bounding the estimator for the expected number of points $\estimatorN[F][][][{\bm \rho}]$ can be transformed to a linear problem. However, the standardization of weights in the H\'ajek estimator and the fact that our estimator is the difference of two linear fractionals forbids us to see the problem of bounding the effect estimator $ \tempeffect[T][F][1][2][1][\bm \rho] $ the same way.

\renewcommand*{\proofname}{\textbf{Proof of \cref{theorem:SA_bound_both}}}
\begin{proof}

We view the problem of bounding $\estimatorN[F][2][][{\bm \rho}]$ as a maximization/minimization problem of a linear fractional with positive denominator. These problems have been previously studied, and it has been shown that they can be transformed to a linear programming problem using the Charnes-Cooper transformation \citep{charnes1962programming}. The theorem states this transformation in the context of our estimator.

For the problem of bounding the effect estimator, the objective can be written as,
\begin{align*}
\tempeffect[T][F][1][2][1][\bm \rho] &=
\frac {\sum_{t = 1}^T \rho_t \ w_t(\interv[F][2]) \ \widetilde N_B(Y_t) }{ \sum_{t = 1}^T \rho_t \ w_t(\interv[F][2]) } -
\frac {\sum_{t = 1}^T \rho_t \ w_t(\interv[F][1]) \ \widetilde N_B(Y_t) }{ \sum_{t = 1}^T \rho_t \ w_t(\interv[F][1]) } \\
&= \estimatorN[F][2][][{\bm \rho}] - \estimatorN[F][1][][{\bm \rho}].
\end{align*}
Thus, maximizing $\tempeffect[T][F][1][2][1][\bm \rho]$ over $\bm \rho \in [\Gamma^{-1}, \Gamma]^T$ is equivalent to maximizing
$\estimatorN[F][2][][{\bm \rho}] - \estimatorN[F][1][][{\bm \rho}]$ over the same region for $\bm \rho$. Since the space $(\bm \rho_1, \bm \rho_2) \in [\Gamma^{-1}, \Gamma]^{2T}$ includes $(\bm \rho_1, \bm \rho_2)$ where $\bm \rho_1 = \bm \rho_2$ as a subspace, we have that
\begin{align*}
\max_{\bm \rho \in [\Gamma^{-1}, \Gamma]^T} \left\{ \estimatorN[F][2][][{\bm \rho}] - \estimatorN[F][1][][{\bm \rho}] \right\}
&\leq
\max_{(\bm \rho_1, \bm \rho_2) \in [\Gamma^{-1}, \Gamma]^{2T}} \left\{ \estimatorN[F][2][][{\bm \rho_2}] - \estimatorN[F][1][][{\bm \rho_1}] \right\} \\
&= \max_{\bm \rho_2 \in [\Gamma^{-1}, \Gamma]^T} \left\{ \estimatorN[F][2][][{\bm \rho_2}] \right\} - \min_{\bm \rho_1 \in [\Gamma^{-1}, \Gamma]^T} \left\{ \estimatorN[F][1][][{\bm \rho_1}] \right\},
\end{align*}
where the last equality holds since $\estimatorN[F][j][][{\bm \rho_j}] \geq 0$. Similarly, we can derive the bound for the minimum of $\tempeffect[T][F][1][2][1][\bm \rho]$.
\end{proof}

Next, we derive similar conservative bounds for the estimators corresponding to the interventions over multiple time periods. Recall that the H\'ajek estimator for the number of events in region $B$ under a stochastic intervention is given by,
\[
\frac {\sum_{t = \lag}^T w_t(\interv[F][][\lag]) \ \widetilde N_B(Y_t) }{ \sum_{t = \lag}^T w_t(\interv[F][][\lag]) },
\quad \text{where} \quad
w_t(\interv[F][][\lag]) = \prod_{j = t - \lag + 1}^t \frac{\intervdistf[][T][j]}{\propscore[j][j][W]}.
\]
So, our sensitivity analysis would search to find the bounds of
\[
\estimatorN[F][][\lag][{\bm \rho}] =
\frac {\sum_{t = \lag}^T \left( \prod_{j = t - \lag + 1}^t \rho_j \right) w_t(\interv[F][][\lag]) \ \widetilde N_B(Y_t) }
{ \sum_{t = \lag}^T \left( \prod_{j = t - \lag + 1}^t \rho_j \right) w_t(\interv[F][][\lag]) }
\]
over $\bm \rho \in [\Gamma^{-1}, \Gamma]$. Since each $\rho_t$ can take a value in $[\Gamma^{-1}, \Gamma]$, the sensitivity analysis weights in the H\'ajek estimator for multiple time periods, $\prod_{j = t - \lag + 1} \rho_j$, take a value in $[\Gamma^{-\lag}, \Gamma^\lag]$. Therefore the set $\{  \bm \alpha \in [\Gamma^{-\lag}, \Gamma^\lag] \}$ includes all vectors of length $T$ whose $t^{th}$ entry can be written as $\prod_{j=t -\lag + 1}^t \rho_t$ for some vector $\bm \rho$. Using the argument similar to that of \cref{theorem:SA_bound_both}, we have
\begin{align*}
\min_{\bm \rho \in [\Gamma^{-1}, \Gamma]} \left\{ \estimatorN[F][][\lag][{\bm \rho}] \right\}
& \geq
\min_{\bm \alpha \in [\Gamma^{-\lag}, \Gamma^\lag]} \left\{ \frac {\sum_{t = \lag}^T \alpha_t \ w_t(\interv[F][][\lag]) \ \widetilde N_B(Y_t) }
{ \sum_{t = \lag}^T \alpha_t \ w_t(\interv[F][][\lag]) } \right\}, \\
\max_{\bm \rho \in [\Gamma^{-1}, \Gamma]} \left\{ \estimatorN[F][][\lag][{\bm \rho}] \right\}
& \leq
\max_{\bm \alpha \in [\Gamma^{-\lag}, \Gamma^\lag]} \left\{ \frac {\sum_{t = \lag}^T \alpha_t \ w_t(\interv[F][][\lag]) \ \widetilde N_B(Y_t) }
{ \sum_{t = \lag}^T \alpha_t \ w_t(\interv[F][][\lag]) } \right\}.
\end{align*}
The quantities on the right can be computed by turning the linear fractional problem to a linear problem via the Charnes-Cooper transformation.  Then, we can use these quantities as the conservative bounds for the minimum and maximum of our target quantities. Based on these bounds, we can again use \cref{theorem:SA_bound_both} to acquire conservative bounds of the effect of changing the intervention for interventions over multiple time periods.

\section{Additional Simulation Results on the Iraq-based scenario}
\label{app_sec:add_iraq_sims}

\subsection{Asymptotic Variance and Bound, and Estimated Variance Bound}
\label{app_subsec:add_sims_iraq_variance}

\begin{figure}[p]
\centering
\includegraphics[width = 0.9 \textwidth,trim = 0 26 0 0, clip]{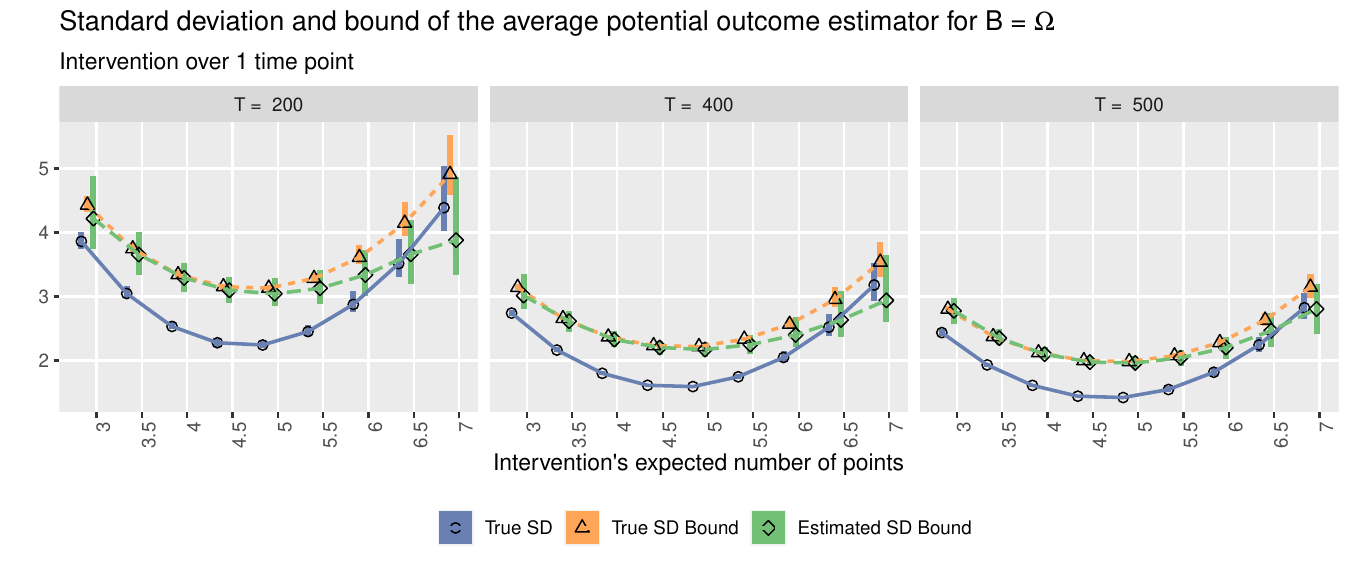} \\[-20pt]
\includegraphics[width = 0.9 \textwidth,trim = 0 26 0 21, clip]{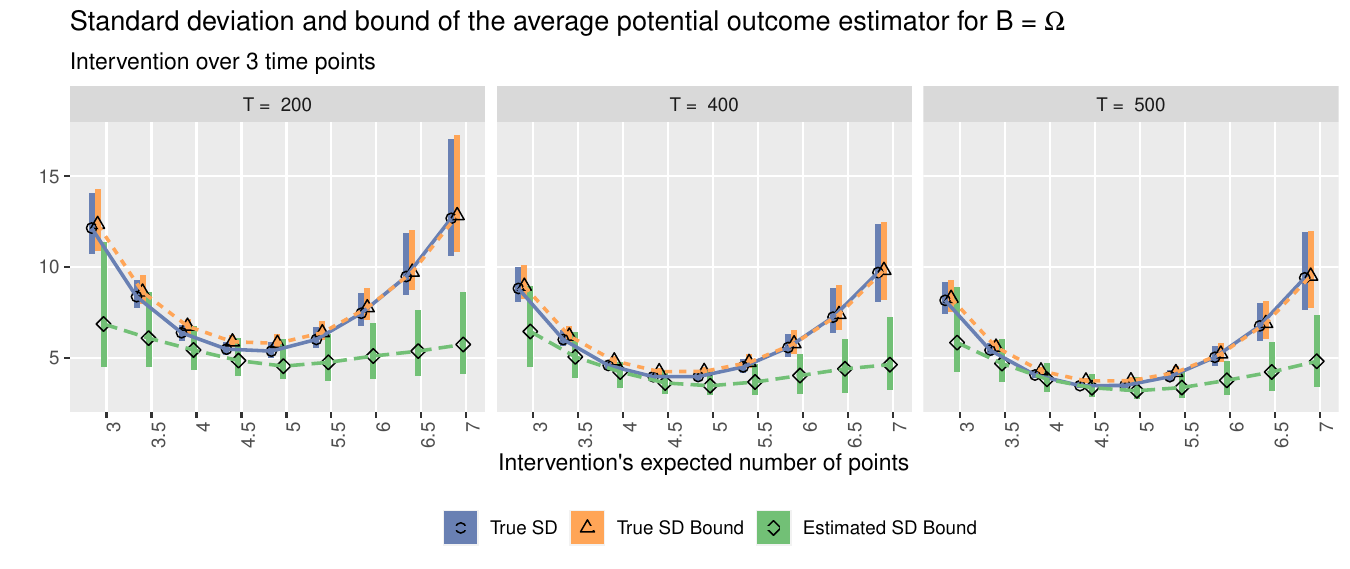} \\[-20pt]
\includegraphics[width = 0.9 \textwidth,trim = 0 0 0 21, clip]{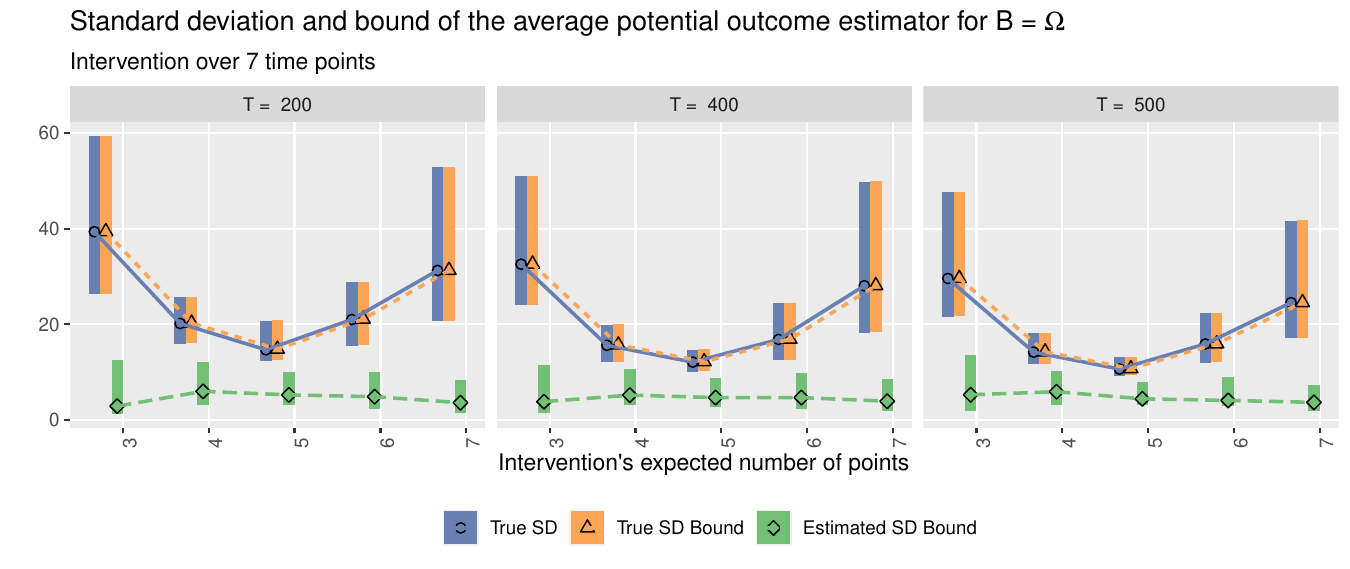}
\caption{Asymptotic Standard Deviation and Bound, and Estimated Bound. This figure shows the true asymptotic standard deviation (blue circles), the true asymptotic bound (orange triangles), and the estimated bound (green rhombuses) of the IPW estimator for the average potential outcome using the true propensity score, under interventions that take place over $M = 1, 3$ and $7$ time periods (rows), and for increasing length of the time series (columns). The horizontal axis shows the intensity of the intervention at each time period. The points show the median value, and the rectangles show the interquartile range over 200 simulated data sets.}
\label{app_fig:sims_iraq_var_y_other_B1}
\end{figure}

\cref{fig:sims_iraq_var_y_int1_B1} shows the average (over 200 simulated data sets) of the true asymptotic standard deviation and true bound as well as the estimated standard deviation bound of the IPW estimator for the average potential outcome using the true propensity score, for interventions taking place over $M \in \{1,  3\}$ time periods. \cref{app_fig:sims_iraq_var_y_other_B1} is a similar plot for the interventions taking place over $M = 1, 3$, and $7$ (rows) time periods, and observed time series of length $T = 200, 400, 500$ (columns). These plots show the median and interquartile range of the asymptotic standard deviation, true bound, and estimated bound over 200 simulated data sets.

We begin by focusing on low uncertainty scenarios, corresponding to the interventions taking place over $M = 1$ or $3$ time periods with the distribution resembling the actual data generating mechanism.  We think that the intervention distribution resembles the data generating mechanism in scenarios where the intervention intensity is close to 5, which is the average number of treatment-active locations for the data generating process. In these scenarios, the asymptotic variance bound is distinctly higher than the true asymptotic variance, indicating that the inference based on the true asymptotic bound would be conservative. We find that in these low uncertainty scenarios, the estimated bound is close to the true bound. For that reason, we would expect the confidence intervals for the IPW estimator based on the estimated bound to have a higher coverage probability than its nominal coverage (see \cref{app_subsec:add_sims_iraq_coverage} for the coverage results).

In contrast, under high uncertainty scenarios such as the interventions over longer time periods, e.g., $M=7$, the asymptotic standard deviation and theoretical bound are essentially indistinguishable. However, under these scenarios, the estimate of the theoretical bound tends to be biased downwards, suggesting that the confidence intervals for the IPW estimator based on the estimated bound would be anti-conservative. Furthermore, we expect it to take a longer time series in order for the estimated bound to converge to its theoretical value when the intervention takes place over a longer time period.

\subsection{Coverage of the Confidence Intervals for the IPW and H\'ajek Estimators}
\label{app_subsec:add_sims_iraq_coverage}

\paragraph{IPW estimator.}
The results in \cref{fig:sims_iraq_var_y_int1_B1} indicate that the coverage of confidence intervals based on the asymptotic variance bound should be similar to those based on the true variance under high uncertainty scenarios, while they should be slightly higher under low uncertainty scenarios. Furthermore, confidence intervals based on the estimated variance bound should yield coverage probability close to (lower than) the coverage achieved using the theoretical bound under low (high) uncertainty scenarios.

\begin{figure}[!t]
\centering
\includegraphics[width = 0.88 \textwidth]{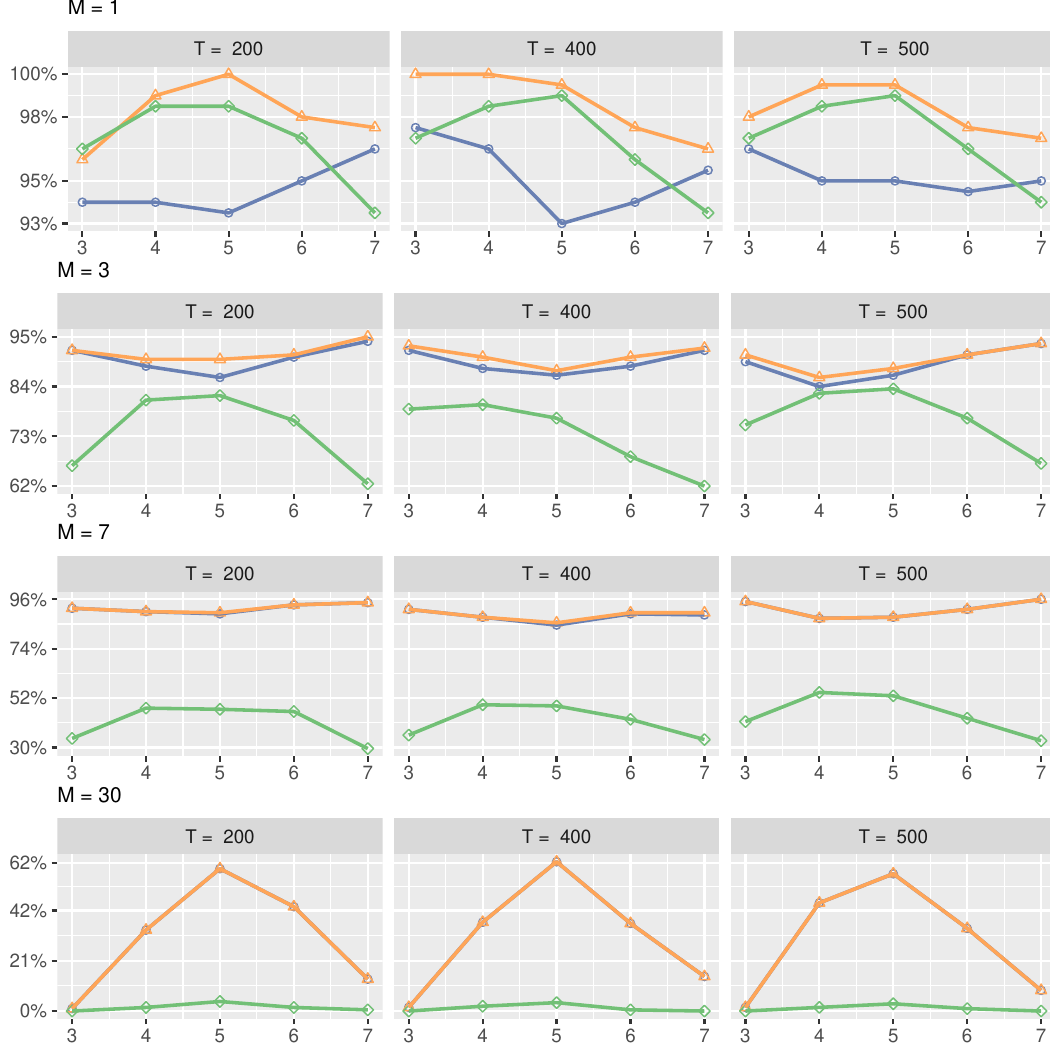}\\
\includegraphics[width = 0.8\textwidth]{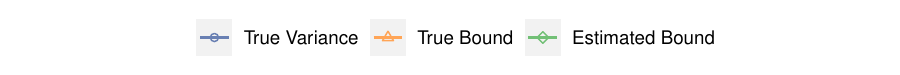}
\caption{Coverage of the IPW Estimator 95\% Confidence Intervals. This figure shows the coverage of 95\% confidence intervals for the average potential outcome over $B = \Omega$ based on the IPW estimator using the true variance (blue lines open circles), the true bound (orange lines with triangles), and the estimated bound (green lines with rhombuses), for interventions taking place over $M \in \{1, 3, 7, 30\}$ time periods (rows) and increasing length of the observed time series (columns).}
\label{app_fig:sims_iraq_ipw_cover}
\end{figure}

These expectations are indeed reflected in the coverage results shown in \cref{app_fig:sims_iraq_ipw_cover}. Except when $M=30$, the confidence interval for the IPW estimator based on either the {\it true} asymptotic variance or the {\it true} variance bound has a coverage of about 80\% or higher. However, when $M=30$, the confidence intervals based on the true asymptotic variance have a coverage below 60\% or less, indicating that for interventions taking place over longer time periods, more data are needed to make use of the asymptotic approximation. However, these results are based on the true variance and variance bound, and instead inference would be based on the estimated variance bound. The under-estimation of the variance bound in high uncertainty scenarios found in \cref{app_fig:sims_iraq_var_y_other_B1} leads to the under-coverage of the confidence intervals based on the IPW estimator when using the estimated variance bound, especially when the interventions take place over long time periods.

\paragraph{H\'ajek estimator.}

Motivated by the good performance of the H\'ajek estimator shown in \cref{fig:5res3_estimates}, we also investigate the coverage probability of the 95\% confidence interval as described in Appendix~\ref{app_sec:hajek}. The rows of \cref{app_fig:sims_iraq_hajek_cover} show the coverage results for increasingly small regions, whereas the columns show the results for increasingly long observed time series ($T = 200, 400, 500$). Different colors correspond to the coverage results under interventions taking place over $M = 1$ (black), $3$ (green), $7$ (red), and $30$ (blue) time periods. We find that the coverage is above 90\% for all combinations of $T$ and $\lag$ for the two largest regions, even when an intervention takes place over 30 time periods. We find that the coverage is lower for the smallest region.

\begin{figure}[!t]
\centering
\includegraphics[width = \textwidth]{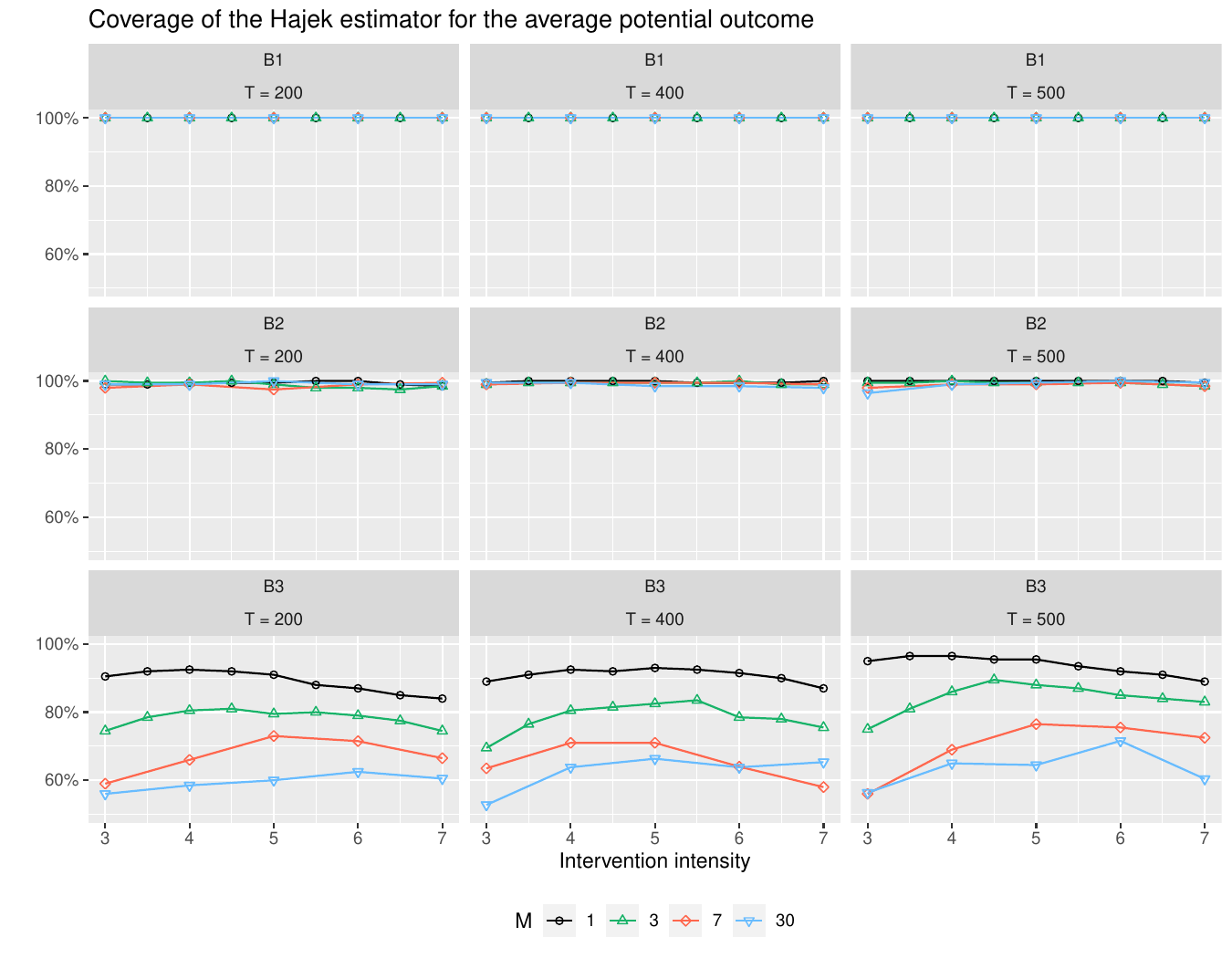}
\caption{Coverage of the H\'ajek Estimator's 95\% Confidence Intervals for the Average Potential Outcomes under Various Interventions.  We vary the intervention intensity $\trtintensity$ (horizontal axis), and the length of intervention $M = 1, 3, 7, 30$ (different lines). Each row represents the coverage for different regions of interest, i.e., $B_1 = [0,1]^2$, $B_2 = [0, 0.5]^2$ and $B_3=[0.75, 1]^2$, whereas each column represents the length of time series, i.e., $T = 200, 400$ and $500$.}
\label{app_fig:sims_iraq_hajek_cover}
\end{figure}

\subsection{Uncertainty Estimates}
\label{app_subsec:add_sims_iraq_uncertainty}

Here, we show that the estimated standard deviation for the H\'ajek estimator outperforms that for the IPW estimator under many simulation scenarios.

\begin{figure}[!p]
\centering
\includegraphics[width = 0.85 \textwidth]{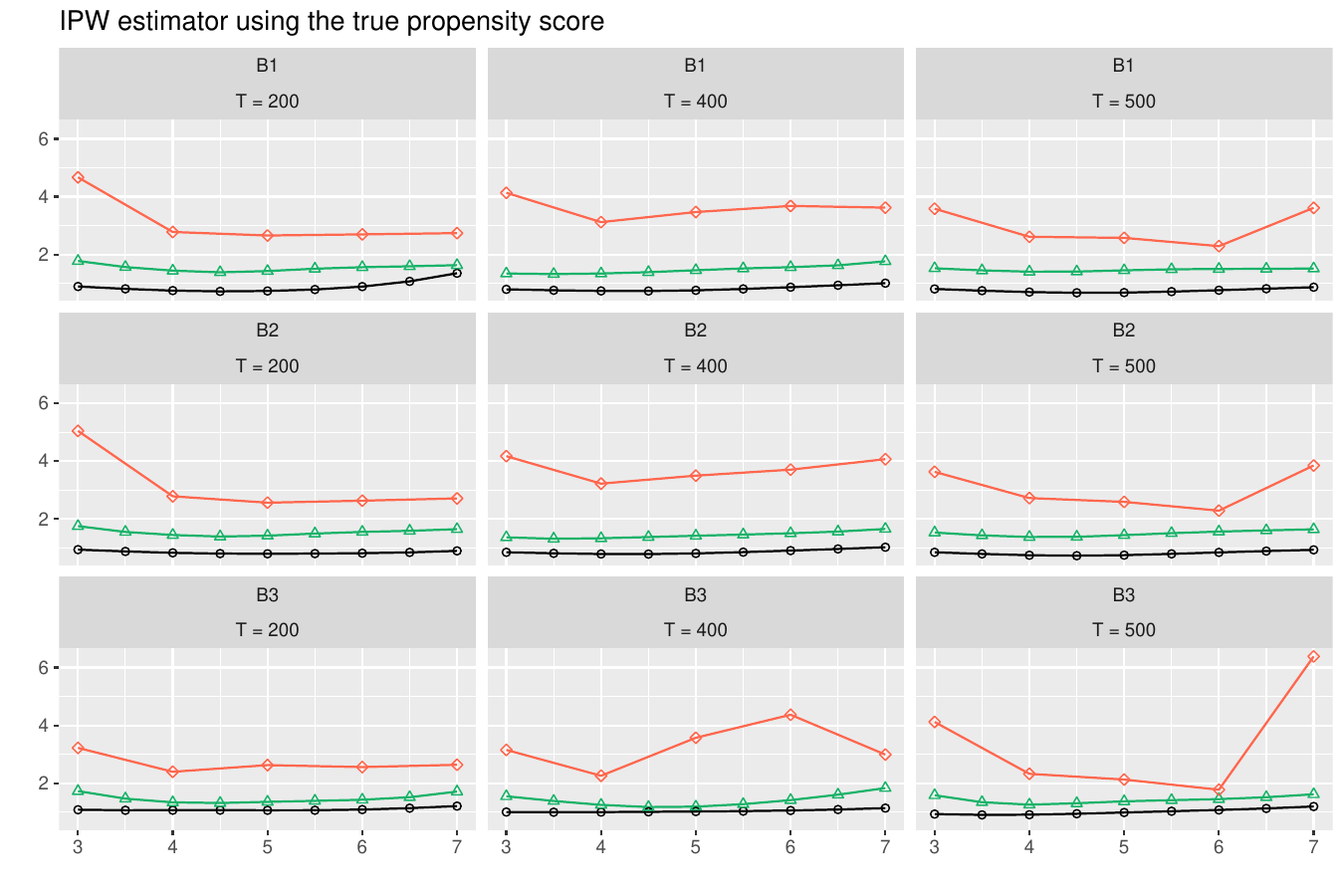} \\[-10pt]
\includegraphics[width = 0.85 \textwidth]{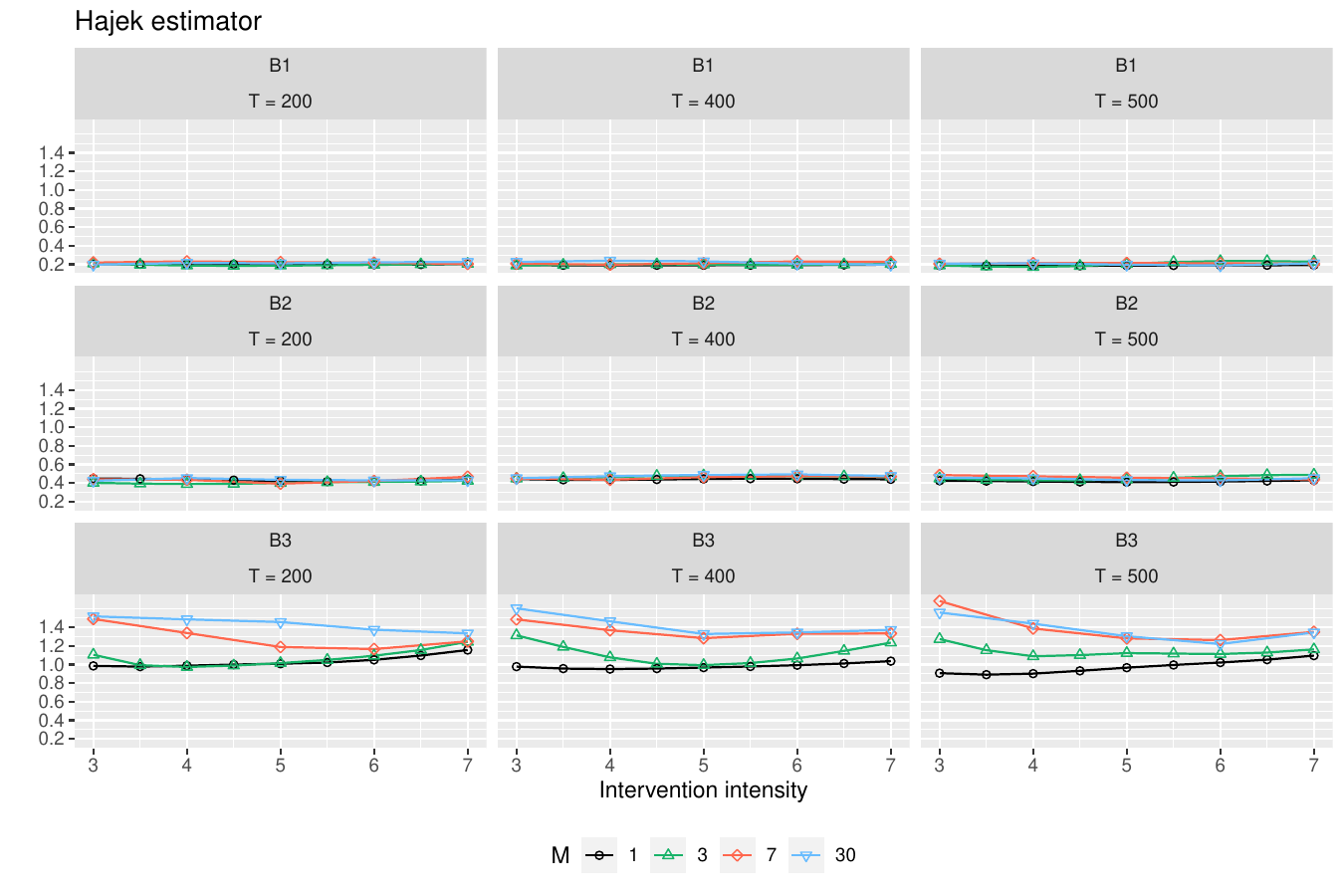}
\caption{\small Comparison of the Estimated and True Uncertainty for the Inverse Probability of Treatment and H\'ajek Estimators. Each plot presents the ratios between the standard deviation of each estimator and the mean estimated standard deviation across simulated data sets.  A value smaller (greater) than 1 implies overestimation (underestimation) of uncertainty. The top (bottom) panel presents the results for the IPW (H\'ajek) estimator with the varying intensity under the intervention (horizontal axis) and for the whole country $B_1$ (first and forth row) and two sub-regions, $B_2$ (second and fifth row) and $B_3$ (third and sixth row).  We also vary the length of intervention, $M = 1, 3, 7$ and $30$ time periods (black, green, red, and blue lines, respectively). The columns correspond to different lengths of the time series $T = 200, 400$ and $500$.}
\label{app_fig:iraq_MCsd_meansd}
\end{figure}

We compute the standard deviation of the estimated average potential outcome across simulated data sets and compare it with the mean of the standard deviations, each of which is used to create the confidence intervals.  The similarity of these two quantities implies the accuracy of our uncertainty estimates. \cref{app_fig:iraq_MCsd_meansd} presents the results as the ratio of these two quantities. A value below (above) 1 indicates that the true variability in our point estimates is smaller (greater) than our uncertainty estimate.

While the ratios are always below 1 for the H\'ajek estimator for the two largest regions $B_1$ and $B_2$, they are almost always above 1 for the IPW estimator (top panel). This is consistent with the above results, showing that we tend to overestimate (underestimate) the uncertainty for the H\'ajek (IPW) estimator.  We find that the confidence interval for the H\'ajek estimator tends to be most conservative when $M$ is small and the region of interest is large.  For the IPW estimator, the degree of uncertainty underestimation decreases as the length of time series $T$ increases but increases as the length of intervention $M$ increases. In fact, when $M=30$, some of the ratios are as large as 20 (hence they are not included in the figure).  The results suggest that in practice the H\'ajek estimator should be preferred over the IPW estimator especially for stochastic interventions over a long time period.

\subsection{Covariate Balance}
\label{app_subsec:add_sims_iraq_balance}

\begin{figure}[!t]
\centering
\includegraphics[width = 0.4 \textwidth]{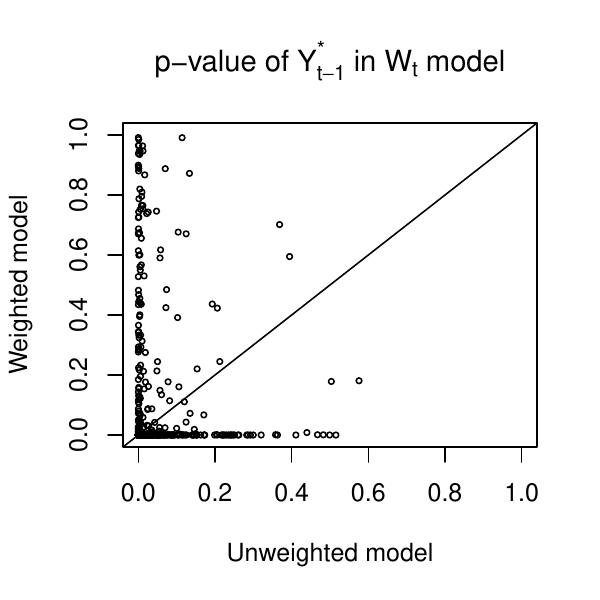}
\caption{Balance of the Previous Outcome-Active Locations in Treatment Model. Each point shows the relative magnitude of the p-value for the previous outcome-active locations in the unweighted propensity score model (horizontal axis) over that of the model weighted by the inverse of the estimated propensity score (vertical axis).}
\label{app_fig:iraq_sims_balance}
\end{figure}

We evaluate the balance of covariates based on the estimated propensity score by comparing their p-values in the propensity score model, and in a model with functional form as in the propensity score model but weighted by the inverse of the estimated propensity score. The left plot of \cref{app_fig:iraq_sims_balance} shows the p-value for the previous outcome-active locations, which are one of the time-varying confounders, across 200 simulated data sets. Evidently, the p-values in the unweighted model are close to 0, indicating that previous outcome-active locations form an important predictor of the treatment assignment. However, in the weighted model, the p-values of the same confounder are more evenly distributed across the $(0, 1)$ range, indicating that this confounder is better balanced in the weighted time series.

\section{Additional simulations on a square geometry}
\label{app_sec:add_square_sims}

\subsection{The Simulation Design}

We also consider a time series of point patterns of length $T \in \{200, 400, 500\}$ on the unit square, $\Omega = [0,1]\times[0,1]$. For each time series length $T$, 200 data sets are generated with the following design.

\subsubsection*{Time-varying and time-invariant confounders.} 
\begin{figure}[!t]
\centering
\subfloat[Time-invariant confounder $\onecov^1(\omega)$]{\includegraphics[width = 0.45\textwidth]{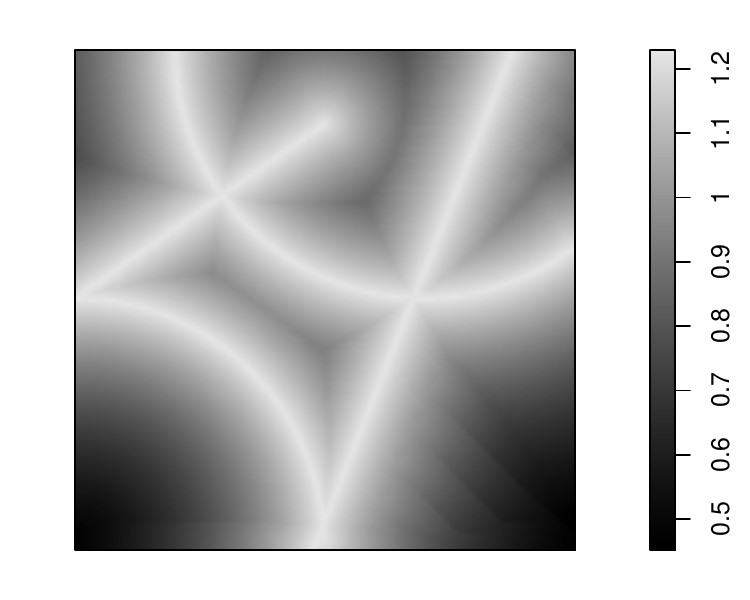} \label{fig:sims_roads}}
\hspace{.2in}\subfloat[Realization of time-varying confounder $\onecov_t^3(\omega)$]{ \includegraphics[width = 0.45\textwidth]{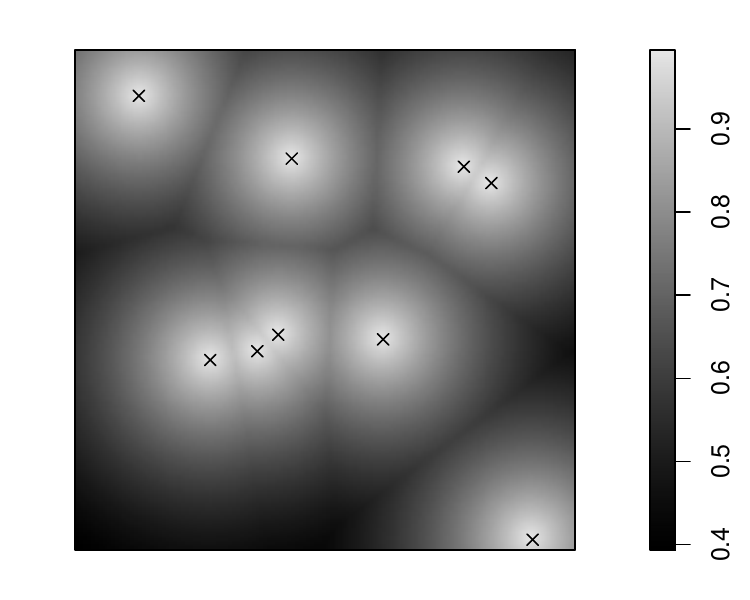} \label{fig:sims_Xt_one}}
\caption{Simulated Confounders. Panel (a) shows one of the two time-invariant confounders representing the exponential decay of distance to the road network.
Panel (b) shows one realization for one of the time-varying confounders. Points $\times$ are generated from a non-homogeneous Poisson process depending on the road network in (a). Then, the time-varying confounder is defined as the exponential decay of distance to the points $\times$.}
\end{figure}

Our simulation study includes two time-invariant and two time-varying confounders. For the first time-invariant confounder, we construct a hypothetical road network on $\Omega$ using lines and arcs, which is highlighted by bright white lines in \cref{fig:sims_roads}.  Then, we define $\onecov^1(\omega)=1.2\exp\{-2D_1(\omega)\}$ where $D_1(\omega)$ is the distance from $\omega$ to the closest line.  The second time-invariant covariate is constructed similarly, as $\onecov^2(\omega)=\exp\{-3D_2(\omega)\}$ where $D_2(\omega)$ is the distance to the closest arc. In addition, the time-varying confounders, $\onecov_t^3(\omega)$ and $\onecov_t^4(\omega)$, are defined based on the exponential decay of distance to the closest point; these points are generated according to a non-homogeneous Poisson point processes with the following intensity function
$$
\lambda_t^{X^j}(\omega) = \exp \big\{ \rho_0^j + \rho_1^j X^1(\omega) \big\}, \ j =3,4,
$$
where $\rho_1^3 = 1$, and $\rho_1^4 = 1.5$. \cref{fig:sims_Xt_one} shows one realization of $\onecov_t^3(\omega)$.

\subsubsection*{Spatio-temporal point processes for treatment and outcome variables.}

We again generate treatment and outcome point patterns from  non-homogeneous Poisson processes that depends on all confounders, and the previous treatment and outcome realizations. The functional specification of the Poisson process intensities is the same as in Section~\ref{sec:simulations}. The model gives rise to an average of 5 observed treatment-active locations and 21 observed outcome-active locations within each time period.

\subsubsection*{Stochastic interventions.}
We consider interventions of the form $\interv[F][][\lag]$ based on a homogeneous Poisson process with intensity $h$ that is constant over $\Omega$ and ranges from 3 to 7.  We consider various lengths of each intervention by setting $M \in  \{1, 3, 7, 30\}$.  The second intervention we consider is defined over the three time periods, i.e., $\interv[T]=\intervdist[F][3] \times \intervdist[F][2] \times \intervdist[F][1]$ with $\lag = 3$. The intervention for the first time period $\intervdist[F][3]$ is a homogeneous Poisson process with intensity $h_3$ ranging from 3 to 7, whereas $\intervdist[F][2] = \intervdist[F][1]$ is a homogeneous Poisson process with intensity equal to 5 everywhere over $\Omega$.  For each stochastic intervention, we consider the region of interest, denoted by set $B$, of three different sizes: $B = \Omega=[0, 1] \times [0,1]$, $B = [0, 0.5]\times[0,0.5]$, and $B = [0.75, 1]\times[0.75, 1]$.

\subsubsection*{Estimand and estimation.}

Approximating the true values of the estimands and estimation is performed as described in Section~\ref{sec:simulations}. In these simulations, for $T = 500$ (the longest time series in our simulation scenario) the spatial smoothing bandwidth is approximately equal to 0.16, smaller than the size of the smallest $B$ (which is equal to $[0.75,1]^2$).

\subsubsection*{Variance and its upper bound.}

We base calculation of the theoretical variance and the variance bound on Theorems~\ref{theorem:normality}~and~\ref{theorem:normality_tau}, and use Monte Carlo approximations to compute these, as in Section~\ref{sec:simulations}. We also use \cref{lemma:consistent_variance} We use \cref{lemma:consistent_variance} to estimate the variance bound.

\subsubsection*{Covariate balance.}

As in Section~\ref{sec:simulations}, we use weighted regression by the estimated propensity score to investigate covariate balance.

\subsection{Simulation Results}

\begin{figure}[p]
\centering
\includegraphics[width=0.98\textwidth]{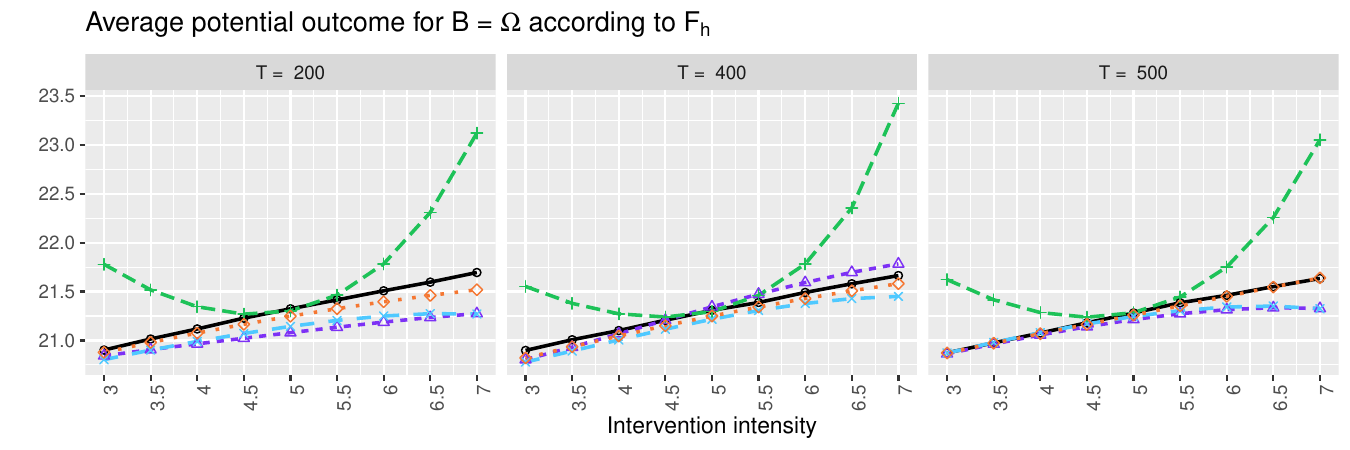}\\
\includegraphics[width=0.98\textwidth]{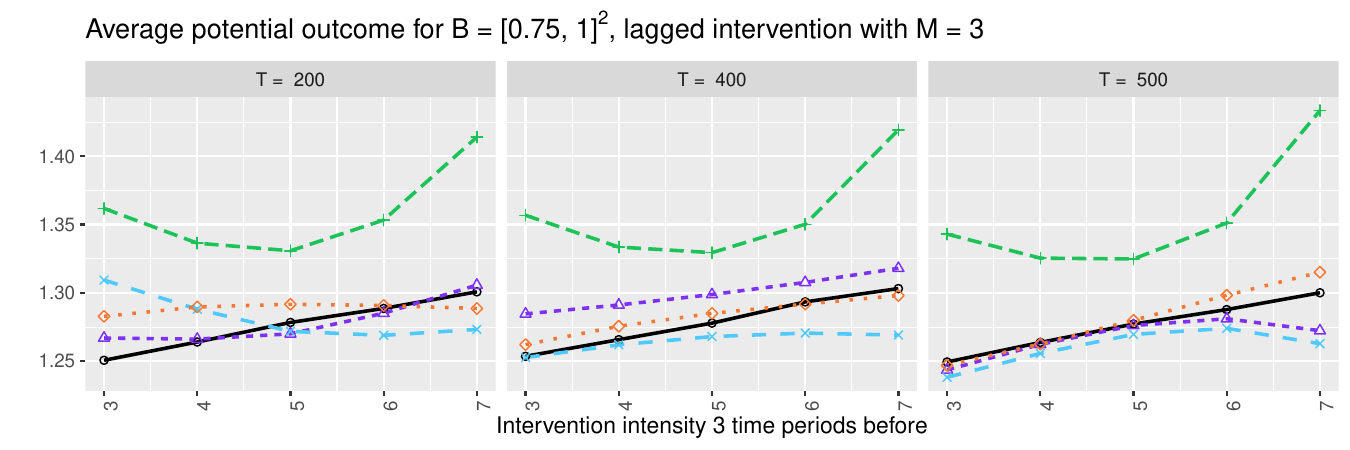}\\
\includegraphics[width=0.8\textwidth]{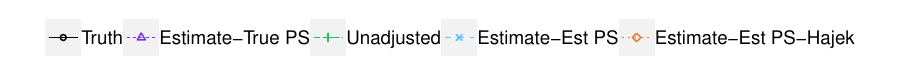}
\caption{Simulation Results for the True and Estimated Average Potential Outcomes. In the top panel, we present the true and estimated average potential outcomes in the entire region $B = \Omega$ under single-time interventions with the varying intensity (horizontal axis). In the bottom panel, we consider the average potential outcome in the sub-region $B = [0.75,1]^2$ for the intervention $\interv[T]$, with $M = 3$, the varying intensity of $\interv[F][3]$ (horizontal axis), and $\interv[F][1], \interv[F][2]$ intensity set to 5. The black lines with solid circles represent the truths, while the other dotted or dashed lines represent the estimates; the estimator based on the true propensity score (purple triangles), the unadjusted estimator (green crosses), the estimator based on the estimated propensity score (blue x's), the H\'ajek estimator based on the estimated propensity score (orange rhombuses).}
\label{fig:4res1_estimates}
\end{figure}

\begin{figure}[!p]
\centering
\includegraphics[width =0.93\textwidth]{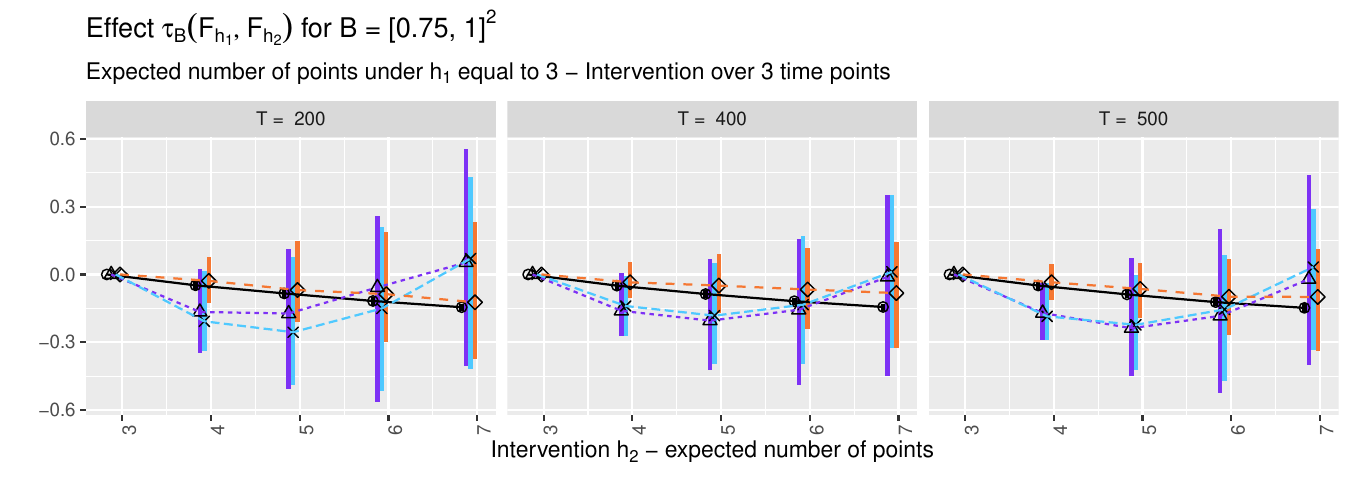}\\
\includegraphics[width =0.93\textwidth]{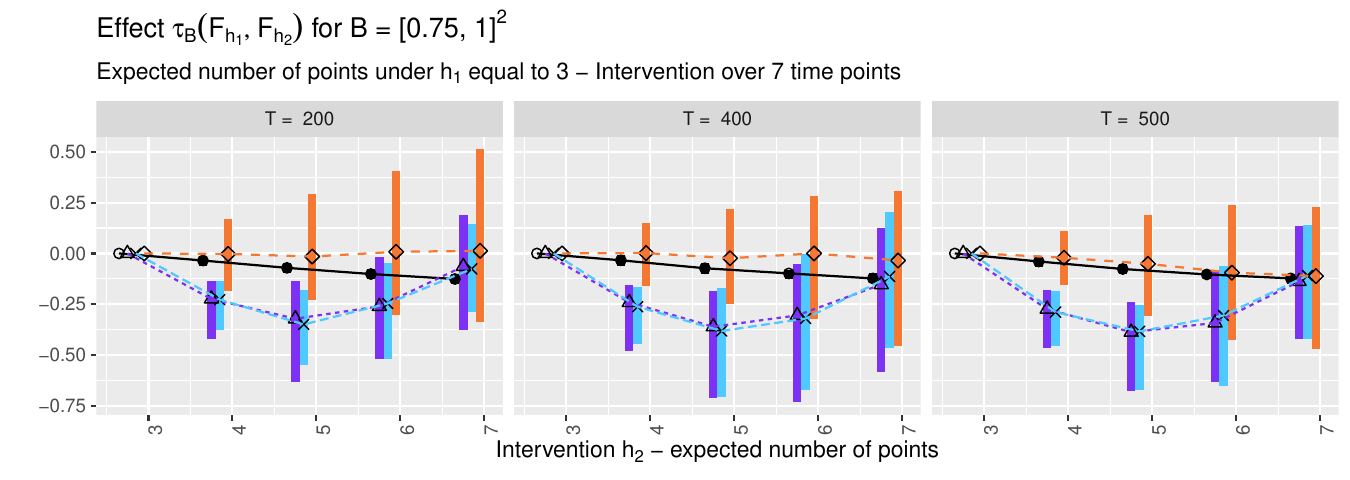}\\
\includegraphics[width =0.93\textwidth]{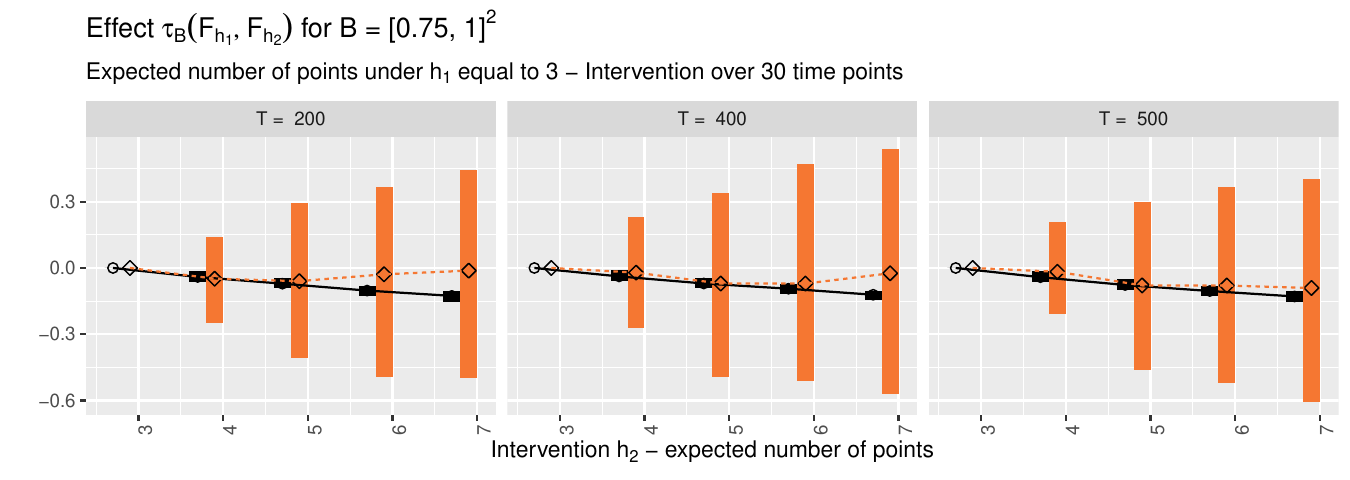}\\
\includegraphics[width=0.85\textwidth]{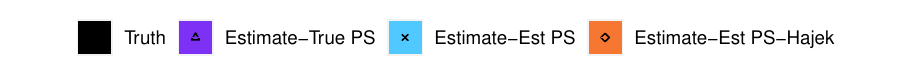}
\caption{Simulation Results for the Interventions of Increasing Time Lengths. Rows correspond to the interventions taking place over $M=3, 7$, and $30$ time periods. Columns correspond to the increasing length of the time series from 200 (left plots) to 500 (right plots). The vertical axis shows the change in the expected number of the outcome active locations over $[0.75, 1]^2$ for a change in the intervention intensity from 3 under $\trtintensity_1$ to the value shown in the horizontal axis under $\trtintensity_2$, for $M$ time periods. The points in the plot show the median estimate over 200 data sets, and the rectangles show the interquartile range of estimates. Only the H\'ajek estimates are shown for $M = 30$ as the extremely small weights arising from a large number of time periods make the estimates from the other estimators close to zero.}
\label{app_fig:res_sims_largeM}
\end{figure}

\paragraph{Estimation.}

Figures~\ref{fig:4res1_estimates}~and~\ref{app_fig:res_sims_largeM} present the results. In \cref{fig:4res1_estimates}, the top panel shows how the (true and estimated) average potential outcomes in the whole region ($B = \Omega$) change as the intensity varies under the single time period interventions. The bottom panel shows how the true and estimated average potential outcomes in the sub-region $[0.75, 1]^2$ change under the three time period interventions when the intensity at three time periods ago ranges from 3 to 7. For both simulation scenarios, we vary the length of the time series from 200 (left plots) to 500 (right plots).

As expected, the unadjusted estimates (green crosses) are far from the true average potential outcome (black solid circles) across all simulation scenarios.  In contrast, and consistent with the results of Theorems~\ref{theorem:normality}~and~\ref{theorem:normality_tau}, the accuracy of the proposed estimator (purple triangles based on the true propensity score, blue x's based on the estimated propensity score) improves as the number of time periods increases.  We note that the convergence is slower when $M=3$ than $M=1$.

\cref{app_fig:res_sims_largeM} shows the performance of the estimators for the interventions over many time periods. The plots show the estimated change in the number of outcome-active locations over the sub-region $B = [0.75, 1]$ for a change in the stochastic intervention from 3 per time period to the value on the horizontal axis. The rows correspond to the interventions over $M = 3 , 7$, and $30$ time periods, respectively, whereas the columns represent the different lengths of time series, i.e., $T = 200, 400$ and $500$. The results are shown for the IPW estimators based on the true propensity score (purple lines with open triangles) and the estimated propensity score (blue lines with x's) as well as the H\'ajek estimator based on the estimated propensity score (orange lines with open rhombuses). Only the H\'ajek estimates are shown for $M = 30$ as the extremely small weights arising from a large number of time periods make the estimates from the other estimators essentially equal to zero.  The lines and points in the plot show the median estimate and the rectangles show the interquartile range of estimates across 200 simulated data sets.

Again, as in the simulations of Section~\ref{sec:simulations}, we find that the H\'ajek estimator performs well across all simulation scenarios, whereas the IPW estimator tends to suffer from extreme weights.

\begin{figure}[p]
\centering
\includegraphics[width = 0.9 \textwidth]{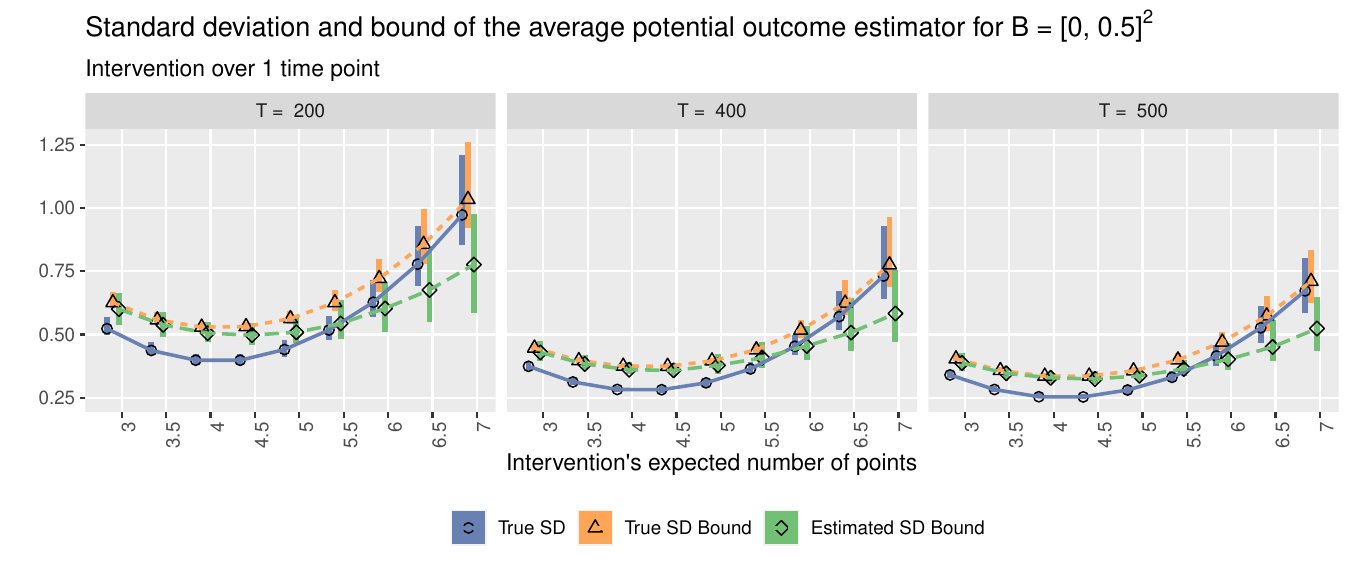} \\[-25pt]
\includegraphics[width = 0.9 \textwidth]{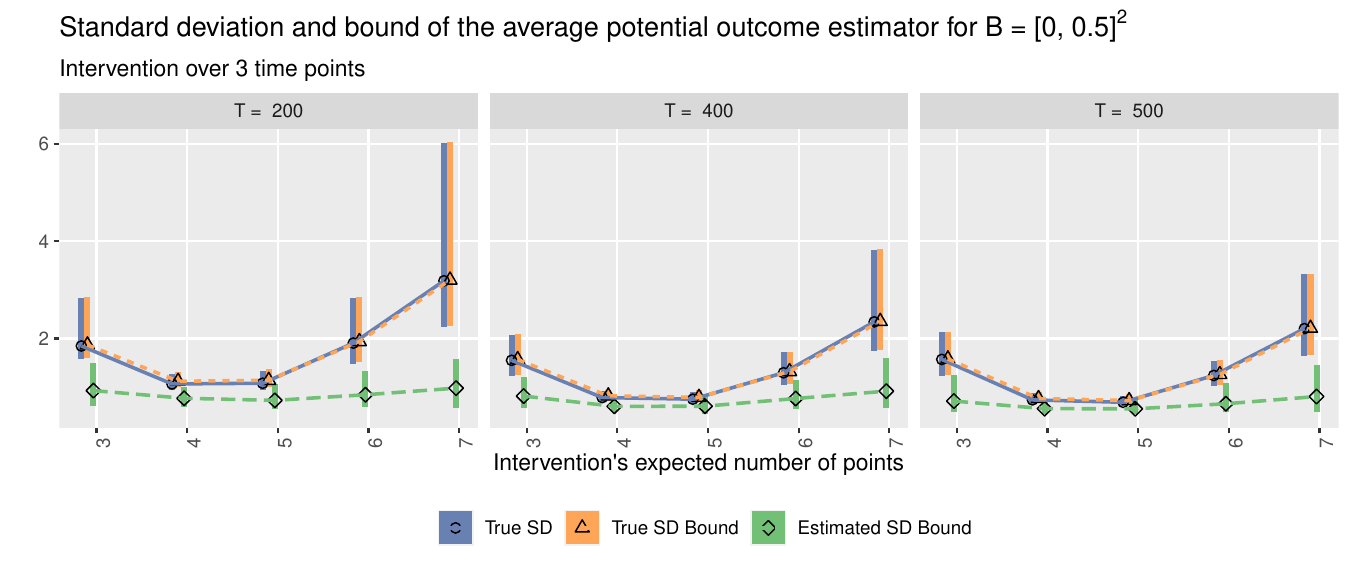} \\[-20pt]
\includegraphics[width = 0.9 \textwidth]{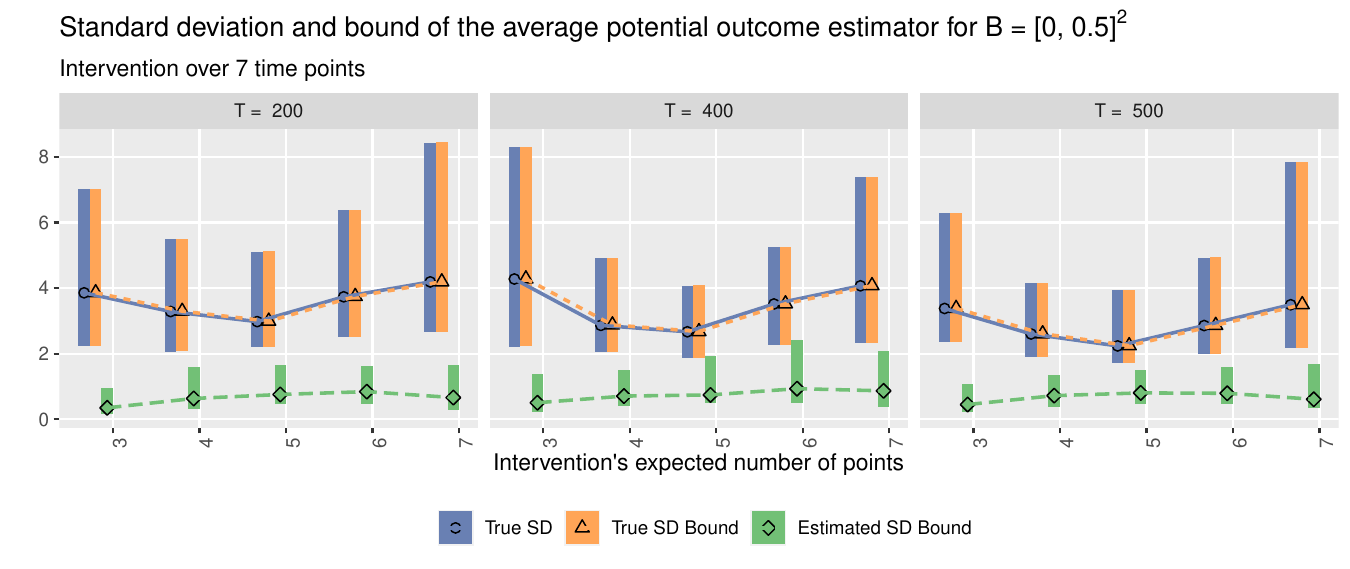}
\caption{Asymptotic Standard Deviation and Bound, and Estimated Bound. This figure shows the true asymptotic standard deviation (blue circles), the true asymptotic bound (orange triangles), and the estimated bound (green rhombuses) of the IPW estimator for the average potential outcome using the true propensity score, under interventions that take place over $M = 1, 3$ and $7$ time periods (rows), and for increasing length of the time series (columns). The horizontal axis shows the intensity of the intervention at each time period. The points show the median value, and the rectangles show the interquartile range over 200 simulated data sets.}
\label{app_fig:var_y_other_B2}
\end{figure}

\paragraph{The variance and its bound.}
Next, we compare the true theoretical variance, $\asymvar / T$, with the variance bound $\asymvar^* / T$ and its consistent estimator (see \cref{lemma:consistent_variance}). We again focus on the proposed estimators with the true propensity score. \cref{app_fig:var_y_other_B2} shows the results of an intervention $\interv[F][][M]$ for $\lag =1, 3$ and $7$, for region $B = [0, 0.5]^2$, and observed time series of length $T = 200, 400, 500$. These plots show the median and interquartile range of the asymptotic standard deviation, true bound, and estimated bound over 200 simulated data sets.

The conclusions are similar to the main manuscript.
As expected, the true variance decreases as the total number of time periods increases.
We start by focusing on low uncertainty scenarios, corresponding to the interventions taking place over $M = 1$ or $3$ time periods with the distribution resembling the actual data generating mechanism.  We think that the intervention distribution resembles the data generating mechanism in scenarios where the intervention intensity is close to 5, which is the average number of treatment-active locations for the data generating process. In these scenarios, the asymptotic variance bound is distinctly higher than the true asymptotic variance, indicating that the inference based on the true asymptotic bound would be conservative. We find that in these low uncertainty scenarios, the estimated bound is close to the true bound. For that reason, we would expect the confidence intervals for the IPW estimator based on the estimated bound to have a higher coverage probability than its nominal coverage.

In contrast, under high uncertainty scenarios such as the interventions over longer time periods, e.g., $M=7$, the asymptotic standard deviation and theoretical bound are essentially indistinguishable. However, under these scenarios, the estimate of the theoretical bound tends to be biased downwards, suggesting that the confidence intervals for the IPW estimator based on the estimated bound would be anti-conservative. As the length of time series increases, the estimated variance bound more closely approximates its theoretical value (consistent with \cref{lemma:consistent_variance}), but we expect it to take a longer time series in order for the estimated bound to converge to its theoretical value when the intervention takes place over a longer time period.

\paragraph{Coverage.}

These results on the asymptotic variance and variance bound lead to similar conclusions with respect to the coverage of 95\% confidence intervals of the IPW estimator.

\begin{figure}[!t]
\centering
\includegraphics[width = 0.88 \textwidth]{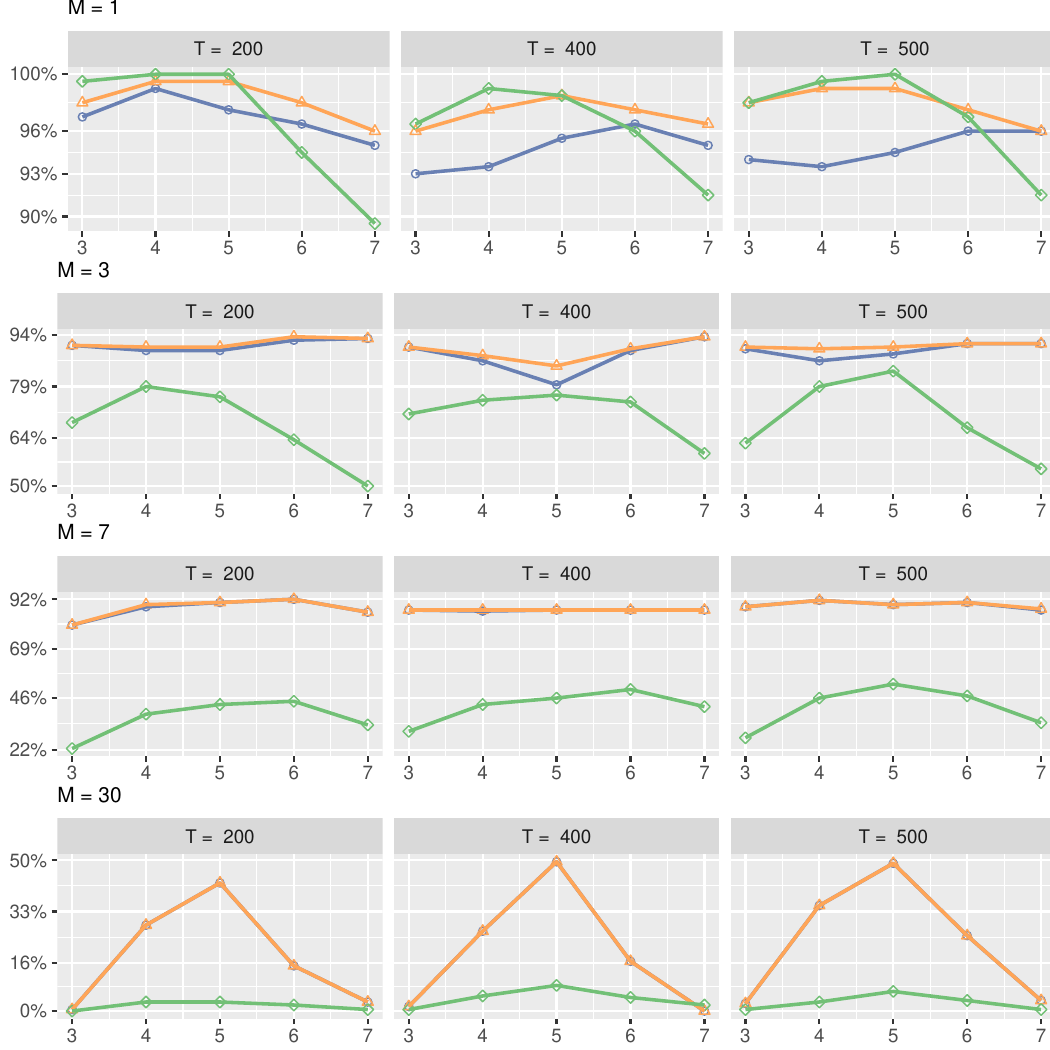}\\
\includegraphics[width = 0.8\textwidth]{4res1_legend3.pdf}
\caption{Coverage of the IPW Estimator 95\% Confidence Intervals. This figure shows the coverage of 95\% confidence intervals for the average potential outcome over $B = \Omega$ based on the IPW estimator using the true variance (blue lines open circles), the true bound (orange lines with triangles), and the estimated bound (green lines with rhombuses), for interventions taking place over $M \in \{1, 3, 7, 30\}$ time periods (rows) and increasing length of the observed time series (columns).}
\label{app_fig:ipw_cover}
\end{figure}

The coverage results are shown in \cref{app_fig:ipw_cover}. We find that, except when $M=30$, the confidence interval for the IPW estimator based on either the true asymptotic variance or the true variance bound has a coverage of about 80\% or higher.  This implies that the asymptotic normality established in \cref{theorem:normality} provides an adequate approximation to the estimator's sampling distribution for small or moderate values of $\lag$.  However, for $M=30$, the confidence interval for the IPW estimator is anti-conservative due to the fact that the weights, which equal the product of ratios across many time periods, become extremely small.  In addition, the underestimation of the variance bound in high uncertainty scenarios found in \cref{app_fig:var_y_other_B2} leads to the under-coverage of the confidence intervals based on the IPW estimator and using the estimated variance bound, especially when the interventions take place over long time periods.

\begin{figure}[!t]
\centering
\includegraphics[width = \textwidth]{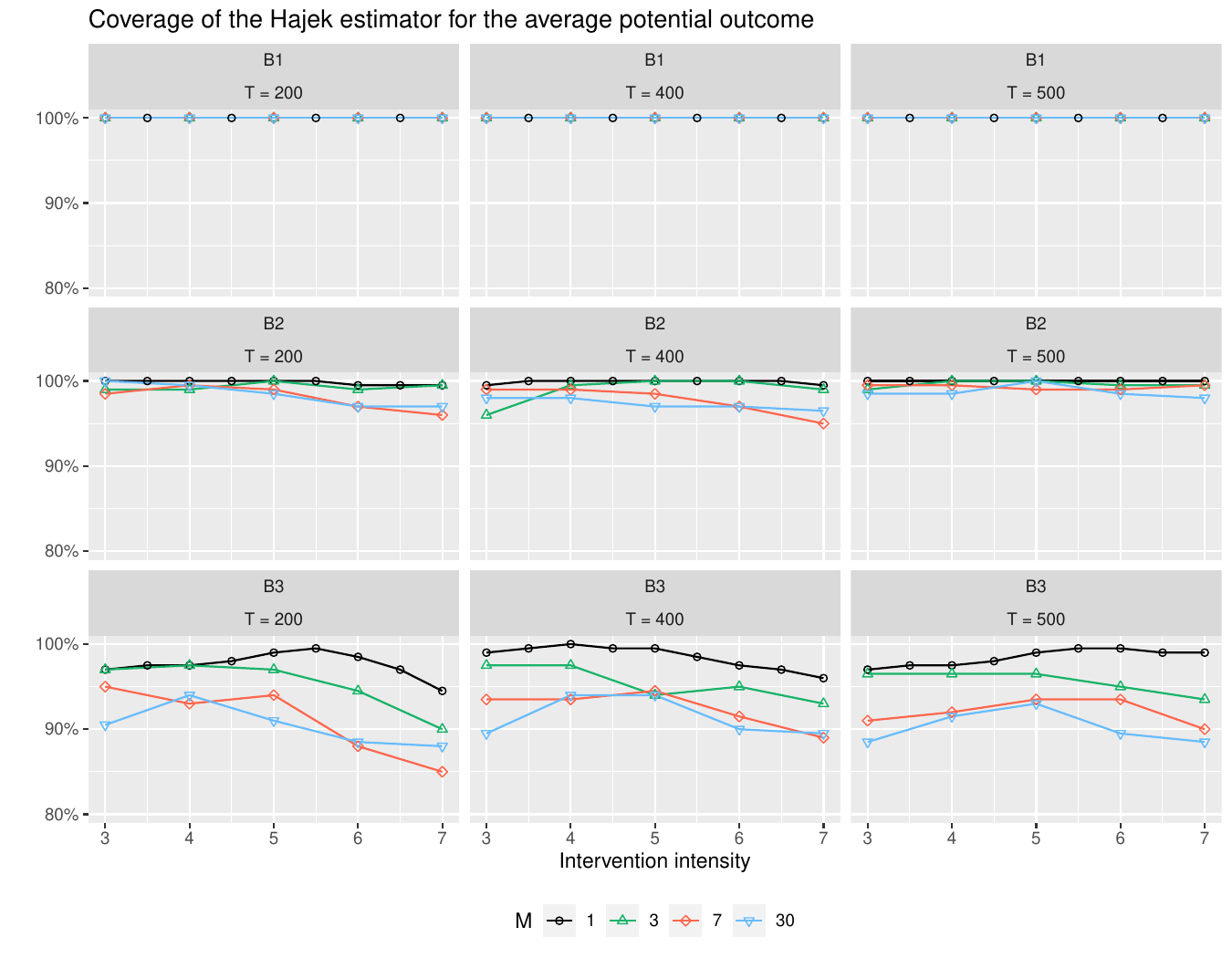}
\caption{Coverage of the H\'ajek Estimator's 95\% Confidence Intervals for the Average Potential Outcomes under Various Interventions.  We vary the intervention intensity $\trtintensity$ (horizontal axis), and the length of intervention $M = 1, 3, 7, 30$ (different lines). Each row represents the coverage for different regions of interest, i.e., $B_1 = [0,1]^2$, $B_2 = [0, 0.5]^2$ and $B_3=[0.75, 1]^2$, whereas each column represents the length of time series, i.e., $T = 200, 400$ and $500$.}
\label{app_fig:hajek_cover}
\end{figure}

We also investigate the coverage probability of the 95\% confidence interval for the H\'ajek estimator. The rows of \cref{app_fig:hajek_cover} show the coverage results for increasingly small regions, i.e., $B_1=[0,1]^2, B_2=[0, 0.5]^2$, and $B_3=[0.75, 1]^2$, whereas the columns show the results for increasingly long observed time series ($T = 200, 400, 500$). Different colors correspond to the coverage results under interventions taking place over $M = 1$ (black), $3$ (green), $7$ (red), and $30$ (blue) time periods. We find that the coverage is above 85\% for all cases, even when an intervention takes place over 30 time periods. As expected, the coverage is higher for smaller values of $M$, since these correspond to lower-uncertainty situations. We also find that the coverage is lower for smaller regions.

\paragraph{Comparison of Monte Carlo and estimated variance}

\begin{figure}[!p]
\centering
\includegraphics[width = 0.85 \textwidth]{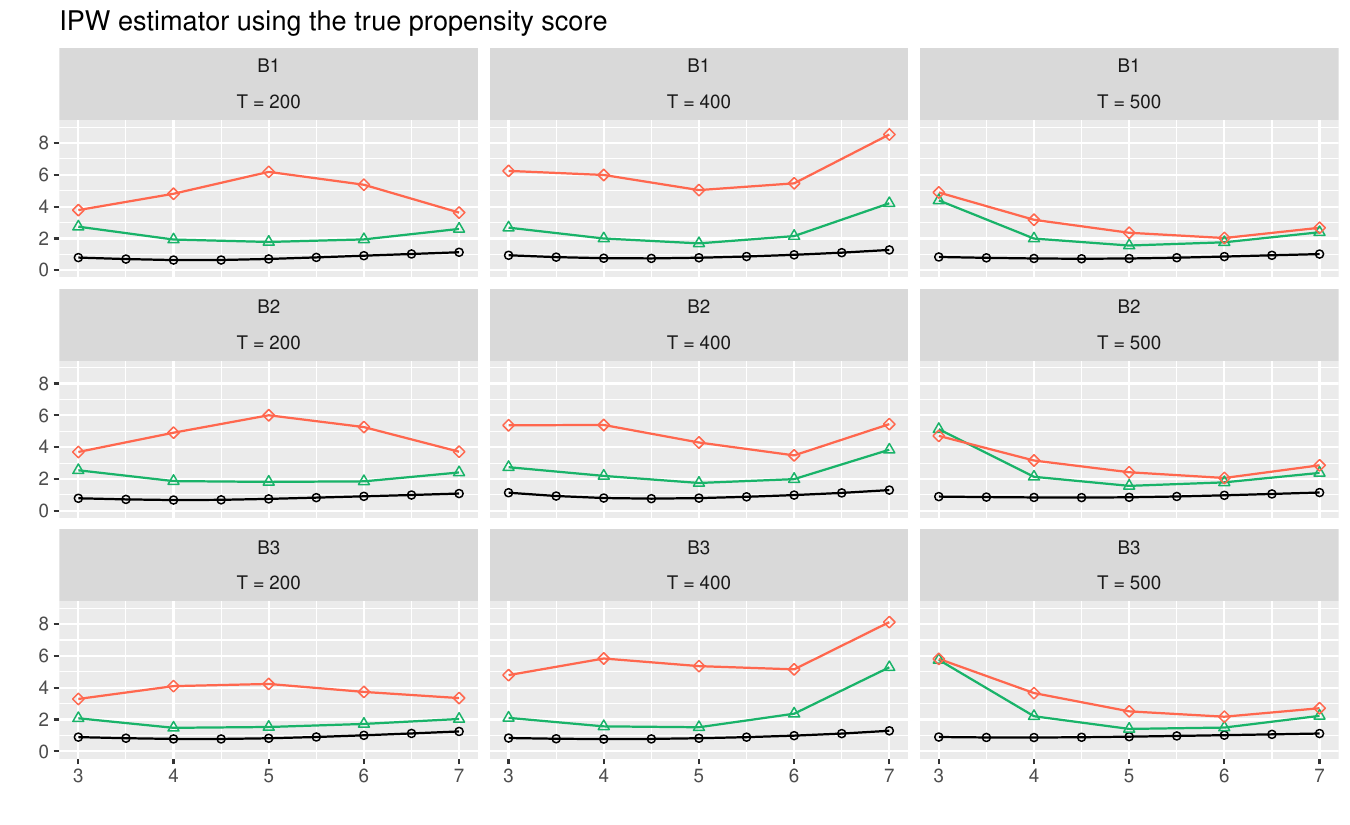} \\[-10pt]
\includegraphics[width = 0.85 \textwidth]{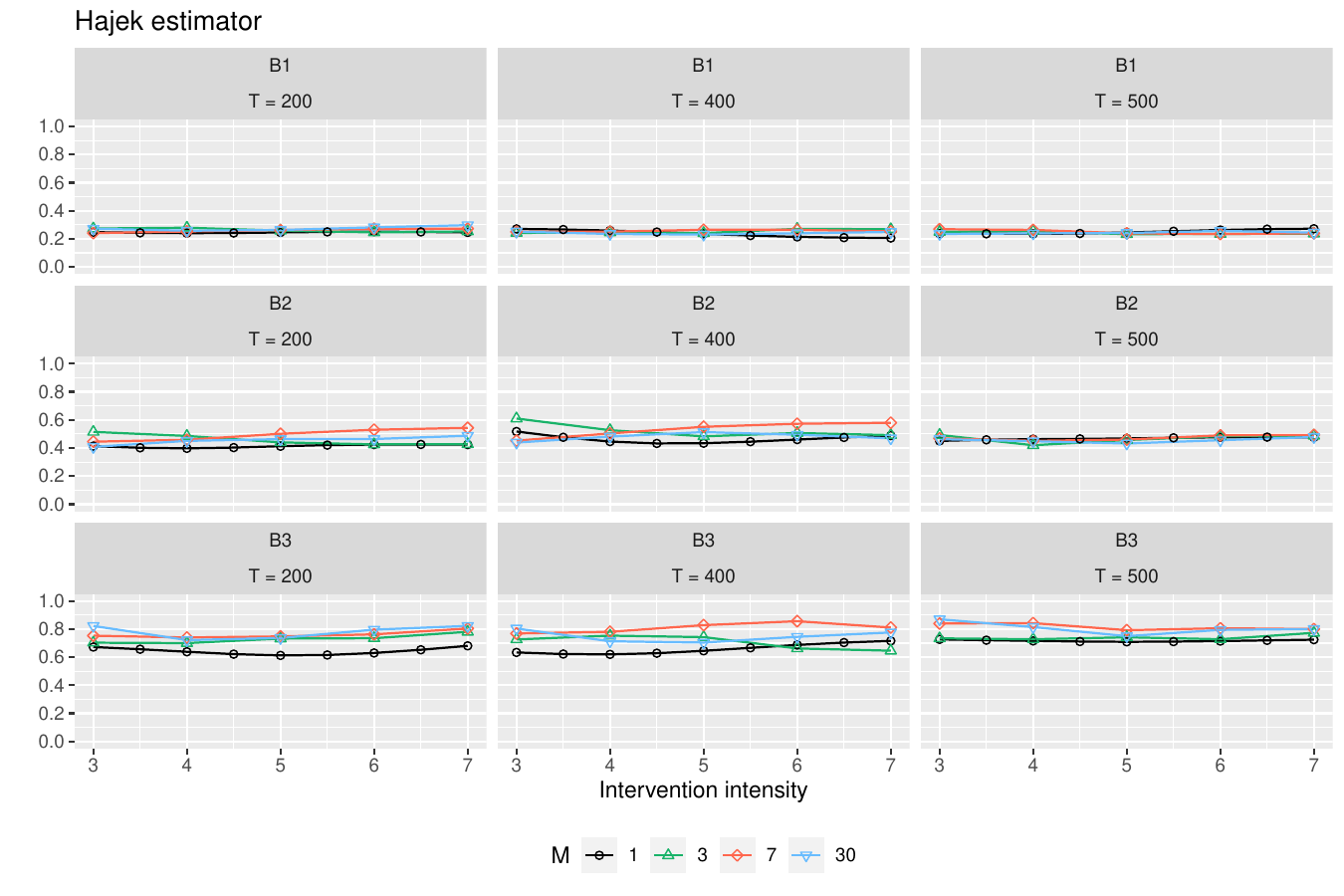}
\caption{\small Comparison of the Estimated and True Uncertainty for the Inverse Probability of Treatment and H\'ajek Estimators. Each plot presents the ratios between the standard deviation of each estimator and the mean estimated standard deviation across simulated data sets.  A value smaller (greater) than 1 implies overestimation (underestimation) of uncertainty. The top (bottom) panel presents the results for the IPW (H\'ajek) estimator with the varying intensity under the intervention (horizontal axis) and for the whole region $B_1$ (first and forth row) and two sub-regions, $B_2 = [0, 0.5]^2$ (second and fifth row) and $B_3 = [0.75, 1]^2$ (third and sixth row).  We also vary the length of intervention, $M = 1, 3, 7$ and $30$ time periods (black, green, red, and blue lines, respectively). The columns correspond to different lengths of the time series $T = 200, 400$ and $500$.}
\label{app_fig:MCsd_meansd}
\end{figure}

We find that the confidence interval for the H\'ajek estimator has a better coverage probability even for the interventions over long time periods. Here, we show that the estimated standard deviation for the H\'ajek estimator outperforms that for the IPW estimator under many simulation scenarios.

\cref{app_fig:MCsd_meansd} shows the ratio of the standard deviation of the estimated average potential outcome across simulated data sets over the mean of the standard deviations. A value below (above) 1 indicates that the true variability in our point estimates is smaller (greater) than our uncertainty estimate. While the ratios are always below 1 for the H\'ajek estimator (bottom panel), they are almost always above 1 for the IPW estimator (top panel).  This shows that we tend to overestimate (underestimate) the uncertainty for the H\'ajek (IPW) estimator. Further, we find that the confidence interval for the H\'ajek estimator tends to be most conservative when $M$ is small and the region of interest is large.  For the IPW estimator, the degree of underestimation decreases as the length of time series $T$ increases but increases as the length of intervention $M$ increases. In fact, when $M=30$, some of the ratios are as large as 20 (hence they are not included in the figure). The results suggest that in practice the H\'ajek estimator should be preferred over the IPW estimator especially for stochastic interventions over a long time period.

\paragraph{Balance.}

\begin{figure}[!t]
\centering
\includegraphics[width = 0.4 \textwidth,trim=0 0 300 0, clip]{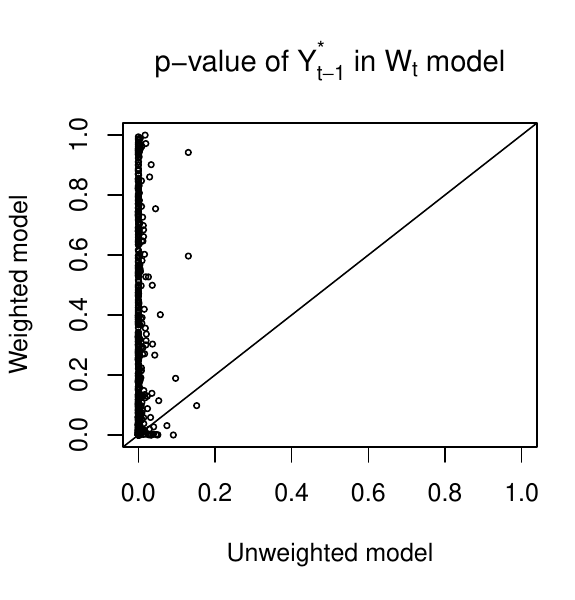}
\caption{Balance of the Previous Outcome-Active Locations in Treatment Model. In the left plot, each point shows the relative magnitude of the p-value for the previous outcome-active locations in the unweighted propensity score model (horizontal axis) over that of the model weighted by the inverse of the estimated propensity score (vertical axis). The right plot shows the distribution of the estimated coefficient of the previous outcome-active locations in the weighted propensity score model.}
\label{app_fig:sims_balance}
\end{figure}

As in the simulations in the main manuscript, we find that the p-values of one of the confounders ($Y_{t - 1}^*$ in Equation\cref{eq:sims_iraq_treatment_intensity}) are substantially greater in the weighted propensity score model than in the unweighted model, where the weights are given by the inverse of the estimated propensity score (shown in \cref{app_fig:sims_balance}).

\section{Additional Empirical Results}
\label{app_sec:study}

\subsection{Visualization}

\begin{figure}[!t]
\centering
\includegraphics[width=0.95\textwidth]{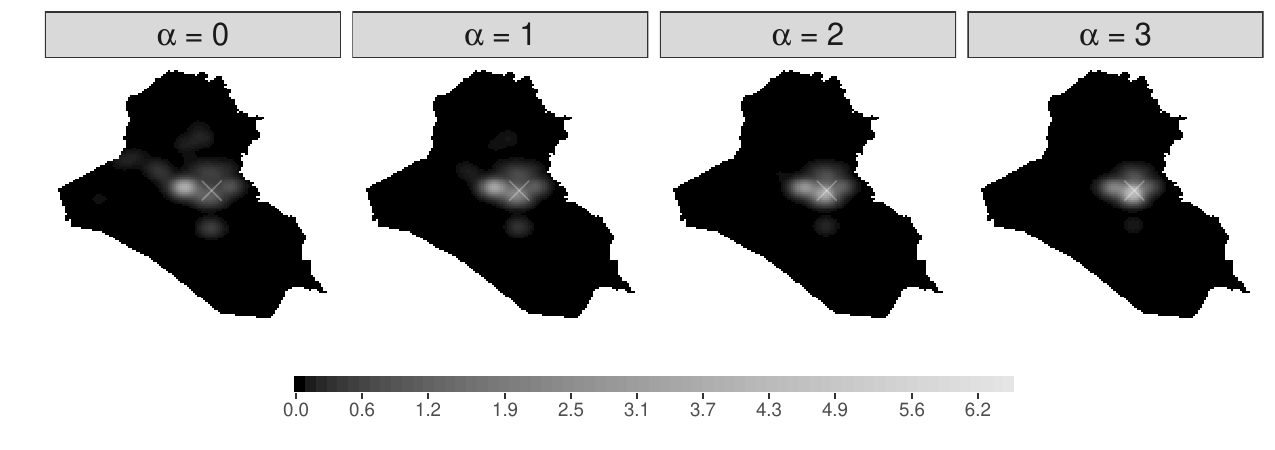}
\caption{Visualization of Intensity under Stochastic Interventions whose Focal Point is the Center of Baghdad.  Across plots, we vary the degree to which the airstrikes are concentrated around the focal point using the precision parameter, while the expected number of airstrikes is held constant at 3 per day.}
\label{fig:type2_interv}
\end{figure}

As discussed in Section~\ref{subsec:study_interventions}, we consider a stochastic intervention whose focal point is the center of Baghdad.  The degree of concentration is controlled by the precision parameter $\alpha$ whose greater value, implying that more airstrikes are occurring near the focal point.  We vary the value of $\alpha$ from 0 to 3, while keeping the expected number of airstrikes constant at 3 per day.  \cref{fig:type2_interv} illustrates intensities for the different values of $\alpha$. The first plot in the figure does not focus on Baghdad at all, representing the baseline spatial distribution $\phi_0$. As the value of $\alpha$ increases, the spatial distribution of airstrikes becomes concentrated more towards the center of Baghdad.

\subsection{Empirical Results}
\label{app_tab:res_tau}

\begin{table}[!t]
\centering
\spacingset{1}
{\small
\begin{tabular}{ccc|| rrr}
Type ($\interv[T]['], \interv[T]['']$) & $\lag$ & Outcome  & Iraq & Baghdad & Outside Baghdad \\ \hline

& 3 & IED & -1.3 (-8.2, 5.7) & -0.2 (-2.2, 1.8) & -1.1 (-6.2, 4) \\
& & SAF & -1.9 (-8.9, 5.2) & -1.2 (-3.8, 1.4) & -0.7 (-5.2, 3.9) \\
& & Other Attack & -1.8 (-19.9, 16.4) & 0.1 (-6.2, 6.5) & -1.9 (-13.9, 10) \\
Increasing the & 7 & IED & 5 (-2.6, 12.7) & 1.2 (-0.8, 3.3) & 3.8 (-1.9, 9.5) \\
intensity & & SAF & {\bf 10 (1.7, 18.2)} & {\bf 3.1 (0.4, 5.8)} & {\bf 6.9 (1.1, 12.6)} \\
(1, 3) & & Other Attack & 14 (-5.2, 33.3) & 5.6 (-0.4, 11.6) & 8.4 (-5.1, 21.9) \\
& 30 & IED & 11 (-1.1, 23) & 3.2 (-0.1, 6.5) & 7.8 (-1.1, 16.6) \\
& & SAF & {\bf 14.8 (2.9, 26.7)} & {\bf 5.9 (1.2, 10.5)} & {\bf 8.9 (1.5, 16.3)} \\
& & Other Attack & {\bf 33.1 (3.3, 62.9)} & {\bf 13.6 (2.7, 24.6)} & {\bf 19.5 (0.3, 38.6)} \\ \hline
& 3 & IED & 2.6 (-7.2, 12.5) & 0.6 (-2.4, 3.5) & 2.1 (-4.9, 9) \\
& & SAF & 2.6 (-6.7, 12) & 0.9 (-2.2, 4.1) & 1.7 (-4.8, 8.1) \\
& & Other Attack & 7.5 (-16, 31) & 3.8 (-4.3, 11.8) & 3.7 (-12, 19.4) \\
Changing the & 7 & IED & 2 (-6.9, 10.8) & 1.1 (-1.7, 3.9) & 0.9 (-5.2, 7) \\
focal points & & SAF & 0.2 (-9.7, 10.1) & 0.6 (-2.8, 3.9) & -0.4 (-7.1, 6.4) \\
(0, 3) & & Other Attack & 3.5 (-18.8, 25.8) & 1.7 (-6, 9.4) & 1.8 (-13, 16.7) \\
& 30 & IED & -1.2 (-15.9, 13.4) & -0.7 (-4.5, 3.1) & -0.5 (-11.6, 10.6) \\
& & SAF & 5.7 (-10, 21.4) & -1.3 (-6.4, 3.8) & 7 (-4.1, 18.1) \\
& & Other Attack & -3.5 (-37.8, 30.7) & -6.6 (-17.5, 4.2) & 3.1 (-21.1, 27.3) \\
\hline

& 3 & IED & -2.3 (-10.1, 5.5) & -0.6 (-2.7, 1.5) & -1.7 (-7.5, 4.1) \\
& & SAF & -1 (-9.9, 8) & -0.7 (-4.5, 3) & -0.2 (-5.5, 5) \\
& & Other Attack & -3.9 (-23.6, 15.8) & -1.2 (-8.2, 5.9) & -2.8 (-15.5, 10) \\
Lagged & 7 & IED & 6.8 (-0.7, 14.3) & 2.2 (-0.2, 4.6) & 4.6 (-0.6, 9.8) \\
effects & & SAF & {\bf 9.4 (1.6, 17.2)} & {\bf 3.6 (1, 6.2)} & {\bf 5.8 (0.4, 11.2)} \\
(1, 5) & & Other Attack & {\bf 20.9 (2.3, 39.4)} & {\bf 8.2 (1.8, 14.6)} & {\bf 12.7 (0.4, 24.9)} \\
& 30 & IED & 1.5 (-3.8, 6.8) & 0.3 (-1, 1.5) & 1.2 (-2.8, 5.3) \\
& & SAF & 2.8 (-1.8, 7.3) & 1.1 (-0.6, 2.8) & 1.6 (-1.2, 4.5) \\
& & Other Attack & 5.8 (-6.2, 17.8) & 2.2 (-1.9, 6.4) & 3.6 (-4.3, 11.4) \\
\hline
\end{tabular}}
\caption{Causal Effect Estimates and 95\% Confidence Intervals for Various Stochastic Interventions. We present the results for three interventions discussed in the main text: increasing the expected number of airstrikes from 1 to 3 per day for $M$ days, changing the focal points of airstrikes from $\alpha = 0$ to $\alpha = 3$ for $M$ days, and the lagged effects of increasing the expected number of airstrikes from 1 to 5 per day $M$ days ago.  The range of $M$ we consider is $\{3, 7, 30\}$. The regions of interest are Iraq, Baghdad, and the area outside Baghdad. The results in bold represent statistically significant estimates.} \label{tab:effectestimates}
\end{table}

\cref{tab:effectestimates} presents the numerical effect estimates and 95\% confidence intervals for various interventions, including those shown in the main text. We also show the effect estimates for the whole Iraq, Baghdad only, and the area outside Baghdad.

\subsection{Single time point adaptive interventions}
\label{app_subsec:adaptive}

Adaptive intervention strategies are often of interest in longitudinal settings, where previous outcomes might drive future treatment assignments. In our setting, these adaptive interventions would correspond to military strategies that depend on the observed history, such as the locations of previous insurgent attacks.  Although we leave full development of adaptive strategies to future research, we consider adaptive strategies that take place over a single time period, and then discuss the challenges of further extending it to the multiple time period interventions.

Here, we design adaptive dosage interventions over a single time period that closely resemble the observed data in terms of the expected number of airstrikes over time and their location. Using the observed number of airstrikes over time, we fitted a smooth function of time to obtain an estimate of the expected number of airstrikes over time, which is denoted by $\widehat n_t$. We used the estimated expected number of points to define adaptive interventions under which (1) the spatial distribution under the intervention is equal to the spatial distribution of airstrikes according to the propensity score, and (2) the expected number of points under the intervention is set to $c \widehat n_t$, with $c$ varying from 0.5 to 2 (representing a change in the number of airstrikes ranging from half to double the observed values). Formally, this intervention that depend on the observed history is given by:
 \[
 h_{t+1}(\omega; \history[t]) = \frac{ c \ \widehat n_t }{\int_\Omega h_{t+1}^{ps}(s; \history[t]) \mathrm{d}s}
 h_{t+1}^{ps}(\omega; \history[t]),
 \]
where $h_{t+1}^{ps}(\omega; \history[t])$ is the estimated propensity score intensity function. This definition of intensity ensures that that expected number of airstrikes at time $t$ is equal to $c \widehat n_t$ using the ratio term, and the relative likelihood of each location $\omega$ being treated is as specified in the estimated propensity score. This approach is related to the incremental propensity score of \cite{kennedy2019nonparametric} who considered non-spatial and non-temporal settings.

\begin{figure}[H]
     \centering
     \includegraphics[width=0.8\textwidth]{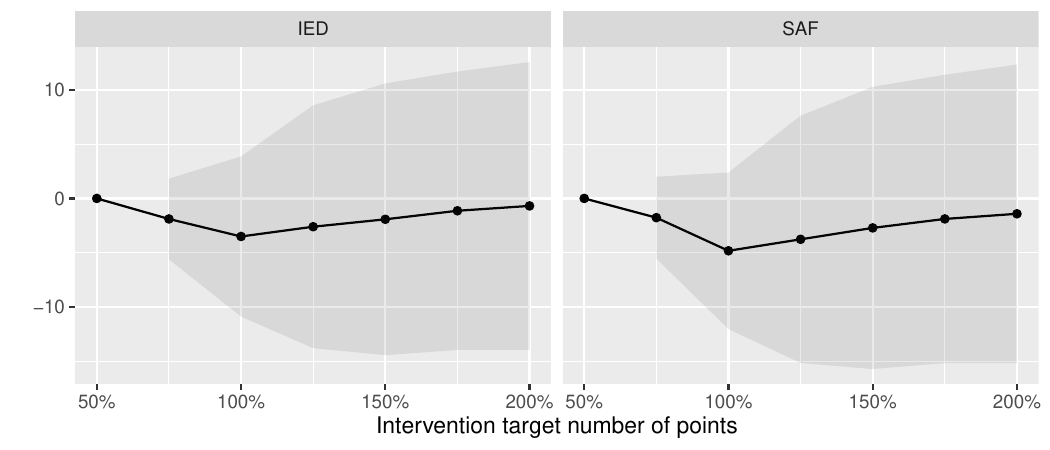}
     \caption{Effect estimates for a change in the expected number of airstrikes from 50\% $\widehat n_t$ to $c \widehat n_t$, for $c$ shown in the x-axis. Left plot shows effect estimates for IEDs and right plot shows effect estimates for SAF attacks in Iraq.}
     \label{fig:incremental_ps_results}
\end{figure}

Figure \ref{fig:incremental_ps_results} shows the effect estimates for number of IED and SAF attacks in Iraq for these interventions.  The result shows that the estimates are too imprecise to lead to a definitive conclusion.

Unfortunately, the evaluation of adaptive strategies over multiple time periods rapidly becomes complicated. Specifically, for interventions over multiple time periods that depend on the most recent history, we would need to have access to intermediate potential outcomes which are unobserved. Therefore, we would have to model the outcome process in order to predict the counterfactual outcomes that would then inform the adaptive treatment assignment in the subsequent time periods.  One advantage of our proposed framework is its ability to incorporate unstructured spillover and carryover effects.  This is possible because our framework does not require researchers to model the outcome process.  Given this difficulty, we will leave the complete investigation of adaptive spatio-temporal treatment strategies to future work.

\end{document}